\documentclass{ar-astro2e}
\usepackage{color}
\usepackage[breaklinks=true,colorlinks=true,linkcolor=blue,citecolor=blue,urlcolor=blue]{hyperref} 
\usepackage{amsmath,amssymb}
\usepackage{ARAstroBib}
\usepackage{graphicx}
\usepackage{longtable}
\usepackage{txfonts}
\usepackage{psfig}
\begin{document}

\jname{Annu. Rev. Astron. Astrophys.}
\jyear{2014}
\jvol{52}
\ARinfo{}

\def\swift{{\it Swift}}
\def\chandra{{\it Chandra}}
\def\simlt{\mathrel{\hbox{\rlap{\hbox{\lower4pt\hbox{$\sim$}}}\hbox{$<$}}}}
\def\simgt{\mathrel{\hbox{\rlap{\hbox{\lower4pt\hbox{$\sim$}}}\hbox{$>$}}}}
\def\nat{Nature}
\def\apj{ApJ}
\def\aap{A\&A}
\def\apjl{ApJ}
\def\mnras{MNRAS}
\def\aj{AJ}
\def\prd{Phys.~Rev.~D}
\def\apjs{ApJS}
\def\physrep{Physics Reports}
\def\araa{ARA\&A}
\def\nar{New Astronomy Reviews}
\def\pasp{PASP}
\def\aaps{A\&AS}
\def\apss{Ap\&SS}

\title{Short-Duration Gamma-Ray Bursts}
\author{Edo Berger
  \affiliation{Harvard-Smithsonian Center for Astrophysics, 60 Garden
    Street, Cambridge, \\ Massachusetts 02139; email:
    eberger@cfa.harvard.edu}}
\markboth{Edo Berger}{Short-Duration Gamma-Ray Bursts}

\begin{keywords}
gamma rays: observations, theory;  gravitational waves;  radiation
mechanisms: non-thermal;  relativistic processes;  stars: neutron
\end{keywords}

\begin{abstract}
  Gamma-ray bursts (GRBs) display a bimodal duration distribution,
  with a separation between the short- and long-duration bursts at
  about 2 sec.  The progenitors of long GRBs have been identified as
  massive stars based on their association with Type Ic core-collapse
  supernovae, their exclusive location in star-forming galaxies, and
  their strong correlation with bright ultraviolet regions within
  their host galaxies.  Short GRBs have long been suspected on
  theoretical grounds to arise from compact object binary mergers
  (NS-NS or NS-BH).  The discovery of short GRB afterglows in 2005,
  provided the first insight into their energy scale and environments,
  established a cosmological origin, a mix of host galaxy types, and
  an absence of associated supernovae.  In this review I summarize
  nearly a decade of short GRB afterglow and host galaxy observations,
  and use this information to shed light on the nature and properties
  of their progenitors, the energy scale and collimation of the
  relativistic outflow, and the properties of the circumburst
  environments.  The preponderance of the evidence points to compact
  object binary progenitors, although some open questions remain.
  Based on this association, observations of short GRBs and their
  afterglows can shed light on the on- and off-axis electromagnetic
  counterparts of gravitational wave sources from the Advanced
  LIGO/Virgo experiments.
\end{abstract}

\maketitle

\section{Introduction}
\label{sec:intro}

Gamma-ray bursts (GRBs) are short, intense, and non-repeating flashes
of $\sim {\rm MeV}$ $\gamma$-rays with a wide range of spectral and
temporal properties.  GRBs were discovered serendipitously by the {\it
  Vela} satellites starting in 1967 \citep{kso73}, leading to an
intense investigation of their origin, energy source, and progenitor
systems that is still on-going at the present.  Despite the broad
diversity in their $\gamma$-ray emission properties (peak energy,
spectral shape, variability timescale, duration), it was recognized
primarily based on data from the Burst and Transient Source Experiment
(BATSE) on-board the Compton Gamma-Ray Observatory that GRBs can be
generally divided into two groups based on their duration and spectral
hardness: the short-hard and long-soft bursts, with a separation at
about 2 sec \citep{ncd+84,dbt+92,kmf+93}.

For both GRB classes, the uniform projected distribution of the bursts
on the sky and the fluence distribution (the so-called ${\rm log}
N-{\rm log}S$ or $V/V_{\rm max}$ distributions) pointed to a
cosmological origin (e.g.,
\citealt{van83,goo86,pac86,shh88,mfw+92,pir92,sch01,gp05}).  Despite
these statistical tests, the distance scale of GRBs continued to be
debated at least into the mid-1990s \citep{lam95,pac95}.  On the other
hand, the non-repeating nature of GRBs, their harder non-thermal
spectra, and non-Euclidean space distribution separated them from the
soft $\gamma$-ray repeaters (SGRs; e.g.,
\citealt{knc+87,lfk+87,nhw+91}), which originate from magnetars in the
Milky Way and nearby galaxies \citep{kds+98}; it is important to note,
however, that some events classified as short GRBs may represent
extragalactic giant SGR flares with a long recurrence timescale
\citep{aaa+08a,omq+08,hrb+10,aaa+12b}.

In the framework of a cosmological origin, the large energy release
(up to an isotropic-equivalent value of $\sim 10^{54}$ erg), short
variability timescale (down to milliseconds), and observed non-thermal
$\gamma$-ray spectra led to the ``compactness problem''; namely, the
observed properties result in an enormous optical depth to
pair-production, and hence an expected thermal emission \citep{rud75}.
A key result in GRB research was the realization that this problem can
be resolved by invoking relativistic expansion with a large bulk
Lorentz factor of $\Gamma\simgt 10^2$ (e.g.,
\citealt{pac86,kp91,feh93,bh97}).  The large Lorentz factor in turn
requires exceedingly clean explosions with ejecta masses of $\simlt
10^{-5}$ M$_\odot$, the so-called baryon loading problem \citep{sp90}.
In addition, it was recognized that the interaction of the
relativistic outflow with the surrounding medium will generate
synchrotron emission ranging from radio to X-rays, with a longer
duration than the prompt $\gamma$-ray emission -- the ``afterglow''
\citep{rm92,mr93,pr93,mr97,spn98}.  The detection of such afterglow
emission became an intense pursuit since it held the key to precise
positions and hence a distance scale to the bursts.

Motivated in part by the $\sim 0.1$ Gyr merger timescale of the
Hulse-Taylor binary \citep{ht75}, seminal studies by \citet{pac86} and
\citet{elp+89} pointed out that neutron star binary mergers may lead
to $\gamma$-ray emission with an energy scale of $\simgt 10^{50}$ erg,
typical of a cosmological GRB.  Such mergers are also expected to
result in an outflow with a low baryon load due to the absence of a
dense surrounding environment, and can therefore drive relativistic
outflows.  A more detailed investigation of the neutron star binary
merger model was carried out by \citet{npp92}, who pointed out that
testable predictions include significant offsets from the host
galaxies due to natal kicks imparted to the binaries at birth, and
gravitational wave emission in the frequency range of the Laser
Interferometer Gravitational Wave Observatory (LIGO).  \citet{lmr+77}
and \citet{elp+89} also proposed these mergers as a potential site for
$r$-process nucleosynthesis.  Subsequent to these initial studies, a
wide range of investigations of neutron star binary (NS-NS) and
neutron star black hole binary (NS-BH) mergers were carried out with
the aim of exploring their gravitational wave emission, the production
of $r$-process enriched ejecta, and the possible production of GRBs.
In parallel, an alternative model of GRBs from core-collapse
supernovae (the collapsar model) was developed \citep{mw99}, with
clear predictions about the resulting population that included an
exclusive origin in star-forming galaxies and association with
core-collapse supernovae.

The major breakthrough in the study of GRBs, their distance and energy
scale, environments, and progenitors came from the eventual discovery
of afterglows from long GRBs starting in 1997
\citep{cfh+97,fkn+97,vgg+97}.  The localization of the bursts to
arcminute scale (in X-rays) and to sub-arcsecond scale (in optical and
radio), led to redshift measurements from optical spectroscopy of the
afterglows and host galaxies, which directly demonstrated a
cosmological origin \citep{mdk+97,kdo+99}.  The afterglow emission
also provided observational evidence for relativistic expansion
\citep{wkf98,tfb+04}, jet collimation with typical opening angles of
$\sim 3-10^\circ$ \citep{hbf+99,sgk+99,fks+01}, a beaming-corrected
energy scale of $\sim 10^{51}$ erg \citep{fks+01,bfs01,bkf03}, and a
typical circumburst density of $\sim 1-10$ cm$^{-3}$ with evidence for
mass loss from the massive progenitor stars in some cases
\citep{cl00,pk02,yhs+03}.  In addition, follow-up studies of the hosts
demonstrated an exclusive location in star-forming galaxies
\citep{bdk+98,dkb+98,chg04,wbp07}, while high resolution imaging with
the {\it Hubble Space Telescope} ({\it HST}) showed that long GRBs
follow the radial distribution expected for star formation in disk
galaxies \citep{bkd02}, and are spatially correlated with bright
star-forming regions in their hosts \citep{fls+06}.  Moreover, long
GRBs were shown to be associated with Type Ic supernovae, based on
both photometric and spectroscopic observations
\citep{gvv+98,hsm+03,smg+03,wb06,hb12}.  Taken together, the
environments and supernova associations indicated that long GRBs arise
from the death of massive stars, and not from compact object binary
mergers.

Despite the impressive pace of discovery for long GRBs, the study of
short GRBs proved much more challenging, although compact object
binary mergers remained an attractive progenitor model.
\citet{kab+01} carried out rapid but shallow optical follow-up of
three short GRBs, with non-detections to a limit of $\sim 13-15$ mag.
\citet{hbc+02} presented deeper optical follow-up (to limits of
$16-20$ mag with a delay of $\sim 2$ d) and the first radio follow-up
(to limits of about $0.5-1.5$ mJy with a similar delay) for four short
GRBs, but none were detected.  In light of the existing afterglow
detections summarized in this review, these early searches were
woefully inadequate.  At the same time, \citet{pkn01} argued that a
lower energy scale, and potentially lower circumburst densities, will
result in dimmer afterglows than for long GRBs, by at least an order
of magnitude (e.g., typical optical brightness of $\sim 23$ mag at 10
hours post-burst).  Similarly, \citet{pb02} argued based on compact
object binary population synthesis models that due to kicks such
mergers will tend to occur in lower density environments than long
GRBs ($\sim 0.1$ cm$^{-3}$), and hence lead to fainter afterglows, if
they indeed produced short GRBs.

\begin{figure}
\centerline{\psfig{figure=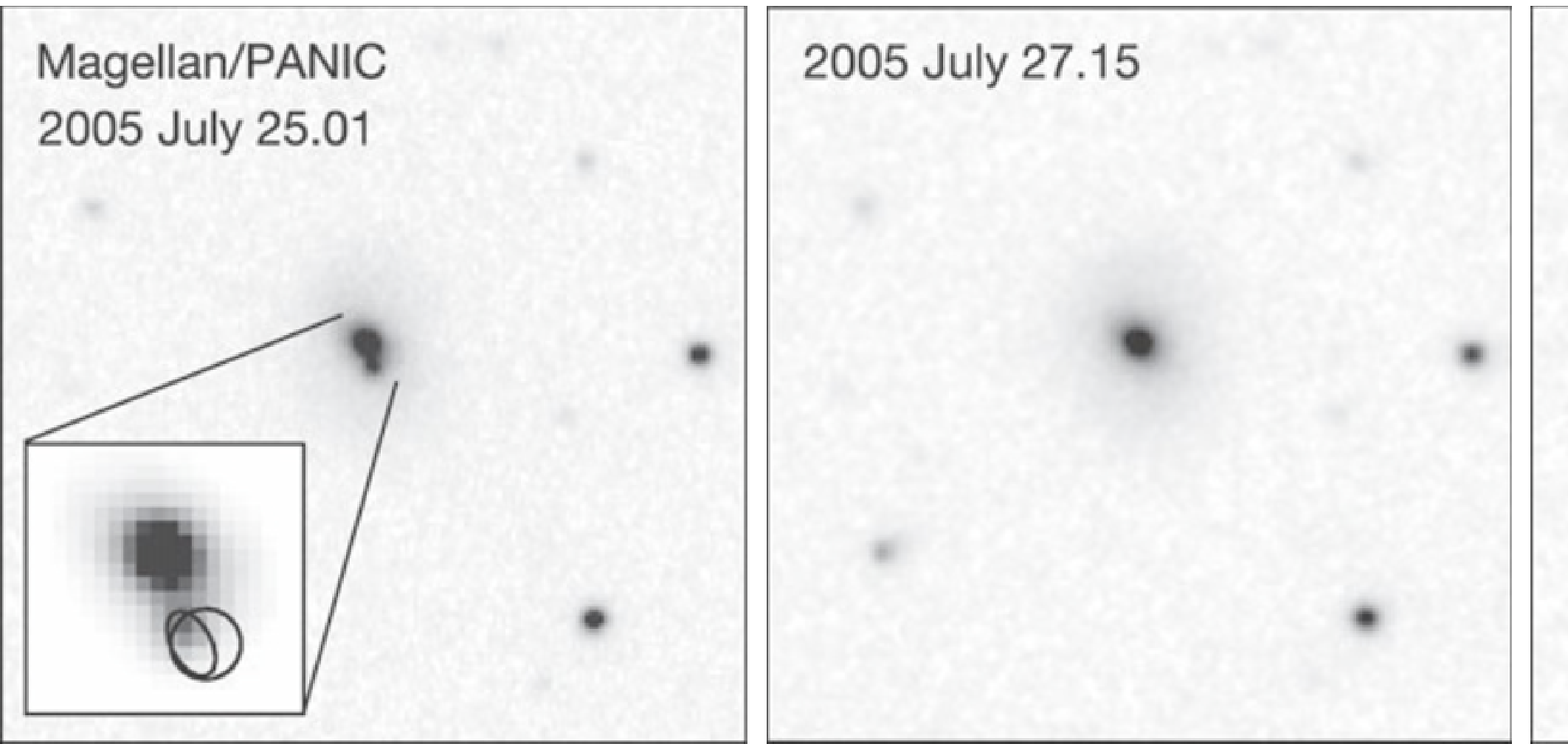,width=\textwidth}}
\caption{Near-infrared images of the afterglow and host galaxy of the
  short GRB\,050724, obtained 0.49 d (left) and 2.63 d (middle)
  post-burst, reveal a fading afterglow (right) on the outskirts of an
  elliptical galaxy.  The inset in the left panel shows the locations
  of the X-ray afterglow detected with {\it Chandra} (circle) and the
  radio afterglow detected with the VLA (ellipse).  This was the first
  short GRB with a definitive association to an elliptical galaxy, and
  a broad-band afterglow spanning radio to X-rays.  From
  \citet{bpc+05}.}
\label{fig:050724}
\end{figure}

As in the case of long GRBs, the watershed moment in the study of
short GRBs came with the discovery of the first afterglows in
May--July 2005, following bursts from the {\it Swift} \citep{gcg+04}
and HETE-2 \citep{rac+03} satellites.  Rapid \swift\ X-ray Telescope
follow-up of GRB\,050509b led to the discovery of the first X-ray
counterpart, with a localization of about $9''$ radius \citep{gso+05}.
Deep follow-up optical observations revealed no afterglow emission,
but instead uncovered a massive elliptical galaxy at $z=0.225$ near
the X-ray error circle, with a chance coincidence probability of $\sim
10^{-3}$ \citep{cdg+05,gso+05,bpp+06}.  Assuming that the association
with the galaxy (and hence the redshift) was correct, it was also
shown that the burst lacked a supernova \citep{hsg+05}.  Shortly
thereafter, HETE-2 discovered the short GRB\,050709 \citep{vlr+05},
and follow-up observations with the {\it Chandra X-ray Observatory}
precisely localized the X-ray afterglow \citep{ffp+05}, with
subsequent observations revealing the first short GRB optical
afterglow \citep{hwf+05}.  The resulting sub-arcsecond position
pinpointed the origin of the burst to the outer regions of a
star-forming galaxy at $z=0.160$ \citep{ffp+05}, while optical
follow-up ruled out the presence of an associated supernova
\citep{hwf+05}.  Finally, the detection of the short GRB\,050724 by
\swift\ \citep{bcb+05} led to the discovery of X-ray, optical/near-IR,
and the first radio afterglow, and a definitive localization in an
elliptical galaxy at $z=0.257$ (Figure~\ref{fig:050724};
\citealt{bpc+05}).  The combination of radio to X-ray afterglow
emission also demonstrated that both the energy and density scale were
lower than for long GRBs \citep{bpc+05}.  Taken together, these three
early events demonstrated that short GRBs are cosmological in origin,
that they produce afterglow emission similar to that of long GRBs but
with a lower energy and density scale, and that their progenitors are
not massive stars.

In the decade following the initial discovery of short GRB afterglows,
over 70 short GRBs have been discovered by \swift\ and other
$\gamma$-ray satellites, with a substantial detection fraction of
X-ray and optical afterglows, and a few radio afterglow detections.
The resulting localizations have also led to the identification of
nearly 40 host galaxies, and have enabled studies of the sub-galactic
locations of short GRBs.  This is the first substantial sample of
short GRBs with extensive afterglow and host galaxy data.  The primary
goal of this review is to utilize and synthesize the information from
short GRB afterglows and environments for the first time in order to:
(i) delineate the basic observational properties of short GRBs and
their environments (from parsec to galactic scales); (ii) use this
information to confront theoretical progenitor models; (iii) determine
some of the basic properties of the progenitors and the resulting
explosions; and (iv) explore the expected electromagnetic counterparts
of Advanced LIGO/Virgo gravitational wave sources in the context of
compact object binary progenitors.

In this vein, the outline of the paper is as follows.  In
\S\ref{sec:prompt} I summarize the basic properties of the prompt
$\gamma$-ray emission, compare these with the $\gamma$-ray properties
of long GRBs, and discuss the issue of short versus long GRB
classification.  In \S\ref{sec:prog} I describe current short GRB
progenitor models.  In \S\ref{sec:sample} I define the existing sample
of short GRBs, including the completeness for afterglow and host
galaxy detections.  In \S\ref{sec:nonmassive} I summarize basic
observational arguments for a distinct origin from the massive star
explosion that produce long GRBs.  In \S\ref{sec:hosts} I provide a
detailed analysis of short GRB host galaxy properties, and use this
information to shed light on the age distribution of the progenitors.
In \S\ref{sec:subgal} I zoom in on the sub-galactic locations of short
GRBs in terms of their radial offset distribution and their spatial
relation to the underlying ultraviolet and optical light distribution
of their hosts, and compare these results with the location of long
GRBs and supernovae, and with predictions from compact object binary
population synthesis models.  In \S\ref{sec:afterglow} I turn to the
afterglow properties in X-rays, optical, and radio, and use a
comparison to both long GRB afterglows and afterglow theory to provide
insight on the energy scale of short GRBs, their collimation, and
their circumburst environments.  In \S\ref{sec:kilonova} I discuss the
expected production and optical/near-IR emission from $r$-process
ejecta in compact object binary mergers, and compare these predictions
to observations of the short GRB\,130603B, the first burst to show
evidence for such emission.  In \S\ref{sec:gwem} I use the inferred
properties of short GRBs and their afterglows to explore on- and
off-axis electromagnetic emission from gravitational wave sources
detected by the upcoming Advanced LIGO/Virgo detectors.  I conclude
with the key results and a discussion of future directions in
\S\ref{sec:conc}.

In addition to the comprehensive review of short GRBs presented here,
I also refer the reader to several previous reviews that provide
additional insight on GRB theory and observations.  These include a
review of GRBs in the BATSE era \citep{fm95}; an early review of long
GRB afterglow discoveries, the emerging GRB-supernova connection, and
afterglow theory \citep{vkw00}; comprehensive reviews of the prompt
emission and afterglow theory \citep{mes02,pir04}; a detailed review
of the long GRB-supernova connection \citep{wb06}; a preliminary
review of short GRBs, focused primarily on theoretical issues, and
including data from the first few well-localized events in 2005
\citep{nak07}; a comprehensive review of short GRB progenitor models,
primarily compact object mergers and collisions \citep{lr07}; a review
of the primary early results from the \swift\ satellite, including in
part the detections of short GRBs \citep{grf09}; and finally, an
initial review of short GRB galactic and sub-galactic environments
\citep{ber11}.

\section{The Prompt Emission and the Short-Long Divide}
\label{sec:prompt}

The defining feature of GRBs among the wide variety of known
astrophysical transients is their dominant, non-repeating, and
non-periodic prompt $\gamma$-ray emission, which displays a rich
phenomenology in terms of duration, variability, spectral parameters,
fluence, peak flux, temporal and spectral evolution, and various
correlations between these parameters.  Within this broad range of
properties, the clearest sub-classes are the short-hard and long-soft
bursts.  It is important to note that generally the short end of the
short GRB duration distribution is affected by detector trigger times
(tens of ms), while at the long end there is some overlap with the
duration distribution of long GRBs, such that some events belonging to
the short burst progenitor population will have durations of $\simgt
2$ s, while some events with duration of $\simlt 2$ s will have long
GRB progenitors (e.g., \citealt{bnp+13}).  In addition, since various
instruments measure durations in different energy bands, it is
possible that the value of $T_{90}$ separating short and long GRBs
varies for each sample; for example, \citet{bnp+13} suggest that for
{\it Swift} the division is at $T_{90}\approx 0.8$ s.  The effect of
cosmological time dilation adds another potential complication, as
indicated for example by the high-redshift GRBs 080913 ($z=6.7$) and
090423 ($z=8.3$) whose observed durations place them in the long GRB
category, while their rest-frame durations of $\approx 1$ s are
nominally in the short GRB range \citep{zzv+09,bhf+10}.

Such effects can be potentially mitigated by the use of auxiliary
information, motivated in part by our knowledge of long GRBs (e.g.,
their association with Type Ic SNe, their exclusive location in
star-forming galaxies), and partly by expectations for short GRB
progenitors.  In this vein, \citet{zzl+07} and \citet{zzv+09} advocate
a new classification scheme into Type I GRBs (compact object mergers)
and Type II GRBs (massive star progenitors), motivated by the
nomenclature for supernovae.  To determine the type of each burst,
they argue for the use of information such as redshift, supernova
association, host galaxy type and star formation rate, burst offset
relative to the host galaxy, the density profile of the circumburst
medium, the $\gamma$-ray and afterglow energy scale, and a combination
of prompt emission properties (see Figure 8 of \citealt{zzv+09}).
\citet{llz+10} instead advocate the use of a new prompt emission
empirical parameter, $E_{\rm \gamma,iso}/ E_p^{5/3}$, with a value of
$\simlt 0.03$ coupled with a rest-frame $T_{90}\simlt 5$ s separating
Type I and II events; here $E_{\rm \gamma,iso}$ is the
isotropic-equivalent $\gamma$-ray energy and $E_p$ is the spectral
peak of the prompt emission.  \citet{bbp08} propose a classification
scheme based on physical properties of the progenitor systems, for
example destructive (Type I) versus non-destructive (Type II) events,
with further separation into sub-classes based on the presence or
absence of a degenerate object (and whether it is a neutron star or
black hole), and membership in a binary system.

While such classification schemes are potentially more powerful than
relying on duration and hardness alone, and in the long run may be
essential for a complete mapping of GRB progenitors, they also carry
the risk of biasing the results.  For example, the requirement of
elliptical hosts for Type I GRBs will obscure the fact that their
presumed compact object binary progenitors occur in both star-forming
and elliptical galaxies, and may therefore impede the ability to
determine properties such as the delay time distribution from the host
galaxy demographics (e.g., \citealt{zr07,ber09,lb10,fbc+13}).
Similarly, the requirement of a lower energy scale or larger physical
offsets for Type I GRBs will restrict our understanding of the full
range of properties spanned by the progenitors and explosions.  In the
case of a progenitor-based classification scheme, it is currently
unclear either from observations or from theory how different systems
manifest in terms of unique observables, particularly in the context
of the prompt $\gamma$-ray emission for which even the radiation
mechanism has not been conclusively identified.  Finally, since
complete knowledge of the prompt emission, distance scale, afterglow
parameters, and host galaxy properties is only available for a small
subset of events, it is currently more profitable to begin the
investigation of short GRBs with a broad sample that may guide future
classification schemes and observational follow-up efforts.
Ultimately, it is important to recognize that while the analogy with
the supernova classification scheme is tempting, that classification
is based on spectroscopic signatures that already provide critical
(though not always definitive) physical insight into the nature of the
explosions (e.g., \citealt{fil97}).  For GRBs, on the other hand, the
starting point for any classification scheme is the prompt emission,
which exhibits many similarities between long and short GRBs (as I
discuss below), and even after four decades of investigation is not
fully understood.

Thus, in this review I pursue the simplest approach of studying a
sample of short GRBs based on a duration cut of $T_{90}\simlt 2$ s.
It is possible that some events in the sample are produced by long GRB
progenitors, and that some events with $T_{90}\simgt 2$ s that belong
to the short GRB progenitor population are not included in the sample.
However, as I discuss in detail in subsequent sections, the overall
differences in the resulting short and long GRB populations in terms
of afterglow and host galaxy properties are sufficiently clear that
they both shed light on the nature of short GRB progenitors (despite
any marginal contamination), and can serve as the basis for future
classification schemes by revealing the most relevant and critical
differences.

\subsection{Comparative Studies of the Prompt Emission}

Within the context of a duration-based classification scheme, there
have been multiple comparative studies of the prompt emission
properties of short and long GRBs, based on the samples from various
instruments, primarily BATSE, {\it Swift}, and most recently {\it
  Fermi}.  These studies have uncovered both differences and
commonalities in the prompt emission of short and long GRBs.  Below I
provide a short overview of these results, and I refer the reader to
\citet{nak07} and \citet{grf09} for a more detailed discussion.

In the framework of the phenomenological Band function which was
constructed to fit the spectrum of the prompt emission \citep{bmf+93},
a key finding is that short GRBs generally have harder spectra than
long GRBs due to a combination of a shallower low-energy spectral
slope, $\alpha\approx -0.4$ versus $\approx -0.9$, and a higher
spectral peak, $E_p\approx 400$ versus $\approx 200$ keV (e.g.,
\citealt{pbp+03,gng+09,ggn11}).  However, these differences are less
evident when the sample is restricted to short GRBs with the highest
peak fluxes \citep{pbp+03,kpb+06}, or when considering only the first
$\sim 2$ s of long GRB light curves \citep{ggc04}.  Similarly, the
variability timescales of short GRBs are comparable to those in the
first $\sim 2$ s of long GRBs \citep{np02}.

Another difference between short and long GRBs is evident in the
distribution of spectral lags, which are generally $\simgt 0$ s for
long GRBs (i.e., they exhibit a hard to soft evolution), while they
are $\sim 0$ s ($\pm 20$ ms) for short GRBs \citep{nb06,ngs11}.  Some
long GRBs also exhibit negligible lags, but those are events with the
highest peak luminosities, which are distinct from the zero lag, low
peak luminosity short GRBs \citep{nb06}.  The difference in spectral
lags between short and long GRBs may ultimately provide another useful
prompt emission discriminator in the context of classification schemes
(e.g., \citealt{gnb+06}).

Finally, it has been noted that short GRBs occupy a distinct region in
the $E_p-E_{\rm \gamma,iso}$ phase-space than long GRBs, namely they
have lower $E_{\rm \gamma,iso}$ values for a given $E_p$ than long
GRBs \citep{aft+02,gng+09,qc13,tyn+13}.  It remains to be seen,
however, whether both sub-classes follow a similar relation if only
the first $\sim 2$ s of long GRBs are considered when determining
their $E_p$ and $E_{\rm \gamma,iso}$ values.

On the other hand, an overall similarity between short and long GRBs
has been found in terms of a common inverse correlation between the
duration and intensity for individual pulses \citep{hp11,ngs11}.  In
addition, short and long GRBs appear to follow the same relation
between $E_p$ and $L_{\rm peak}$ \citep{ymn+04,gng+09}.

To date only two short GRBs have been detected at $\sim {\rm GeV}$
energies with the {\it Fermi} Large Area Telescope.  GRB\,081024B was
detected at $\sim 0.1-3$ GeV with a slight delay and a duration of
about 3 s, compared to $T_{90}\approx 0.8$ s at MeV energies
\citep{aaa+10a}.  Even more extreme GeV emission (up to $\sim 30$ GeV)
was observed in GRB\,090510, with a delay of about 0.6 s and extending
to $\sim 200$ s \citep{aaa+09a,aaa10b}.  The delayed onset and longer
duration of the GeV emission in both long and short GRBs have been
interpreted as evidence for a hadronic origin \citep{agm09}, a
synchrotron self-Compton origin \citep{aaa10b}, or an afterglow origin
\citep{aaa10b,ggn10,kb10}.

\subsection{Extended Emission}

An interesting feature of some short GRBs is the presence of extended
$\gamma$-ray emission that is softer than the prompt spike, lasts
$\sim 10-100$ s, and sometimes exhibits a delayed onset.  The presence
of such an emission component was first noted in a stacking analysis
of BATSE short GRBs by \citet{lrg01} who demonstrated that the
emission was softer than both the prompt spike and long GRBs of
similar durations.  These authors interpreted the soft extended
component as the onset of the X-ray afterglow.  Subsequent studies
confirmed the presence of extended emission in at least some BATSE and
Konus short bursts \citep{con02,fag+04}.

The presence of soft extended emission in individual bursts garnered
attention with the discovery of the first short GRB afterglows.
GRB\,050709 exhibited an initial spike with $T_{90}\approx 0.07$ s in
the $30-400$ keV band, followed by a pronounced bump in the $2-25$ keV
range with a duration of about 130 s, and a larger fluence by about a
factor of 2.5 compared to the initial spike \citep{vlr+05}.  A similar
component was observed in the GRB\,050724, with emission in the
$15-25$ keV band lasting about 100 s, but accounting for only 10\% of
the fluence of the initial spike \citep{bcb+05}.  A particularly
extreme example is GRB\,080503, in which the extended emission
dominated over the fluence of the prompt spike by a factor of 32
\citep{pmg+09}.

A study by \citet{nb06} identified eight BATSE short bursts with
individual detections of extended emission, with a counts ratio
relative to the initial spike ranging up to $\sim 40$, and an
estimated overall dynamic range of this ratio of $\sim 10^4$.  These
authors also noted that short GRBs with and without extended emission
share a negligible spectral lag in the initial spike.  A follow-up
study of extended emission in {\it Swift} short bursts by
\citet{ngs10} indicated that about one-quarter of the events clearly
exhibit extended emission, while the bulk of the remaining events lack
such a component.  However, I note that not all of the bursts in their
sample have a genuine initial short spike, and the incidence rate of
extended emission may be closer to $\sim 15\%$.  Building on this
sample, \citet{ngs11} argued that bursts with extended emission have
durations and pulse profiles in the prompt spike that are $\sim 2-3$
times longer than events lacking extended emission, potentially
indicative of differences in the central engine.

Two other note-worthy events in the context of extended emission and
the $T_{90}$ divide between long and short bursts are GRBs 060505 and
060614, with durations of 4 and 102 s, respectively, low redshifts of
0.089 and 0.125, and yet no supernova associations to limits of about
$6-7$ mag fainter than typical long GRB-SNe
\citep{dcp+06,fwt+06,gfp+06,gnb+06}.  The prompt emission of
GRB\,060505 exhibited a significant spectral lag, typical of long
GRBs, and the burst coincided with a low metallicity and young
star-forming region of its late-type host galaxy
\citep{ocg+07,tfo+08}, matching the properties of long GRB
environments.  The $\gamma$-ray emission of GRB\,060614 exhibited an
initial 5-sec spike with negligible spectral lag, typical of short
GRBs, followed by extended emission with significant variability,
somewhat softer spectrum, and a larger fluence than the prompt spike
by about a factor of 5 \citep{gnb+06}.  The host galaxy of GRB\,060614
has a low luminosity typical of long GRB hosts, but a lower than
average specific star formation rate \citep{gfp+06}.  Moreover, the
burst itself was located at a significant offset from the host in a
region with little evidence for underlying ultraviolet emission
\citep{gfp+06}.  Thus, GRBs 060505 and 060614 present a challenge to
any GRB classification scheme, with their prompt emission sharing
properties typical of both long and short GRBs, their host
environments similarly spanning locations typical of long and short
GRBs, and the lack of supernova association suggesting a non-massive
star progenitor.

There have been attempts to discern other differences between the
population of short GRBs with and without extended emission.
\citet{ngs11} claim that short bursts with extended emission have
larger X-ray fluxes at $\sim 100$ s post-burst and longer lasting
X-ray afterglows.  However, it is not obvious that the X-ray emission
at such early times is not directly related to the soft extended
emission itself.  \citet{tko+08} used a sample of short GRBs up to
mid-2007 to argue that events with extended emission have
systematically smaller spatial offsets from their hosts, and hence
distinct progenitors.  However, subsequent studies of the offset
distribution indicate that short GRBs with extended emission span as
wide an offset range as events with no extended emission
\citep{fbf10,fb13}; a Kolmogorov-Smirnov (K-S) test comparing the two
offset distributions leads to a $p$-value of 0.9, indicating that
there is no evidence for two distinct populations \citep{fb13}.  In
addition, \citet{fbc+13} find that the distributions of host galaxy
types for short GRBs with and without extended emission are currently
indistinguishable.

Motivated by the discovery of extended emission, several attempts have
been made to provide a theoretical explanation.  \citet{mqt08} propose
that short GRBs with extended emission are produced by the formation
of a rapidly rotating proto-magnetar through accretion-induced
collapse of a white dwarf, the merger of a white dwarf binary, or the
merger of a neutron star binary.  In this framework, the prompt spike
is due to the standard picture of accretion onto the central object,
while the extended emission is powered by a relativistic wind that
extracts the proto-magnetar's rotational energy.  \citet{bmt+12}
discuss a slight variation of the magnetar model, in which the delay
in the onset of extended emission is due to a breakout of the
relativistic outflow through a baryon-loaded wind from the
proto-magnetar.  \citet{maq+10} suggest instead that the gap between
the prompt spike and extended emission may be due to heating from
$r$-process nucleosynthesis, which momentarily halts fall-back
accretion onto the central object.  In this scenario, events lacking
extended emission are due to a timescale for the $r$-process heating
of $\simgt 1$ s, which leads to a complete cut-off in fallback
accretion, while those with extended emission resume fall-back
accretion after a delay.  \citet{mrz05} propose a model in which the
short GRBs with extended emission are due to the accretion-induced
collapse of a neutron star (generating the prompt spike) followed by
interaction of the relativistic outflow with the non-degenerate
stellar companion (leading to a delay followed by soft extended
emission).

\subsection{X-ray Flares}

The discovery of early X-ray flares following the prompt $\gamma$-ray
emission was enabled by the rapid follow-up capabilities of {\it
  Swift} \citep{brf+05,nkg+06,cmr+07,fmr+07}.  The X-ray flares
generally follow a fast rise exponential decay profile with $\Delta
t/t\ll 1$, and share spectral and temporal properties with the prompt
emission, pointing to a common origin in activity from the central
engine \citep{cmm+10,mgc+10}; here, $\Delta t$ is the flare duration
and $t$ is the time since the onset of the burst.  In the case of
short GRBs, \citet{mcg+11} carried out a systematic search for flares,
and found eight {\it Swift} events that exhibit flaring activity.  The
sample includes bursts with and without extended emission, and in both
early- and late-type host galaxies.  The resulting flares follow some
of the relations of long GRB flares, for example a positive
correlation between the flare width and time since the burst.  On the
other hand, the short GRB flares are generally weaker in terms of flux
contrast relative to the underlying X-ray emission (with $\Delta
F/F\simlt 2$), have about 1\% of the peak luminosity and
isotropic-equivalent energy of long GRB flares, and like the prompt
emission, do not follow the lag-luminosity relation of long GRB
flares.  The lower peak luminosity compared to long GRB flares remains
true even when scaled by the peak luminosity of the prompt emission,
but short and long GRB flares track the same negative correlation
between the ratio of peak flare to prompt luminosity as a function of
the ratio of flare peak time to burst duration.  Potentially related
to this point, the ratio of flare to prompt isotropic-equivalent
energy is similar for both short and long GRBs.  Given these results,
\citet{mcg+11} conclude that flares in both short and long GRBs are
related to the prompt emission.

Flares and re-brightenings at late time ($\sim {\rm hours}$) have also
been detected in some short GRBs, with distinct properties from the
early flares described above.  GRB\,050724 exhibited a significant
X-ray flare centered at $\approx 14$ hr with $\Delta F/F\approx 40$
and $\Delta t/t\approx 1$, with apparent associated emission in the
optical/near-IR \citep{ctl+06,gbp+06,mcd+07}.  The fact that the flare
is superposed on an underlying single power law decline suggests a
distinct, though currently not understood origin from the afterglow.
A delayed peak was found in the optical (and potentially X-ray)
emission from GRB\,080503 at $\sim 1$ day, although the subsequent
evolution matched typical afterglow emission rather than a distinct
flare \citep{pmg+09}.  Finally, excess X-ray emission at $\simgt 1$ d
was found in GRB\,130603B, with no corresponding emission in the
optical or radio bands, and an apparent single power law behavior
compared to the early X-ray emission \citep{fbm+13}.  Thus, the late
flares and re-brightenings are likely to have a distinct origin from
the early rapid X-ray flares.

Several ideas have been proposed to explain flares in short (and long)
GRBs.  \citet{paz06} suggest that fragmentation in the outer accretion
disk, caused by gravitational instabilities, could lead to large
amplitude variations of the central engine.  \citet{ros07} propose
that the flares can be caused by fall-back accretion of material
launched into eccentric orbits during a compact object merger (see
also \citealt{lrl09}).  \citet{pz06} suggest that the accretion can be
episodically stopped and re-started through variations in the
accumulated magnetic flux around the accreting central engine.
Finally, \citet{gia06} propose that the flares can be produced by
delayed magnetic reconnection as a strongly magnetized outflow
decelerates through interaction with the circumburst medium; unlike
the other proposed models, this scenario does not require a
re-activation of the central engine.

\subsection{Precursors}

Precursor $\gamma$-ray emission, preceding the main event by tens of
seconds, has been found in $\sim 20\%$ of long GRBs \citep{laz05}, and
has been attributed to either the transition of the expanding fireball
to the optically thin regime, or to the breakout of the relativistic
jet from the surface of the progenitor star (e.g., \citealt{lb05}).
If the former scenario is correct, precursors should also be present
in short GRBs.  In addition, precursors in short GRBs may be produced
by other mechanisms, for example magnetospheric interaction between
the compact objects \citep{hl01} or resonant shattering of the neutron
star crusts \citep{trh+12} prior to the merger.  \citet{trg10}
conducted a search for precursors in 49 {\it Swift} short GRBs with
and without extended emission, and uncovered 5 potential candidates
(with $2-5.5\sigma$ significance), of which 3 candidates (GRBs
081024A, 090510, and 091117) had likely independent detections in
other satellite data.  In those 3 cases, the precursor candidates
occurred $0.5-2.7$ s prior to the main spike, and it is therefore not
clear if these emission episodes are indeed distinct precursors, or a
part of the prompt emission complex.

\section{Short GRB Progenitor Models}
\label{sec:prog}

The bimodality of GRB durations is highly suggestive of two dominant
progenitor populations.  The short durations of short GRBs, down to
tens of milliseconds, point to compact progenitor systems with a
dynamical timescale of milliseconds.  In this context, the most
popular progenitor model is the merger of compact object binaries
comprised of two neutron stars or a neutron star and a black hole
(NS-NS/NS-BH; \citealt{elp+89,npp92}).  The merger occurs due to
angular momentum and energy losses by gravitational wave radiation.
In the binary neutron star case, the expected remnant is a black hole
surrounded by a hyper-accreting disk of debris, while a NS-BH merger
can lead to the same configuration if the neutron star is tidally
disrupted outside of the black hole's horizon.  The resulting
interplay of high accretion rate and rapid rotation can lead to energy
extraction via neutrino-antineutrino annihilation or
magnetohydrodynamic processes (e.g., \citealt{bz77,rr02,lr07}), which
in turn drive a collimated relativistic outflow.  As discussed in
\S\ref{sec:intro}, NS-NS/NS-BH mergers were proposed as GRB
progenitors prior to the discovery of the duration bimodality since
they provided a known source population with roughly the correct event
rate, the requisite rapid release of a large energy reservoir, and a
clean environment to avoid significant baryon loading.

The compact object merger model leads to several testable
observational predictions.  First, the delay time between the binary
formation and eventual merger is expected to span a wide range that
depends on the initial separation and consituent masses, $\tau_{\rm
  GW}\propto a^4/(\mu M^2)$, where $a$ is the initial binary
separation, $M\equiv M_1+M_2$ is the total binary mass and $\mu\equiv
M_1M_2/M$ is the reduced mass.  As a result of the wide delay time
distribution, the resulting short bursts will occur in both early- and
late-type galaxies (e.g., \citealt{bpb+06,zr07}).  This is indeed the
case for the small observed population of Galactic NS-NS binaries,
which have coalescence timescales of tens of Myr to much longer than a
Hubble time (e.g., \citealt{bdp+03,clm+04}).  Second, natal kicks
imparted to the binary system during the supernova explosions that
gave rise to the neutron stars and/or black hole, coupled with the
broad range of merger timescales, should lead to some mergers at large
offsets from their birth sites and host galaxies (tens to hundreds of
kpc; \citealt{npp92,bsp99,fwh99,pb02,bpb+06}).  A broad spatial
distribution is also expected if some short GRBs result from
dynamically-formed binaries in globular clusters \citep{gpm06,lrv10}.
Third, the mergers will be accompanied by strong gravitational wave
emission, detectable with the Advanced LIGO/Virgo detectors to about
200 Mpc for NS-NS mergers \citep{aad+92,har10,aaa+11a}.  Fourth, the
mergers will produce neutron-rich ejecta, which will in turn lead to
$r$-process nucleosynthesis; the decay of the resulting radioactive
elements may be detectable at optical/near-IR wavelengths
\citep{ls76,lp98,mmd+10,bk13}.  Finally, the mergers will not be
accompanied by supernova explosions.  Most of these predictions can
now be tested with existing observations of short GRBs, as delineated
in subsequent sections of this review.

In principle, some of these properties may differ between NS-NS and
NS-BH mergers.  For example, numerical simulations suggest that
mergers with a larger mass ratio (i.e., NS-BH) may produce larger
ejecta masses, with more pronounced asymmetry that may lead to
brighter optical/near-IR counterparts \citep{pnr13}.  Similarly, since
neutron star disruption outside the horizon of a typical black hole
with $\sim 10$ M$_\odot$ requires significant black hole spin, the
resulting accretion disk will undergo significant Lense-Thirring
precession that may imprint detectable variability on the prompt
emission light curve \citep{slb13}.  Finally, it is possible that the
larger masses of NS-BH systems may lead to systematically smaller
effective kicks, and hence smaller offsets than for NS-NS mergers
(e.g., \citealt{bpb+06,dmr+07}).  However, it is important to keep in
mind that to date no NS-BH binaries have been identified in nature,
and it is therefore not clear if they contribute to the short GRB
population at all, whether any of the predicted differences match the
actual properties of NS-BH binaries, and if the differences relative
to NS-NS binaries are significant enough to be discernible in the
existing short GRB sample.  These questions may be resolved with joint
gravitational wave and short GRB detections (\S\ref{sec:gwem}).

In the NS-NS merger model it is generally assumed that a black hole is
rapidly formed, with the resulting accretion powering the short GRB
and its afterglow.  However, the recent discovery of neutron stars
with masses of about 2 M$_\odot$ \citep{dpr+10,afw+13}, suggest that
some mergers may lead instead to a transitory or stable
rapidly-spinning and highly-magnetized neutron star (a magnetar;
\citealt{dt92,mqt08}).  The magnetar can in turn power a short GRB
through its spin-down energy.  In the transitory case, as the magnetar
spins down it will eventually collapse to a black hole when
differential rotation can no longer support its large mass.  This
transition could in principle lead to an observable signature; for
example, it has been recently claimed that short-lived X-ray plateaus
in some short GRBs are due to magnetar spin-down energy injection and
the subsequent collapse to a black hole \citep{rom+13}.  However,
fall-back accretion may also explain the plateaus \citep{ros07}.
Similarly, it is not clear if a stable magnetar can be distinguished
from an accreting black hole engine in terms of the resulting prompt
and afterglow emission.

Given the potential ability of magnetars to provide the required
energy source, it has also been proposed that such objects can form,
and power short GRBs or soft $\gamma$-ray repeaters, following the
accretion-induced collapse of a white dwarf or white dwarf binary
mergers \citep{lwc+06,mqt08}.  Unlike formation through core-collapse,
such delayed magnetar formation will lead to events in both early- and
late-type galaxies, similar to the prediction for NS-NS/NS-BH mergers.
However, such systems are not expected to experience significant natal
kicks, to be accompanied by gravitational wave emission in the
Advanced LIGO/Virgo band, or to produce $r$-process radioactive
elements.

Another proposed progenitor model is accretion-induced collapse of a
neutron star to a black hole \citep{qwc98,mrz05}, although this model
has not been explored in detail.  Finally, \citet{lmb10} suggested
that short GRBs with extended emission may be produced from the same
massive star progenitors as long GRBs, but with a wide off-axis
viewing angle that leads to predominant emission from the cocoon
surrounding the jet.  In this model, short GRBs with extended emission
will occur only in star-forming galaxies and will be accompanied by
Type Ic supernovae.  These requirements are violated in several events
(e.g., \citealt{bpc+05,hwf+05}).

\section{Defining the Sample}
\label{sec:sample}

In this review I address the properties of short GRBs, their
afterglows, and environments, by focusing on a sample of $70$ events
discovered primarily by the \swift\ satellite \citep{gcg+04} in the
eight-year period spanning January 2005 to January 2013, supplemented
by the recent well-studied benchmark event GRB\,130603B
\citep{bfc13,cpp+13,dtr+13,tlf+13}.  The sample includes three short
bursts discovered by HETE-2 (GRBs 050709 and 060121;
\citealt{ffp+05,hwf+05,vlr+05,dcg+06,ltf+06}) and INTEGRAL
(GRB\,070707; \citealt{mfm+08,pdc+08}) for which afterglows and host
galaxies have been identified.  This is the first sample of short GRBs
with a substantial fraction of afterglow detections, enabling studies
of both the burst properties and the host galaxies.  A basic summary
of the prompt and afterglow emission of these GRBs is provided in
Table~\ref{tab:afterglows}.

The most detailed information, both in terms of explosion properties
and host galaxy and sub-galactic environments, comes from events
localized to sub-arcsecond precision through an optical afterglow
detections or a \chandra\ detection of the X-ray afterglow (e.g.,
\citealt{ffp+05,hwf+05,bpc+05,fbm+12,mbf+12}); radio detections are
also a route to sub-arcsecond positions \citep{bpc+05}, but to date
all short GRBs detected in the radio have also been detected in the
optical.  \swift/XRT positions alone, generally with a precision of
$\sim 1.5-5''$, are also useful for host identifications
\citep{bpc+07,bfp+07,fbc+13}.  However, for these bursts the
availability of only X-ray data limits the extraction of afterglow
properties, and the positions are not precise enough to locate the
bursts within their hosts.

In this vein, of the $67$ \swift\ short GRBs considered here, 53
events were rapidly followed up with the on-board X-ray Telescope
(XRT) on a timescale of $\simlt 100$ s, leading to 47 detections
($89\%$).  This is only slightly lower than the detection rate for
long GRBs.  It is important to note that the 6 undetected events had a
lower than average $\gamma$-ray fluence of $\langle F_\gamma\rangle
\approx 2.5\times 10^{-8}$ erg cm$^{-2}$ and may therefore simply be
fainter events.  Two of the three non-\swift\ events (GRBs 060121 and
070707) were also followed up with the XRT and both were detected,
enabling the subsequent identification of optical afterglows.  The
third non-\swift\ event (GRB\,050709) was detected in the X-rays with
\chandra\ \citep{ffp+05}.  Of the 50 total X-ray detections, $28$
events ($56\%$) exhibit long-term X-ray emission beyond $\sim 10^3$ s,
while the remaining 22 events rapidly fade below the \swift/XRT
detection threshold at $\simlt 10^3$ s.  The \swift/XRT positions span
a range of $1.4-5.5''$ radius, with a median value of $1.8''$ ($90\%$
confidence\footnotemark\footnotetext{\tt
  http://www.swift.ac.uk/xrt\_positions/}).  These positions are
generally sufficient for a robust identification of host galaxies with
a probability of chance coincidence of $\sim 1-10\%$
\citep{bpc+07,bfp+07,fbc+13}.  Finally, the 14 bursts lacking rapid
XRT follow-up include 8 events with \swift\ observing constraints, and
6 events with delayed follow-up ($\simgt 1$ hr post-burst).  Thus, the
lack of X-ray detections for these events is not expected to bias the
sample considered here.

Deep optical follow-up observations were obtained for $37$ of the $47$
\swift\ bursts with XRT positions, leading to $21$ detections ($57\%$)
with a range of about $21-26$ mag at $\delta t\approx 0.4-30$ hr.  It
is remarkable that the optical afterglow detection fraction is
comparable to that for long GRBs despite the relative faintness of
short GRB afterglows.  This is a testament to the concerted follow-up
effort undertaken by the community over the past decade, utilizing the
largest ground-based telescopes.  The three non-\swift\ events were
also detected in the optical.  Of the 10 events lacking deep optical
follow-up, 6 events were located along constrained Galactic
sight-lines with large extinction and/or contaminating bright stars,
while the other 4 events lack deep afterglow and host galaxy searches
at the present.  Thus, as in the case of missing X-ray follow-up,
these events are not expected to bias the sample considered here.  The
16 bursts with deep follow-up and no detected optical afterglows
typically have limits of $\sim 23$ mag at $\sim 1-20$ hr, comparable
to the median brightness of the detected afterglows (e.g.,
\citealt{ber10}).  As I demonstrate in \S\ref{sec:afterglow} these
events do not appear to be distinct from those with detected
afterglows, suggesting that the searches were likely too shallow.  In
addition, 2 of the 16 events with optical non-detections (GRBs 111020A
and 111117A) were subsequently detected with \chandra\, leading to
sub-arcsecond positions \citep{fbm+12,mbf+12}.

Finally, radio observations have been obtained for $28$ short GRBs
(including some non-\swift\ bursts and \swift\ bursts with only
$\gamma$-ray positions) leading to only three detections to date
($11\%$; GRBs 050724A, 051221A, and 130603B:
\citealt{bpc+05,sbk+06,fbm+13}).  This is a low detection fraction
compared to long GRBs (with $\sim 30\%$; \citealt{cf12}), but it is
not surprising considering that radio follow-up is generally
sensitivity-limited even for long GRBs.  As I discuss in
\S\ref{sec:afterglow}, despite the low detection fraction, the radio
limits are generally useful for placing constraints on the circumburst
densities of short GRBs.

In addition to afterglow follow-up observations, there has been an
intensive effort to characterize the environments of short GRBs from
sub-galactic to galaxy cluster scales (e.g.,
\citealt{bsm+07,ber09,dmc+09,fbf10,ber11,fbc+13,fb13}).  The sample
presented in this paper includes nearly 40 identified host galaxies,
with about 30 redshift measurements.  A subset of these galaxies have
detailed measurements that include morphologies, stellar masses,
stellar population ages, star formation rates, and metallicities (see
Table~\ref{tab:hosts}).  In addition, for about 20 events (those with
sub-arcsecond afterglow localizations) there are precise measurements
of the sub-galactic environments based on {\it HST} observations,
including projected offsets and the brightness at the GRB location
relative to the underlying light distribution of the host galaxy
\citep{ber10,fbf10,fb13}.

\section{A Non-Massive Star Origin}
\label{sec:nonmassive}

Two key observations helped to establish the connection between long
GRBs and the deaths of massive stars (the collapsar model;
\citealt{mw99}): (i) the association of long GRBs with Type Ic
core-collapse supernovae \citep{gvv+98,hsm+03,smg+03,wb06}; and (ii)
the exclusive locations of long GRBs in star-forming galaxies
\citep{bdk+98,dkb+98,chg04,wbp07}, as well as their spatial offsets
relative to their hosts' overall light distribution and their
coincidence with bright star-forming regions within their hosts
\citep{bkd02,fls+06}.  The SN associations are based on spectroscopic
observations, mainly at $z\simlt 0.5$ where the spectroscopic
signatures can be discerned with sufficient significance (e.g.,
\citealt{hsm+03,smg+03,pmm+06,bch+11}), and on photometric
re-brightenings on a timescale of $\sim 15-20$ d post-burst that match
Type Ic SN light curves in both brightness and color (to $z\sim 1$;
e.g., \citealt{bkd+99,bkp+02,wb06}).  The observed distribution of
long GRB-SN peak magnitudes is relatively narrow, spanning only about
$1$ mag for the bulk of the sample \citep{wb06,hb12}.  Moreover, long
GRB-SNe are generally more luminous than normal Type Ib/c SNe
\citep{dsg+11}, although there is some overlap in the distributions.
These results are summarized in Figure~\ref{fig:sne} in which I plot
the peak absolute magnitudes of GRB-SNe and Type Ib/c SNe relative to
the canonical SN\,1998bw associated with GRB\,980425.  In terms of
spatial locations within their hosts, long GRBs radially track an
exponential light distribution, typical of star formation in disk
galaxies, with a median offset of about one half-light radius
\citep{bkd02}.  Moreover, long GRBs are spatially correlated with
bright star-forming regions, even in comparison to normal
core-collapse SNe \citep{fls+06,slt+10}.

\subsection{Lack of Supernova Associations}

\begin{figure}
\centerline{\psfig{figure=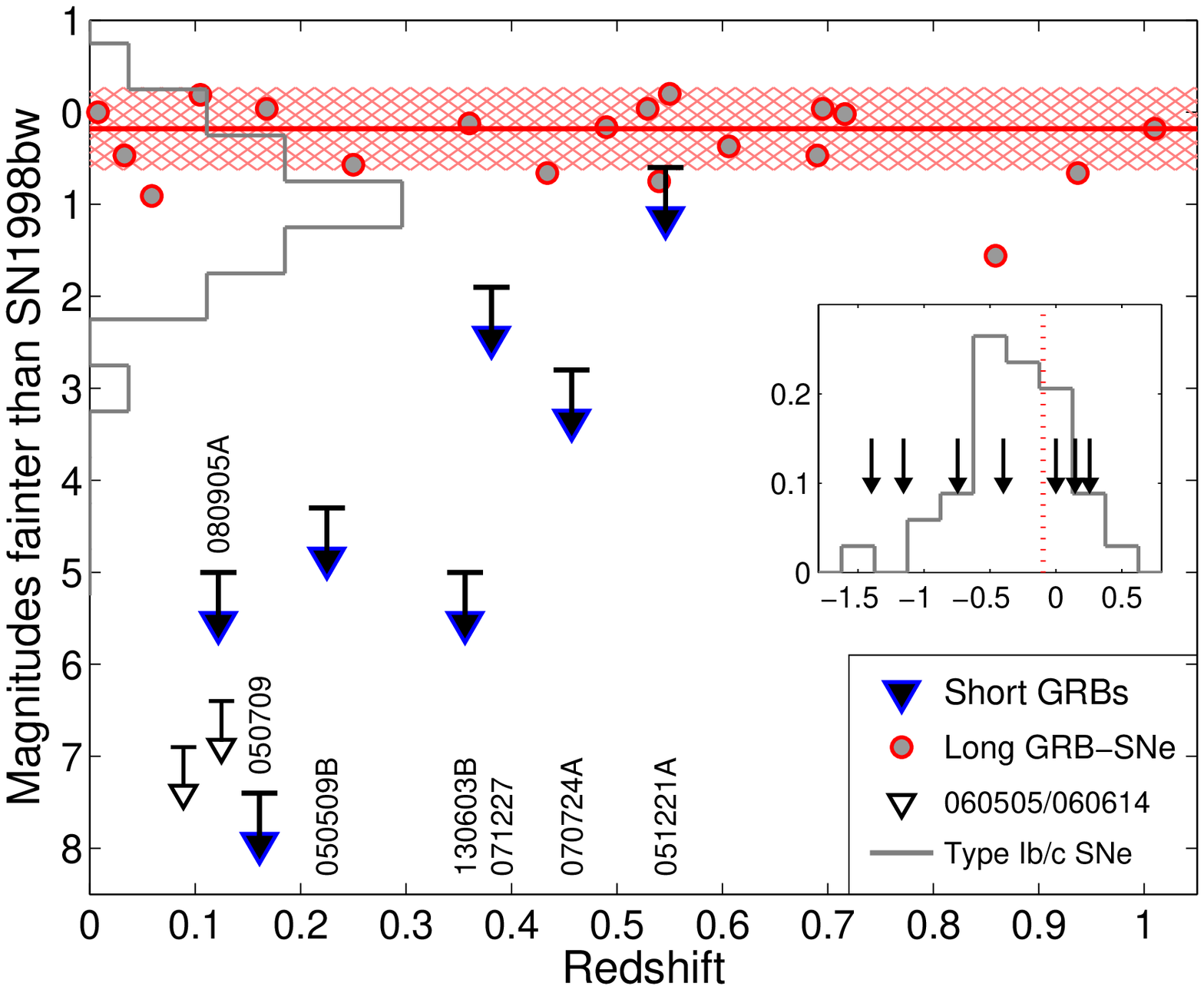,width=\textwidth}}
\caption{Limits on supernovae associated with short GRBs (filled
  triangles) relative to the peak absolute magnitude of the canonical
  long GRB-SN\,1998bw.  Also shown are the distribution of SN peak
  magnitudes for long GRBs (filled circles; hatched region marks the
  median and standard deviation for the population; \citealt{hb12}),
  local Type Ib/c SNe (histogram; \citealt{dsg+11}), and two unusual
  long GRBs that lacked associated SNe (060505 and 060614;
  \citealt{dcp+06,fwt+06,gfp+06,gnb+06}).  The latter may represent a
  long duration or extended emission tail of the short GRB population.
  With the exception of GRB\,050509b, all short bursts with limits on
  associated SNe occurred in star-forming galaxies, indicating that
  despite the overall star formation activity, the short GRB
  progenitors were not massive stars.  The inset shows the overall
  duration distribution of the short GRBs considered in this review
  (histogram), with the durations of the 7 short GRBs with SN limits
  marked by arrows.  The dotted vertical line marks the claimed
  duration separating \swift\ non-collapsar and collapsar progenitors
  according to the analysis of \citet{bnp+13}, and yet three of the
  short GRBs lacking SN associations have longer durations.}
\label{fig:sne}
\end{figure}

In the short GRB sample defined in \S\ref{sec:sample} there are
several events at sufficiently low redshifts to allow clear detections
of associated SNe, yet none have been found to date
\citep{ffp+05,hsg+05,hwf+05,sbk+06,dmc+09,ktr+10,rwl+10,bfc13}.  In
Figure~\ref{fig:sne} I plot all existing upper limits for associated
SNe (for 7 short bursts), measured relative to the peak absolute
magnitude of the canonical long GRB-SN\,1998bw.  I also plot the
limits for the peculiar events GRBs 060505 and 060614.  As can be seen
from the figure, SNe associated with long GRBs span a narrow peak
brightness range, with a median and standard deviation relative to
SN\,1998bw of $+0.18\pm 0.45$ mag.  On the other hand, the upper
limits on SN associations for short GRBs range from 0.6 to 7.4 mag
fainter than SN\,1998bw; GRBs 060505 and 060614 have limits of 6.9 and
6.4 mag fainter than SN\,1998bw, respectively.  Thus, in all cases an
associations with SNe that are drawn from the same distribution as
long GRB-SNe can be ruled out.  This demonstrates that short and long
GRBs do not share a common progenitor system, and that at least the
short GRBs with deep SN limits are not produced by massive star
explosions.  It is also important to note that of the 7 short bursts
with limits on associated SNe, 6 events are located in star-forming
galaxies, indicating that while the hosts exhibit on-going star
formation activity, the short GRB progenitors themselves do not belong
to a young population of massive stars.  I return to this point in
\S\ref{sec:subgal} when discussing the locations of short GRB within
and around their host galaxies.

The dominance of star-forming hosts in the sample of short GRBs
lacking SN associations indicates that the use of galaxy type as an
indicator of progenitor type (as advocated for example by
\citealt{zzl+07,zzv+09}) can severely bias the resulting GRB
classification.  In particular, a star-forming host does not indicate
a young massive star progenitor.  In a similar vein, I note that two
of the short bursts with SN non-detections (GRBs 051221A and 070724A)
are claimed to be likely collapsars according to the duration-hardness
analysis of \citet{bnp+13}, and a third event (GRB\,080905A) has a
duration that is longer than their nominal divide between short and
long GRBs (see inset of Figure~\ref{fig:sne}).  The fact that these
events lack associated SNe casts doubt on the claimed statistical
significance of a collapsar origin assigned to individual events in
their analysis.

\subsection{A Mix of Ellipticals and Spirals}

\begin{figure}
\centerline{\psfig{figure=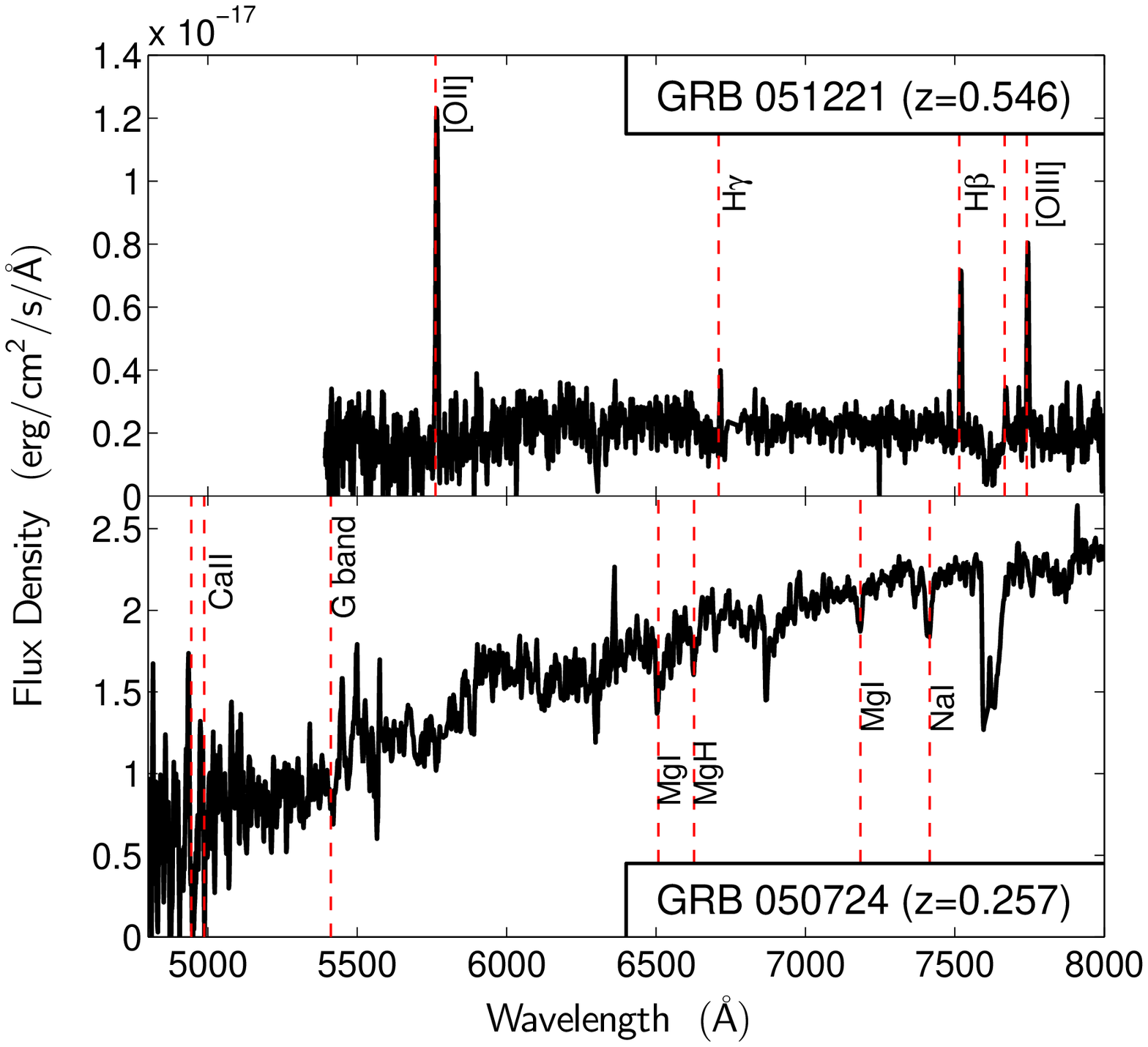,width=\textwidth}}
\caption{Representative optical spectra of a late-type short GRB host
  galaxy exhibiting emission lines typical of star-forming galaxies
  (top: GRB\,051221 at $z=0.546$) and an early-type short GRB host
  galaxy with no evidence for on-going star formation activity
  (bottom: GRB\,050724 at $z=0.257$).  Primary absorption and emission
  lines seen in the spectra are marked.  The spectra are from
  \citet{sbk+06} and \citet{bpc+05}, respectively.}
\label{fig:spec}
\end{figure}

The second clear distinction between the short and long GRB
populations is the occurrence of some short GRBs in elliptical
galaxies (Figure~\ref{fig:spec}).  The current short GRB sample
includes two secure cases of elliptical host galaxies based on
sub-arcsecond afterglow positions and spatially coincident hosts (GRBs
050724A and 100117A; \citealt{bpc+05,fbc+11}), two additional cases
with sub-arcsecond afterglow positions and likely elliptical hosts
with large projected offsets (GRBs 070809 and 090515;
\citealt{ber10}), and four additional likely cases (probabilities of
about $1-5\%$) based on \swift/XRT positions alone (GRBs 050509b,
060502b, 070729, 100625A; \citealt{gso+05,bpp+06,bpc+07,fbc+13}).
Overall, about 20\% of short GRBs are associated with early-type host
galaxies \citep{fbc+13}.  In nearly all cases, the identification of
the hosts as early-type galaxies is based on spectroscopic
observations that reveal no star formation activity (to $\simlt 0.1$
M$_\odot$ yr$^{-1}$), optical/near-IR spectral energy distributions
that are matched by a single stellar population with an age of $\simgt
1$ Gyr, and/or morphological information based on {\it HST}
observations.  I explore the host galaxy demographics distribution,
and its implications for the progenitor population, in the next
section, but it is clear from the occurrence of at least some short
GRBs in elliptical galaxies that he progenitors belong to an old
stellar population.

\section{Short GRB Galaxy-Scale Environments}
\label{sec:hosts}

Having established that the progenitors of short GRBs are generally
distinct from those of long GRBs based on the lack of SN associations
and their occurrence in elliptical galaxies, I now turn to the
question of what the progenitors are, and what we can infer about
their nature from the properties of the host galaxies.  In general,
the galaxy-scale and local environments of astrophysical transients
provide critical insight into the nature of their progenitors, and
this has been used to establish the progenitor properties of various
supernova types and long GRBs (e.g.,
\citealt{vt91,bkd02,fls+06,lcl+11}).  In this section I contrast the
environments of short and long GRBs to further demonstrate their
distinct origins, and compare the short GRB environments to field
galaxy populations to establish some of the basic properties of their
progenitors.  I also provide a comparison with the hosts of Type Ia
supernovae since these white dwarf thermonuclear explosions are known
to arise from an evolved stellar population, with a broad range of
delay times \citep{mdp06,slp+06,mml+11}, potentially similar to
compact object binary mergers.

A summary of the key properties of short GRB host galaxies is provided
in Table~\ref{tab:hosts}.  The data are primarily drawn from
\citet{ber09}, \citet{lb10}, \citet{ber10}, \citet{fbc+13} and
references therein.

\subsection{The Redshift Distribution}
\label{sec:z}

\begin{figure}
\centerline{\psfig{figure=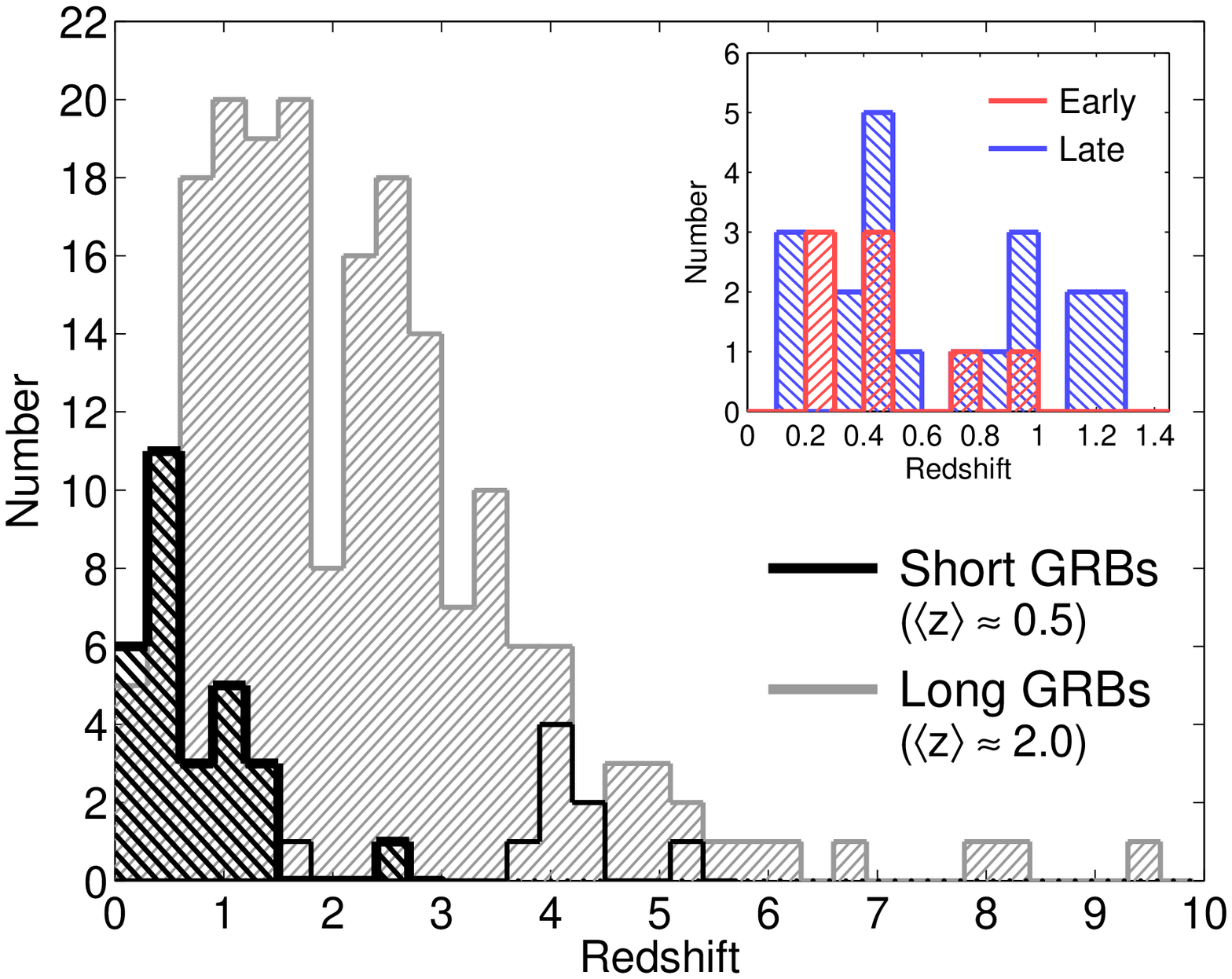,width=\textwidth}}
\caption{The redshift distribution of short GRBs (black) and long GRBs
  (gray).  The open histogram marks redshift upper limits based on the
  lack of a Lyman-$\alpha$ break in afterglow and/or host galaxy
  optical detections.  The inset shows the redshift distribution of
  short GRBs separated by host galaxy type, which exhibits no
  discernible difference between early-type (red) and late-type (blue)
  hosts.}
\label{fig:z}
\end{figure}

At the most basic level, host galaxy associations are critical since
essentially all short GRB redshifts to date (spectroscopic or
photometric) have been obtained from the associated hosts
(Figure~\ref{fig:spec}; e.g.,
\citealt{ffp+05,bpc+05,bpp+06,ber09,ber11}).  The sole exceptions are
GRB\,090426 at $z=2.609$ \citep{adp+09,lbb+10} and GRB\,130603B at
$z=0.356$ \citep{cpp+13,dtr+13} for which redshifts have been
determined from afterglow absorption spectra.  The bulk of the
measured redshifts span $z\approx 0.1-1.3$, but it is likely that at
least some of the faintest host galaxies (with optical magnitudes of
$\approx 24-27$), which lack redshift measurements are located at
$z\simgt 1$ \citep{bfp+07}.  The redshift distributions of short and
long GRBs are shown in Figure~\ref{fig:z}.  The median redshift of the
short GRB population with established redshifts is $\langle
z\rangle\approx 0.48$, while the addition of faint hosts that are
presumably located at $z\simgt 1$ increases the median to $\langle
z\rangle\approx 0.63$; the use of redshift upper limits from afterglow
and/or host galaxy detections in optical bands (i.e., the lack of a
Lyman break) leads to an upper bound on the median redshift of
$\langle z\rangle\simlt 0.83$.  Since only 6 short bursts lack XRT
positions despite rapid follow-up (\S\ref{sec:sample}), and hence the
ability to identify hosts, the observed redshift distribution robustly
represents the redshifts of \swift\ short GRBs.  Assuming that the
median redshift is $\langle z\rangle\sim 0.5-0.8$, and that it is not
affected by the sensitivity threshold of \swift\, the resulting median
progenitor age relative to the peak of the cosmic star formation
history ($z\sim 3$) is $\tau\sim 5$ Gyr.  Since the observed sample is
almost certainly limited by the detector sensitivity, this inferred
age can be used as an upper bound.

There is no clear trend between redshift and host galaxy type, with
both early- and late-type hosts spanning the same redshift range with
similar median values (Figure~\ref{fig:z}; \citealt{fbc+13}).  Since
the progenitors are expected to be systematically older in early-type
galaxies and hence to occur at lower redshifts (e.g., \citealt{zr07}),
the similar redshift distributions for early- and late-type galaxies
suggest that the \swift\ sensitivity threshold indeed plays a role in
the redshift distribution.  The short GRB population is substantially
more local than \swift\ long GRBs, which extend to $z\sim 9.4$
\citep{clf+11} with a median value of $\langle z\rangle\approx 2$
\citep{bkf+05,jlf+06}.  This is partly a reflection of the lower
energy scale of short GRBs, but also a result of longer delay times
between star formation activity and the occurrence of short GRBs.

\subsection{Demographics}
\label{sec:demographics}

An additional constraint on the progenitors and their age distribution
is provided by the demographics of the host galaxy sample.  While long
GRBs only occur in star-forming galaxies \citep{bkd02,fls+06,wbp07},
as expected for young massive star progenitors, short GRBs occur in a
mixed population of early-type and star-forming galaxies
\citep{bpc+05,ffp+05,bpp+06,fbc+11,fbc+13} indicating that their
progenitors span a wide range of ages.  More broadly, since at $z\sim
0-1$, the redshift range relevant for the existing short GRB
population, a roughly equal fraction of the cosmic stellar mass
density is in early-type and star-forming galaxies
\citep{bmk+03,isl+10}, an exclusively old progenitor population (i.e.,
tracking stellar mass alone) will also exhibit an equal fraction of
early- and late-type hosts.  On the other hand, a progenitor
population skewed to relatively short delay times relative to star
formation activity (tens to hundreds of Myr) will exhibit an
over-abundance of late-type hosts due to their recent star formation
\citep{zr07,obk08}.  Studies of the first few short GRB host galaxies
from 2005, led several groups to conclude that early-type galaxies
dominate the sample at a ratio of about 2:1, and that the progenitors
are therefore exceedingly old, with a characterstic age of $\sim 5-10$
Gyr \citep{pbc+06,nak07,gno+08}.

\begin{figure}
\centerline{\psfig{figure=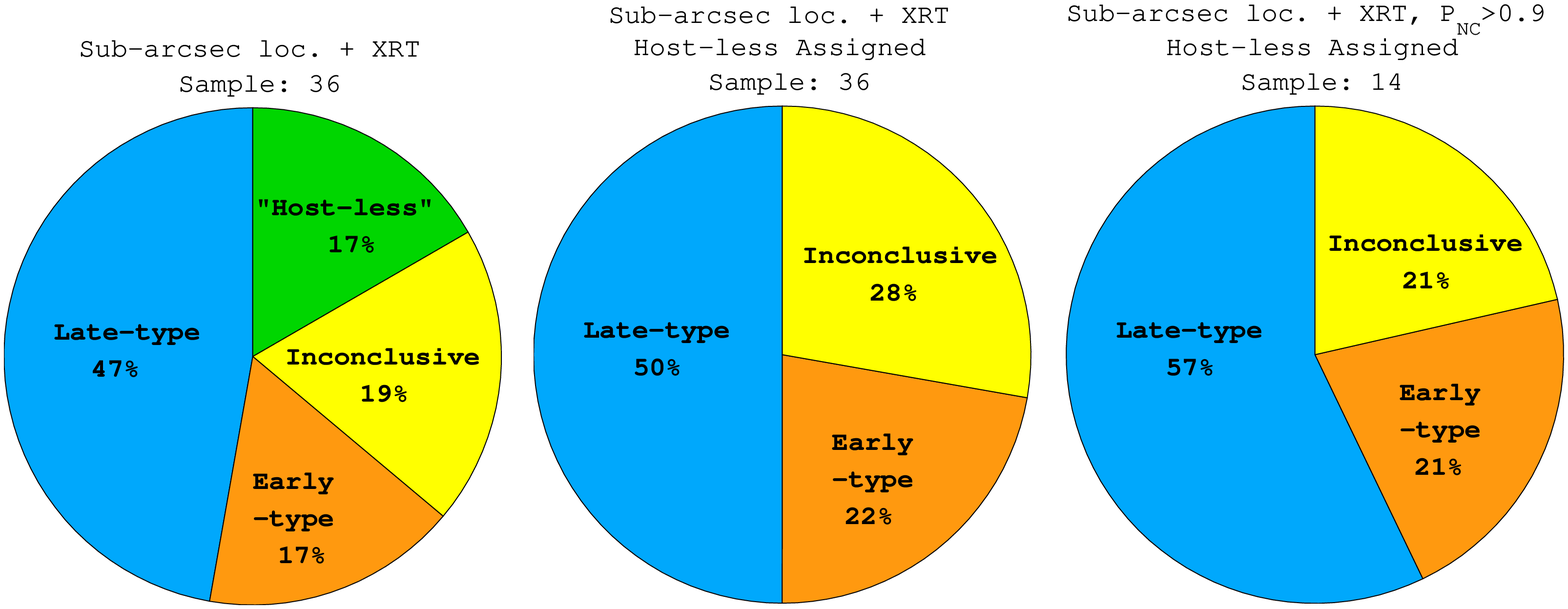,width=\textwidth}}
\caption{Demographics of the galaxies hosting short GRBs.  {\it Left:}
  A breakdown into late-type (blue), early-type (orange), host-less
  (green), and inconclusive (yellow) for all identified hosts based on
  sub-arcsecond positions and \swift/XRT positions
  (Table~\ref{tab:hosts}).  {\it Middle:} Same as the left panel, but
  with the host-less events assigned to the other categories based on
  the galaxies with the lowest probability of chance coincidence in
  each case \citep{ber10,fb13}.  {\it Right:} Same as the middle
  panel, but for short GRBs with a probability of a non-collapsar
  origin of $P_{\rm NC}\simgt 0.9$ based on the analysis of
  \citet{bnp+13}.  Regardless of the sample selection, late-type
  galaxies dominate the host sample.  This indicates that star
  formation activity plays a role in the short GRB rate.  Adapted from
  \citet{fbc+13}.}
\label{fig:demographics}
\end{figure}

A different conclusion is reached by \citet{fbc+13} in a study of the
short GRB host galaxy demographics using a much larger sample of 36
events.  These authors find that regardless of various cuts on the
sample, only $\sim 20\%$ of the hosts are early-type galaxies
(Figure~\ref{fig:demographics}; see also \citealt{lb10}).  This result
is robust when considering only events with sub-arcsecond positions,
when including those with XRT positions, or when restricting the
sample according to the duration-hardness analysis of \citet{bnp+13}.
The dominance of late-type galaxies indicates that the short GRB rate
does not depend on stellar mass alone, and is instead influenced by
recent star formation activity.  Comparing the observed relative
fraction of early- and late-type hosts to the theoretical predictions
of \citet{zr07} leads to a power law delay time distribution of
$P(\tau)\propto \tau^{-1}$.

An additional clue to the role of star formation in the short GRB rate
is the identification of two short GRB hosts as luminous and
ultra-luminous infrared galaxies (U/LIRG).  GRB\,100206A is associated
with a dusty LIRG at $z=0.407$ \citep{pmm+12}, while GRB\,120804A is
associated with a dusty ULIRG at a photometric redshift of $z\approx
1.3$ \citep{bzl+13}.  The expected U/LIRG fraction for a progenitor
population that tracks stellar mass alone is $\sim 1\%$
\citep{cmd+06}, while for progenitors that track only star formation
it is $\sim 25\%$ \citep{lpd+05,cly+07}.  Since the observed fraction
in the short GRB sample is $\sim 5-10\%$ percent, it suggests that the
progenitor population is influenced by both stellar mass and star
formation activity, indicative of a broad age distribution.

\subsection{Stellar Masses and Stellar Population Ages}
\label{sec:mass}

\begin{figure}
\centerline{\psfig{file=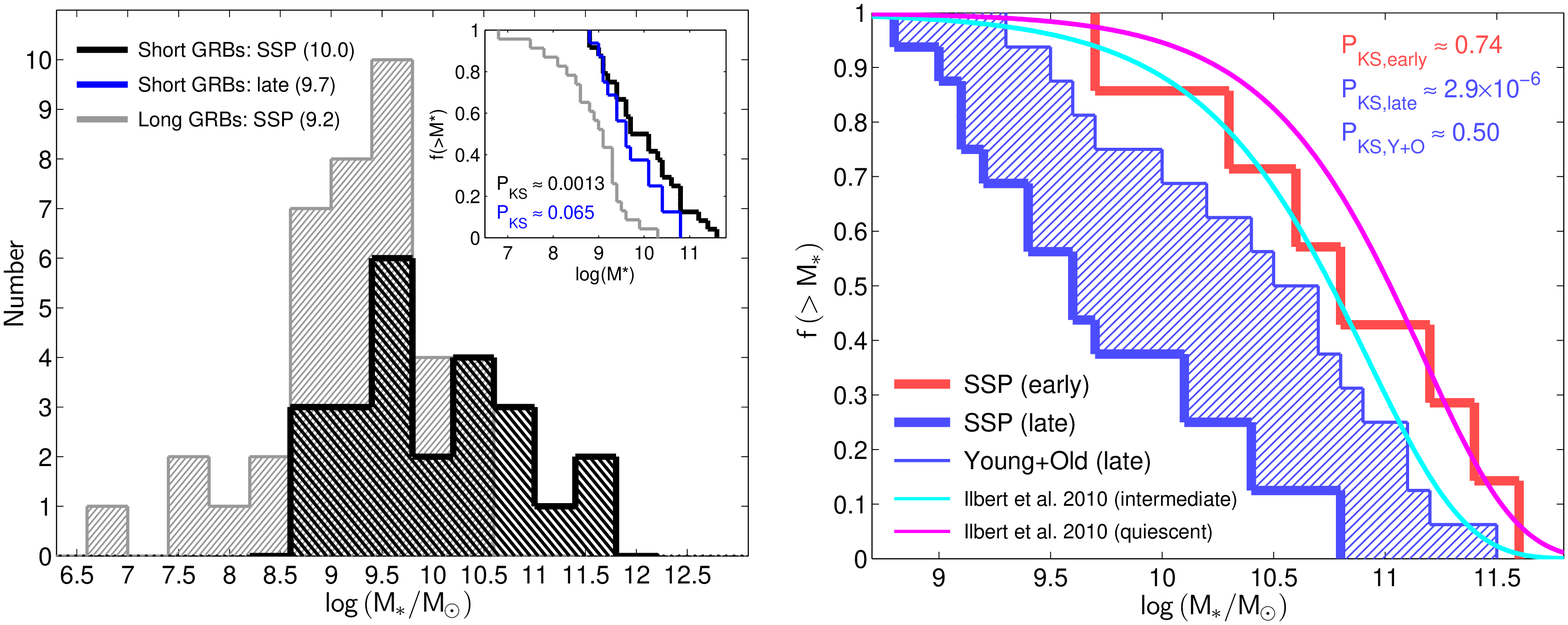,width=\textwidth}}
\caption{{\it Left:} Histogram of host galaxy stellar masses for short
  GRBs (black) and long GRBs (gray).  Median values for each
  population (and separately for short GRB late-type hosts) are quoted
  in parentheses.  The inset shows the cumulative distributions along
  with K-S probabilities that the short and long GRB hosts are drawn
  from the same parent population.  {\it Right:} Comparison of the
  cumulative distributions of stellar masses for late-type (blue) and
  early-type (red) short GRB hosts to the expected distributions for a
  mass-weighted selection from the field galaxy mass function (cyan
  and magenta lines, respectively).  The resulting K-S probabilities
  indicate that the early-type hosts are consistent with pure mass
  selection, while the late-type hosts have lower than expected
  stellar masses.  This indicates that star formation activity plays a
  role in the short GRB rate. Adapted from \citet{lb10}.}
\label{fig:mass}
\end{figure}

The distribution of host galaxy stellar masses and stellar population
ages can also shed light on the progenitor age distribution.  The
stellar masses inferred from modeling of the host optical/near-IR
spectral energy distributions with single stellar population models
span $M_*\approx 10^{8.5-11.8}$ M$_\odot$.  The median for the full
sample is $\langle M_*\rangle\approx 10^{10.0}$ M$_\odot$, while for
the star-forming hosts alone it is $\langle M_*\rangle\approx
10^{9.7}$ M$_\odot$ (Figure~\ref{fig:mass}; \citealt{lb10}).  The
stellar masses of long GRB hosts are substantially lower, with a
median value of about $10^{9.2}$ M$_\odot$ \citep{sgl09,lb10}.  This
indicates that even the star-forming hosts of short GRBs are typically
more massive than the hosts of long GRBs, pointing to a more dominant
role of stellar mass in determining the rate of short GRBs.

A comparison to the mass function of field galaxies is even more
illuminating.  In Figure~\ref{fig:mass} I compare the cumulative
distributions of stellar masses for the early- and late-type hosts of
short GRBs with the expected distributions for mass-selection from the
field galaxy mass function \citep{bmk+03,isl+10}.  For a progenitor
population that tracks stellar mass alone, we expect that the observed
stellar mass distribution of short GRB hosts will closely track the
mass-weighted mass distribution of field galaxies.  While this is
indeed the case for the early-type mass function, the late-type hosts
of short GRBs have systematically lower stellar masses than expected
from mass-selection alone \citep{lb10}.  This indicates that in
late-type galaxies the short GRB rate per unit stellar mass is higher
than in early-type galaxies, due to the presence of star formation
activity.  This agrees with the observed over-abundance of late-type
galaxies in the short GRB host population.  Quantitatively, for a
range of host stellar mass determinations, the short GRB rate per unit
stellar mass in late-type galaxies is about $2-9$ times higher than in
early-type galaxies \citep{lb10}.

\begin{figure}
\centerline{\psfig{figure=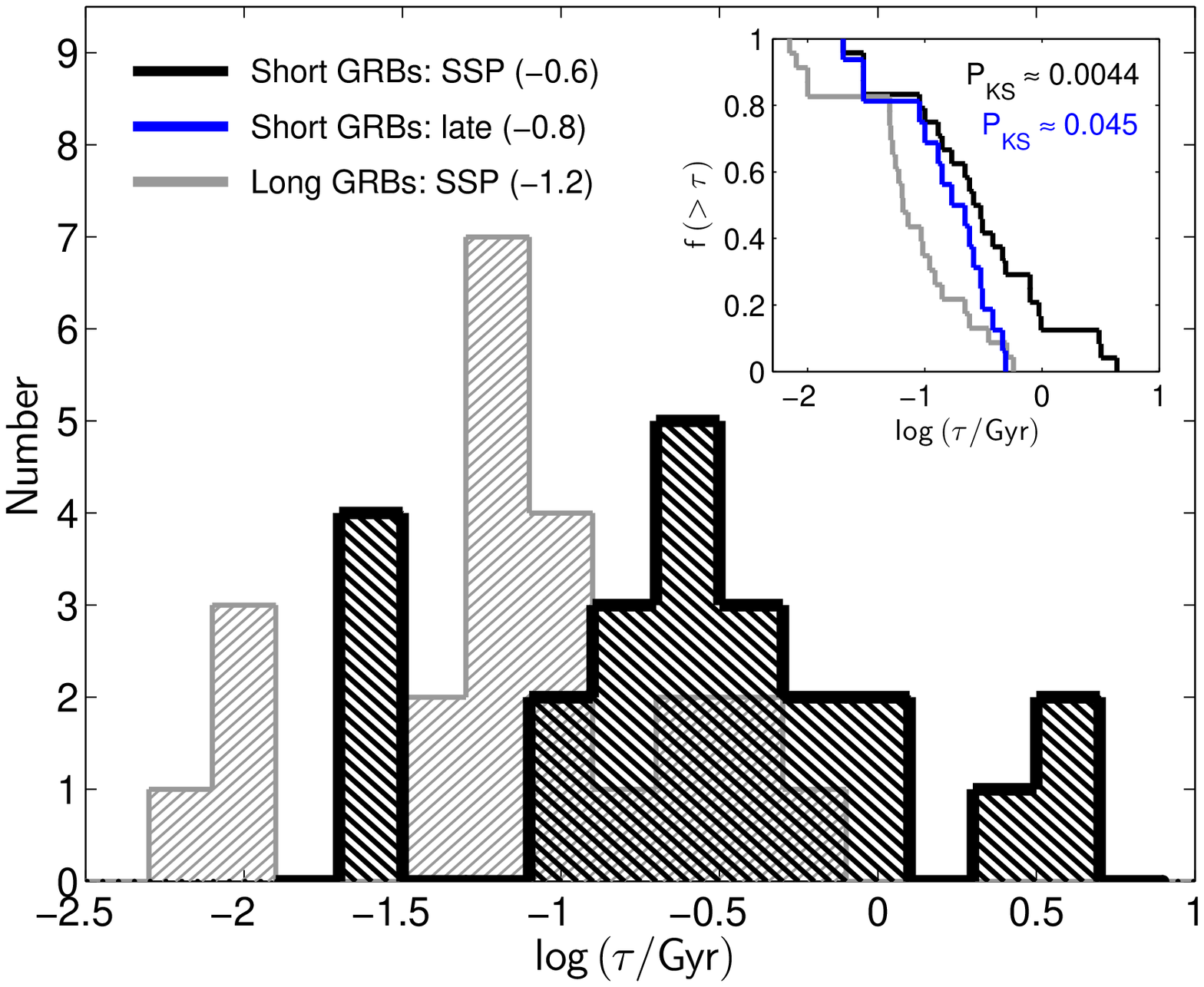,width=\textwidth}}
\caption{Histogram of host galaxy stellar population ages for short
  GRBs (black) and long GRBs (gray).  Median values for each
  population (and separately for short GRB late-type hosts) are quoted
  in parentheses.  The inset shows the cumulative distributions along
  with K-S probabilities that the short and long GRB hosts are drawn
  from the same parent population.  The results indicate that short
  GRB hosts, even the late-type galaxies, have systematically older
  stellar population than long GRB hosts.  Adapted from \citet{lb10}.}
\label{fig:age}
\end{figure}

The difference between short and long GRB hosts is also evident in the
distribution of stellar population ages.  The distribution for short
GRB hosts spans a wide range, from a few tens of Myr to about 4 Gyr,
with a median value of $\langle\tau_*\rangle\approx 0.25$ Gyr, while
the long GRB hosts have a median stellar population age of
$\langle\tau_*\rangle\approx 60$ Myr (Figure~\ref{fig:age};
\citealt{lb10}).  Coupled with the stellar mass distribution discussed
above, this indicates that the short GRB rate per $10^{10}$ M$_\odot$
in late-type hosts (with $\langle\tau_*\rangle\approx 0.15$ Gyr) and
early-type hosts (with $\langle\tau_*\rangle\approx 1.6$ Gyr) is $0.7$
and $0.075$, respectively, leading to a power law delay time
distribution of $P(\tau)\propto \tau^{-0.9}$, in good agreement with
the distribution inferred from the early- to late-type fraction.
Similarly, separating the sample into galaxies with stellar population
ages younger and older than $\sim 0.3$ Gyr I find that the typical
ages are about 0.1 and 1 Gyr, respectively, with a resulting rate per
$10^{10}$ M$_\odot$ of 0.8 and 0.12, respectively, leading to a delay
time distribution of $P(\tau)\propto\tau^{-0.8}$.  The inferred delay
time distribution is in good agreement with that of Galactic neutron
star binaries with a power law index of $n\sim -1$ (e.g.,
\citealt{pir92}).

\subsection{Specific Star Formation Rates}
\label{sec:ssfr}

\begin{figure}
\centerline{\psfig{figure=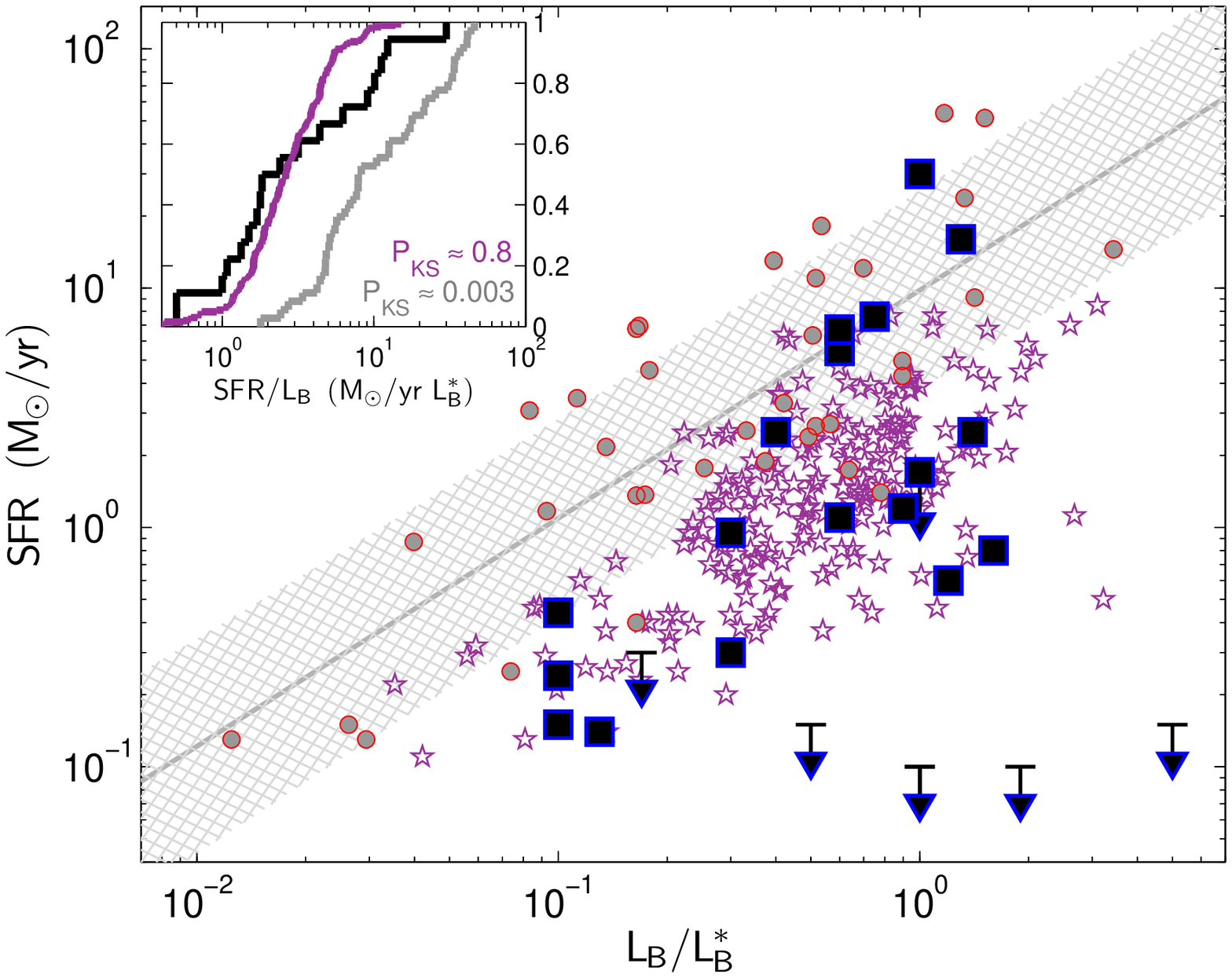,width=\textwidth}}
\caption{Star formation rate as a function of rest-frame $B$-band
  luminosity for the host galaxies of short GRBs (squares), long GRBs
  (circles), and field star-forming galaxies at similar redshifts to
  short GRB hosts (pentagrams; \citealt{kk04}).  The gray line and
  hatched region delineate the correlation (and standard deviation)
  for long GRB host galaxies.  Short GRB host galaxies have
  substantially lower star formation rates as a function of luminosity
  than long GRBs hosts (i.e., they have lower specific star formation
  rates), but they closely track the field galaxy population (inset).}
\label{fig:sfr}
\end{figure}

The interplay of stellar mass and star formation activity in short GRB
hosts is also evident in the distribution of specific star formation
rates (SSFR), relative to both long GRB hosts and field star-forming
galaxies.  In Figure~\ref{fig:sfr} I plot the star formation rates
(SFR) as a function of rest-frame optical $B$-band luminosity ($L_B$)
for short and long GRB hosts.  For the short GRB hosts the
luminosities span $L_B\approx 0.1-5$ $L_B^*$, and the star formation
rates span $\simlt 0.1$ to $\approx 5$ M$_\odot$ yr$^{-1}$, with the
upper limits corresponding to the early-type hosts; here $L_B^*$ is
the characteristic luminosity in the Schechter function.  However,
while both short and long GRB hosts exhibit a clear trend between SFR
and $L_B$ (at least for the star-forming short GRB hosts), the overall
normalization (i.e., the SSFR) is lower for the short GRB hosts by
about a factor of 5, with $\langle {\rm SFR}/L_B\rangle \approx 2$
M$_\odot$ yr$^{-1}$ $L_B^*$.  The K-S test gives a $p$-value of 0.003
for the null hypothesis that the short and long GRB hosts are drawn
from a galaxy population with the same underlying distribution of
SSFR. 

Since the specific star formation rate is the inverse of the
characteristic timescale to build up a galaxy's stellar mass, the
lower values for short GRB hosts indicates that while the majority are
star-forming galaxies, their star formation activity is more moderate,
and has a longer characteristic timescale than in long GRB hosts.
Thus, the massive star progenitors of long GRBs track recent star
formation (tens of Myr), while the progenitors of short GRBs track
star formation with a delay of hundreds of Myr to several Gyr.
Moreover, the short GRB hosts clearly track the distribution of
specific star formation rates in field galaxies of similar
luminosities \citep{kk04}, with a K-S $p$-value of 0.8, indicating
that short GRBs select galaxies from the general field sample.

\subsection{Metallicities}
\label{sec:Z}

\begin{figure}
\centerline{\psfig{figure=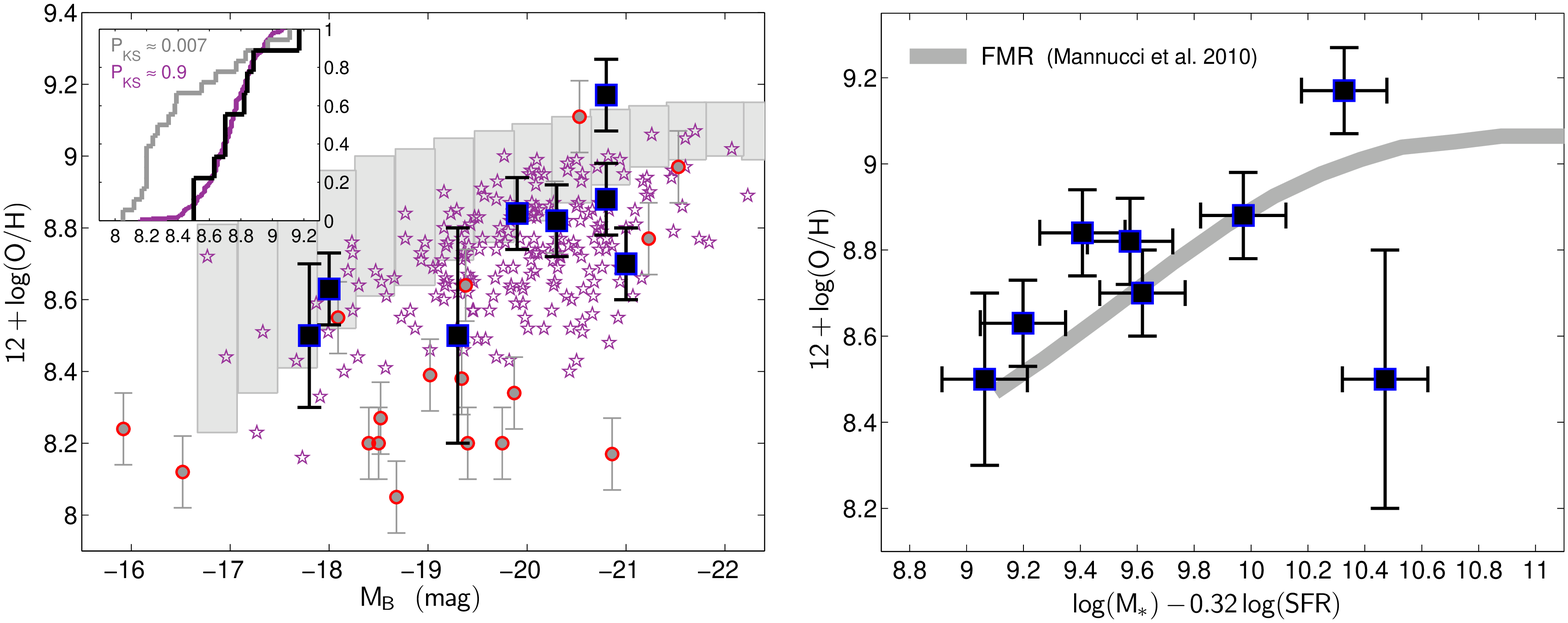,width=\textwidth}}
\caption{{\it Left:} Metallicity as a function of host galaxy
  rest-frame $B$-band luminosity for short GRBs (squares), long GRBs
  (circles), field galaxies at similar redshifts to short GRB hosts
  (pentagrams; \citealt{kk04}), and the Sloan Digital Sky Survey
  luminosity-metallicity relation \citep{thk+04}.  Short GRB host
  galaxies have higher metallicities than long GRBs hosts, but closely
  track the luminosity-metallicity relation for the field galaxy
  population (inset).  Adapted from \citet{ber09}.  {\it Right:}
  Comparison of short GRB hosts to the fundamental metallicity
  relation of star-forming galaxies (grey line; \citealt{mcm+10}),
  which uses a combination of stellar mass and star formation rate,
  ${\rm log}(M_*)-0.32{\rm log}({\rm SFR})$.  The short GRB hosts
  follow the relation along a wide span of the combined stellar mass
  and star formation parameter.}
\label{fig:metallicity}
\end{figure}

Short GRB hosts span a wide range of metallicities, with $12+{\rm
  log}({\rm O/H})\approx 8.5-9.2$, and a median value of $\langle
12+{\rm log}({\rm O/H})\rangle \approx 8.8\approx 1$ Z$_\odot$
(Figure~\ref{fig:metallicity}; \citealt{ber09,dmc+09}).  In addition,
the measured metallicities track a positive trend with host galaxy
luminosity, as generally observed for field galaxies \citep{thk+04}.
The metallicities of short GRB hosts are significantly higher than
those of long GRB hosts, for which the median value is $\langle
12+{\rm log}({\rm O/H})\rangle\approx 8.3$
\citep{sgb+06,mkk+08,lkb+10}.  On the other hand, short GRB hosts
track the metallicity distribution of field star-forming galaxies at
similar redshifts and with similar luminosities \citep{kk04}.
Combining the host stellar masses and star formation rates according
to the fundamental metallicity relation of \citet{mcm+10}, $\mu={\rm
  log}(M_*)-0.32\,{\rm log}({\rm SFR})$, I find that short GRB hosts
track the relation for field galaxies (Figure~\ref{fig:metallicity}).
Thus, while the sample of short GRB hosts with metallicities is still
small, it demonstrates that the hosts follow the relations for field
galaxies, but are clearly distinct from long GRB hosts.  Based on this
result, it does not appear that short GRB progenitors are directly
affected by metallicity.

\subsection{Galaxy Clusters}
\label{sec:clusters}

The first short GRB with an afterglow detection, GRB\,050509b, was
most likely associated with a galaxy cluster at $z=0.225$
\citep{gso+05,peh+05,bpp+06,dsl+13}.  This association led to several
investigations of galaxy clusters hosting short GRBs.  \citet{gmg+06}
cross-correlated BATSE short GRBs with catalogs of X-ray selected
clusters to $z\sim 0.45$ and claim an association at a $2\sigma$
confidence level on an angular scale of $\simlt 3^\circ$.
\citet{bsm+07} searched for galaxy clusters in association with 4
\swift\ short GRBs using multi-object optical spectroscopy, and in
association with 15 \swift\ short GRBs using stacked X-ray
observations from \swift/XRT.  The search yielded a galaxy cluster at
$z=0.165$, based on redshift clustering and diffuse X-ray emission, in
the field of the short GRB\,050911 (BAT-only position) with a
probability of chance coincidence of $0.1-1\%$.  The limit on galaxy
clusters in the fields of the remaining bursts is $M\simlt 5\times
10^{13}$ M$_\odot$ based on the lack of diffuse X-ray emission
\citep{bsm+07}.

Since the stellar populations in galaxy clusters are systematically
older than in field galaxies, the fraction of short GRBs in clusters
can be used as an indicator of the age distribution \citep{sb07}.  At
present, the fraction of short GRBs in clusters appears to be $\sim
10\%$ \citep{bsm+07}, at the low end of the fraction of cosmic stellar
mass in clusters of about $10-20\%$ \citep{fhp98,ebc+05}.  There is
therefore no indication from galaxy cluster associations for an
exceedingly old progenitor population.

\subsection{Comparison to the Host Galaxies and Delay Time
  Distribution of Type Ia Supernovae}
\label{sec:Ia}

Type Ia SNe result from the thermonuclear explosions of CO white
dwarfs that approach the Chandrasekhar mass in single- or
double-degenerate binary systems.  Unlike core-collapse SNe, Type Ia
events are found in both early- and late-type galaxies (e.g.,
\citealt{van90,vt91,lcl+11}), reflecting the wide range of timescales
for the accretion or merger process that eventually leads to a
supernova.  As a result, the environments and delay time distribution
of Type Ia SNe provide a useful comparison to short GRBs, particularly
in the context of compact object merger progenitors.  \citet{mdp+05}
found that for nearby Type Ia SNe, about 30\% occur in E/S0 galaxies,
while the remaining events occur in spirals (63\%) and irregular
galaxies (7\%).  This distribution is similar to the short GRB
demographics (Figure~\ref{fig:demographics}; \citealt{fbc+13}).  The
Type Ia SNe also exhibit a clear increase in the rate per unit stellar
mass from early- to late-type galaxies, ranging from about 0.044
(E/SO) to 0.065 (Sa/b) to 0.17 (Sc/d) per $10^{10}$ M$_\odot$ per
century \citep{mdp+05}, similar to the inference for short GRBs
(\S\ref{sec:mass}; \citealt{lb10}).

The Type Ia SN rate as a function of galaxy type and properties
(stellar mass, star formation rate) led \citet{mdp06} and
\citet{slp+06} to propose a bimodal or two-component delay time
distribution with a prompt population ($\sim 0.1-0.3$ Gyr) whose rate
depends primarily on the host galaxy star formation rate, and a
delayed population ($\sim 3$ Gyr with a broad range) whose rate
depends on the host galaxy stellar mass (see also
\citealt{mb10,mml+11}).  Indeed, \citet{slp+06} find that the Type Ia
SN rate per unit stellar mass increases by at least an order of
magnitude with higher SSFR (over the range of $0.02-1$ Gyr$^{-1}$).
They conclude that the Type Ia SN rate per year is about $5\times
10^{-4}\, (10^{10}\,{\rm M}_\odot)^{-1}+4\times 10^{-4}\,({\rm
  M}_\odot/{\rm yr})^{-1}$.  More recent work on the delay time
distribution point instead to a continuous, rather than bimodal
distribution, following a power law of about $\tau^{-1.1}$
\citep{mmb12}.  Thus, the results for short GRBs and Type Ia SNe
exhibit close similarities, in terms of both the distribution of host
galaxy types, and the enhanced rate in late-type galaxies, with a
typical delay of $\sim 0.1-0.3$ Gyr.  This indicates that short GRBs
and Type Ia SN progenitors share a similar delay time distribution.

\section{The Locations of Short GRBs in and Around Their Host
  Galaxies}
\label{sec:subgal}

As demonstrated in the previous section, the properties of short GRB
hosts help to reveal the short GRB redshift distribution, to shed
light on the progenitor age distribution, and to establish the
differences between long and short GRB progenitors.  Studies of short
GRB sub-galactic environments can further this insight by providing
information on their spatial association (or lack thereof) with star
formation or stellar mass.  Similar studies of long GRBs have
established that their projected spatial offsets relative to the host
galaxy centers follow the expected distribution for star formation in
exponential disk galaxies, with a median value of $\approx 1$ $r_e$
\citep{bkd02}, where $r_e$ is the host galaxy half-light radius.
Moreover, studies of the rest-frame UV brightness at the locations of
long GRBs relative to the overall UV light distribution of the hosts
have shown that long GRBs tend to occur in unusually bright
star-forming regions, significantly more so than core-collapse SNe
\citep{fls+06,slt+10}.

What is the expectation for the locations of short GRBs?  As I already
demonstrated from the lack of SN associations, the occurrence of some
short GRBs in elliptical galaxies, and the systematically older
stellar population of short GRB hosts compared to long GRB hosts,
short GRBs are not expected to coincide with young star-forming
environments.  On the other hand, based on the distribution of host
galaxy stellar masses, and the host galaxy demographics I concluded
that short GRB progenitors have a broad delay time distribution and
that their rate is influenced in part by star formation activity
(i.e., some systems have delays of only $\sim 0.1$ Gyr).  In addition,
one of the key predictions of the compact object merger model is
systemic natal kicks that may lead to substantial offsets between the
birth and explosion sites of these systems.  Indeed, offset
distributions predicted from population synthesis models indicate a
typical median projected offset of $5-7$ kpc, with about $10-25\%$ of
events occurring at offsets of $\simgt 20$ kpc, and up to $\sim 10\%$
extending beyond 100 kpc \citep{bsp99,fwh99,pb02,bpb+06}.
Observationally, short GRBs at such large offsets will appear to be
``host-less'' since their projected locations will extend much beyond
the visible extent of typical galaxies.  A progenitor population of
young magnetars, on the other hand, will exhibit a spatial coincidence
with star-forming regions and offsets that track an exponential disk
distribution.  Delayed magnetars will not track young star-forming
regions, but will track the overall light distribution of their hosts
due to the lack of natal kicks.  Finally, a dominant population of
dynamically formed NS-NS binaries in globular clusters will track the
overall globular cluster distribution around galaxies, extending to
tens of kpc \citep{sdc+10}.

\subsection{The Offset Distribution}

\begin{figure}
\centerline{\psfig{file=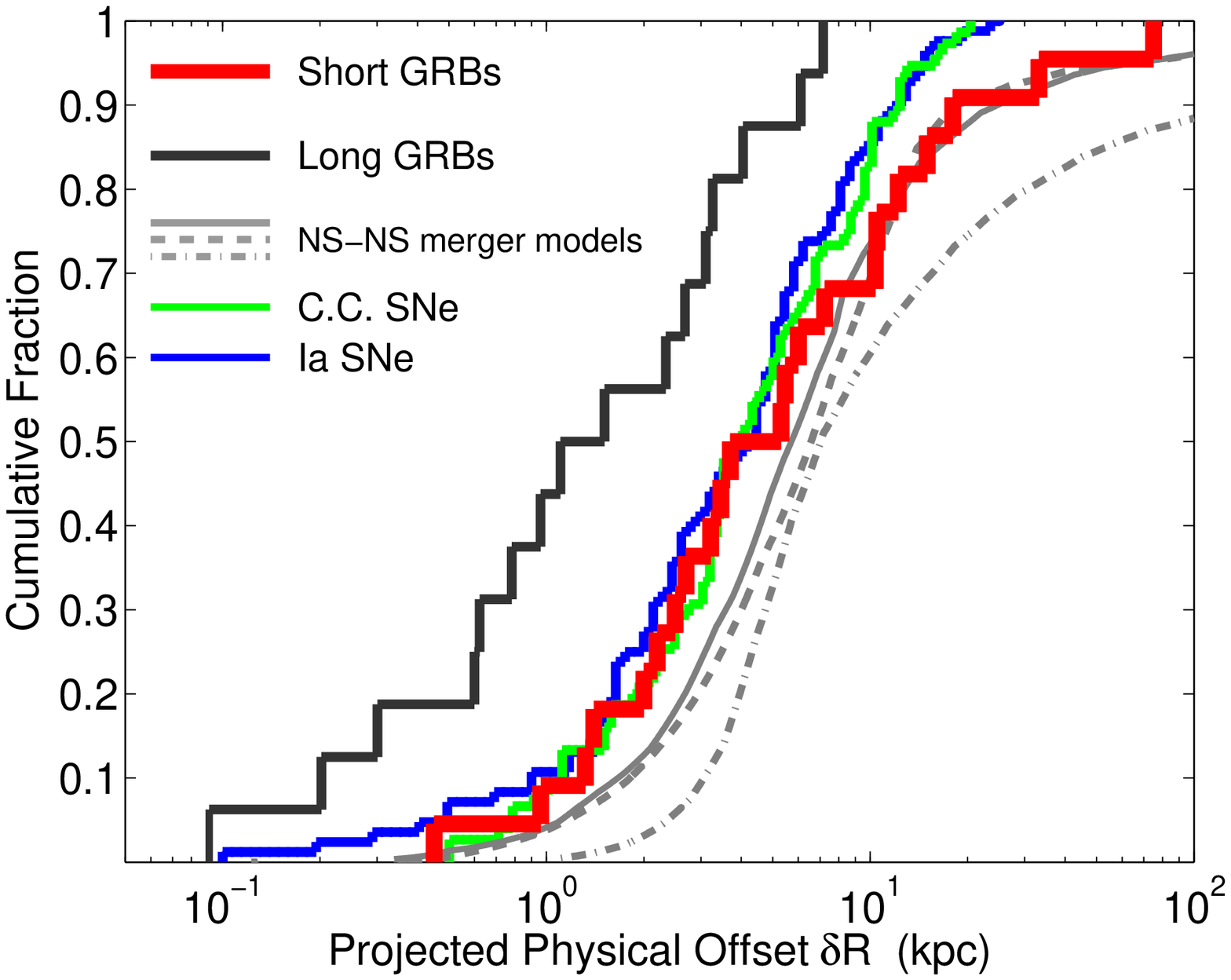,width=\textwidth}}
\caption{Cumulative distribution of projected physical offsets for
  short GRBs with sub-arcsecond positions (red; \citealt{fbf10,fb13}),
  compared to the distributions for long GRBs (black;
  \citealt{bkd02}), core-collapse SNe (green; \citealt{psb08}), Type
  Ia SNe (blue; \citealt{psb08}), and predicted offsets for NS-NS
  binaries from population synthesis models (grey;
  \citealt{bsp99,fwh99,bpb+06}).  Short GRBs have substantially larger
  offsets than long GRBs, and match the predictions for compact object
  binary mergers.  From \citet{fb13}.}
\label{fig:offset1}
\end{figure}

To determine the locations of short GRBs relative to their host
centers (offsets) and relative to the underlying light distribution in
the rest-frame optical (stellar mass) and UV (star formation) requires
the high angular resolution and superior depth of the {\it HST}.  A
comprehensive study based on {\it HST} observations of 32 short GRB
host galaxies was carried out by \citet{fbf10} and \citet{fb13}; see
also \citet{cld+11}.  The {\it HST} data, combined with ground-based
optical afterglow observations (and in two cases, \chandra\
observations), provide accurate offsets at the sub-pixel level, and
reveal the broad distribution shown in Figure~\ref{fig:offset1}.  The
projected offsets span $0.5-75$ kpc with a median of about 5 kpc.
This is about 4 times larger than the median offset for long GRBs
\citep{bkd02}, and about 1.5 times larger than the median offsets of
core-collapse and Type Ia supernovae \citep{psb08}.  In addition,
while no long GRBs, and only $\sim 10\%$ of SNe have offsets of
$\simgt 10$ kpc, the fraction of short GRBs with such offsets is about
$25\%$.  For offsets of $\simgt 20$ kpc, the fraction of short GRBs is
about $10\%$, but essentially no SNe exhibit such large offsets.

Equally important, the observed offset distribution is in remarkable
agreement with the population synthesis predictions for compact object
mergers, particularly the fraction of events with large offsets.  I
note that the observed distribution is mainly based on short GRBs with
optical afterglows, which may be skewed on average to higher
circumburst densities, and hence to smaller offsets.  It is therefore
conceivable that events with only XRT or BAT positions (14 and 6
\swift\ short GRBs, respectively; \S\ref{sec:sample}), whose precise
offsets cannot be ascertained, in reality have systematically larger
offsets than a few kpc.

\begin{figure}
\centerline{\psfig{figure=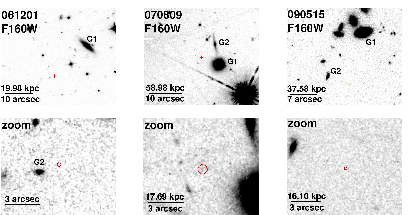,width=\textwidth}}
\caption{{\it Hubble Space Telescope} Wide-Field Camera 3 images of
  the locations of three short GRBs with sub-arcsecond positions and
  no coincident host galaxies (host-less bursts).  In each case the
  top panel shows a wide field and the bottom panel is zoomed on the
  GRB location (red circle).  The galaxies marked ``G1'' and ``G2'' in
  represent the objects with the lowest and second lowest
  probabilities of chance coincidence in each field.  The magnitude
  limits at the GRB positions are $m_{\rm F160W}\simgt 26$ mag, ruling
  out the presence of galaxies typical of short GRB hosts at $z\simlt
  3$.  Adapted from \citet{fb13}.}
\label{fig:hst}
\end{figure}

The subset of short GRBs with offsets of $\simgt 20$ kpc require
specific mention since these offsets are larger than the typical
visible extent of galaxies.  These bursts lack coincident host
galaxies at angular separations of $\simlt 1''$ in deep optical and
near-IR {\it HST} observations, and have been termed ``host-less''.
Systematic studies of the host-less bursts, their afterglows, and
environments was carried out by \citet{ber10} and \citet{fb13},
although individual cases were noted previously
\citep{bpc+07,sdp+07,pmg+09}.  A few examples based on deep {\it HST}
near-IR observations are shown in Figure~\ref{fig:hst}.  In all cases
there is a clear absence of coincident galaxies to optical limits of
$\simgt 27$ mag and near-IR limits of $\simgt 26$ mag
\citep{ber10,fb13}.  Any undetected coincident host galaxies with
properties similar to the known host galaxy sample
(Figure~\ref{fig:sfr}) will have to reside at $z\simgt 3$ to evade
detection, leading to an unexpected bimodal redshift distribution
(Figure~\ref{fig:z}; \citealt{ber10,fb13}).  Similarly, if we impose a
similar redshift range for any undetected hosts, then the resulting
host luminosities will be $\simlt 0.01$ L$^*$, at least an order of
magnitude below the known host population (Figure~\ref{fig:sfr}),
making this explanation equally unlikely.

A more likely alternative is that these bursts are associated with
nearby galaxies in their field, with resulting projected offsets of
tens of kpc, as expected for a subset of compact object mergers.  This
possibility can be assessed by analyzing the probabilities of chance
coincidence for field galaxies in the {\it HST} images.  The probability for
a given galaxy of brightness $m$ to be located at a separation $\delta
R$ from a short GRB position is given by \citep{bkd02,ber10}:
\begin{equation}
P_{\rm cc}=1-e^{-\pi(\delta R)^2\Sigma(\le m)},
\label{eqn:pcc}
\end{equation}
where the galaxy number counts are given by $\Sigma(\le m)= 1.3\times
10^{0.33(m-24)-2.44}$ arcsec$^{-2}$ \citep{hpm+97,bsk+06}.  Applying
this approach, \citet{ber10} and \citet{fb13} find that the host-less
short GRBs exhibit nearby field galaxies with low chance-coincidence
probabilities of $P_{\rm cc}\approx {\rm few}$ percent, indicating a
likely association with resulting projected separations of $\sim
10''$.  A similar analysis for short GRBs with coincident host
galaxies does not reveal a similar effect \citep{ber10}, demonstrating
that the offset galaxies with a low probability of chance coincidence
in the fields of host-less short GRBs are indeed the hosts.  While a
definitive association will benefit from independent redshift
measurements for the afterglow and galaxy, the observed excess of
galaxies at moderate separations is highly indicative of a progenitor
population capable of occurring at large offsets.  The redshifts and
properties of these likely host galaxies reveal that they are similar
to the known population of short GRB hosts, and the resulting
projected physical offsets are tens of kpc \citep{ber10}.

\begin{figure}
\centerline{\psfig{file=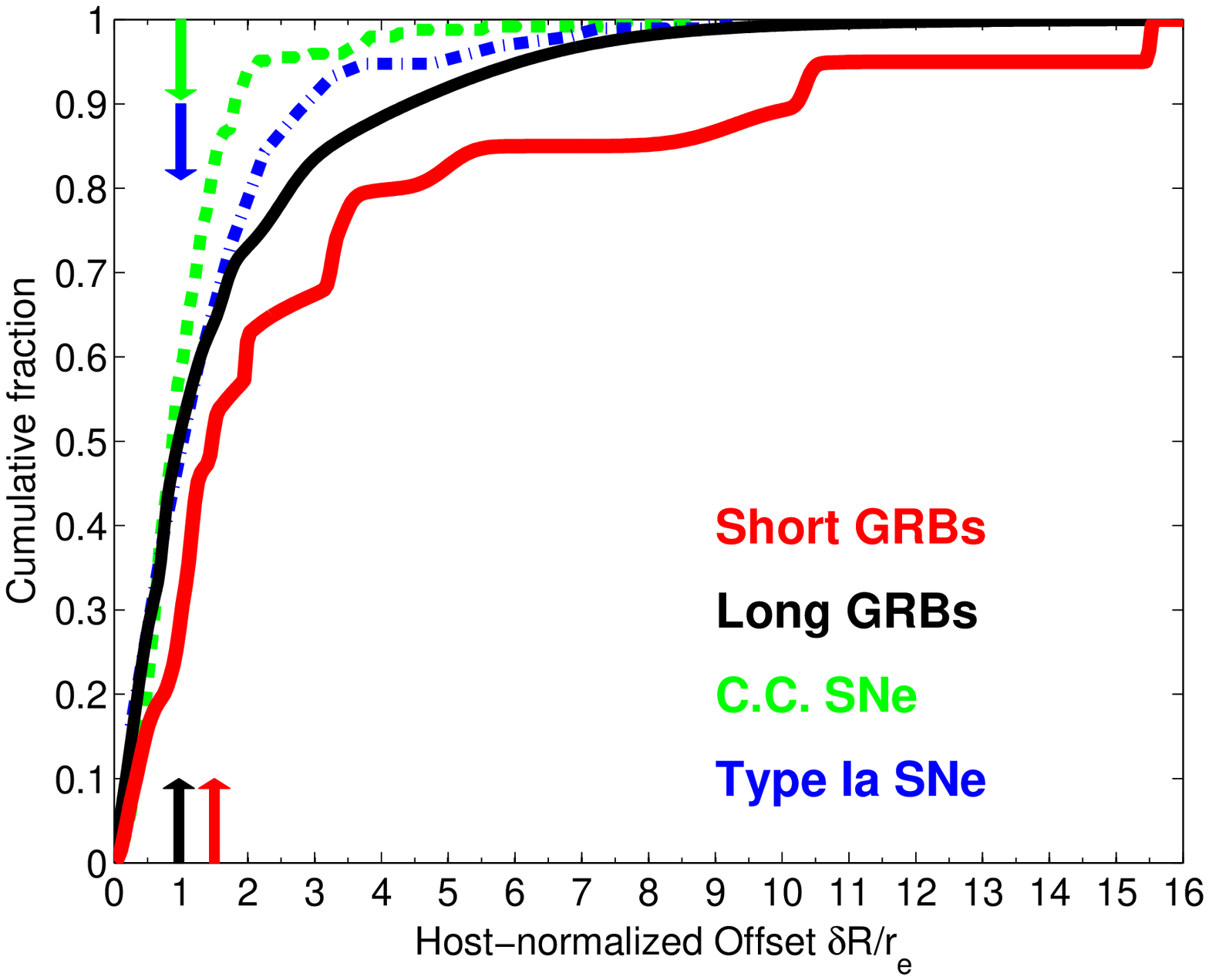,width=\textwidth}}
\caption{Cumulative host-normalized offset distributions for short
  GRBs, long GRBs, core-collapse SNe, and Type Ia SNe (colors are as
  in Figure~\ref{fig:offset1}).  The host-normalized offsets are
  determined relative to the effective radius of each host, and the
  cumulative distributions take into account the uncertainty in each
  offset measurement.  The distributions for core-collapse SNe are
  from \citet{kk12}, while those for Type Ia SNe are from
  \citet{gmo+12}.  The arrows mark the median host-normalized offset
  for each type of transient, with long GRBs and SNe matching the
  expected value of about 1 $r_e$.  Short GRBs, on the other hand,
  have a median offset of about 1.5 $r_e$ and only 20\% of events
  occur within 1 $r_e$.  From \citet{fb13}.}
\label{fig:offset2}
\end{figure}

While the projected physical offsets are already indicative of compact
object mergers, it is important to account for the range of host
galaxy sizes, and any systematic trends in these sizes between the
various GRB and SN populations.  This can be accomplished by
normalizing the projected offsets in units of host galaxy effective
radii.  Short GRB hosts tend to be larger than long GRB hosts,
commensurate with their larger luminosities and stellar masses
\citep{fbf10,fb13}.  The cumulative distributions of host-normalized
offsets for short GRBs, long GRBs, core-collapse SNe, and Type Ia SNe
is shown in Figure~\ref{fig:offset2}.  The median host-normalized
offsets for long GRBs and for both Type Ia and core-collapse SNe is
$\delta R/r_e\approx 1$ \citep{fb13}, as expected from the definition
of the half-light radius and the fact that long GRB and SN progenitors
do not migrate from their birth-sites. Short GRBs, on the other hand,
have a median offset of $\delta R/r_e\approx 1.5$, with $20\%$ of the
population occurring at $\delta R/r_e\simgt 5$ (compared to only a few
percent of stellar light) , and only $20\%$ of the bursts located at
$\delta R/r_e\simlt 1$.  A K-S test relative to the Type Ia SN
population yields a $p$-value of about $10^{-3}$ for the null
hypothesis that both populations are drawn from the same underlying
distribution of host-normalized offset \citep{fb13}.  Thus, the offset
distribution of short GRBs indicates that they are located at larger
distances than expected for a progenitor population that radially
tracks stellar light, particularly when accounting for their host
galaxy sizes.  The offsets are larger than even those of Type Ia SNe,
but are in good agreement with compact object population synthesis
predictions that include kicks.

\subsection{Relation to the Underlying Ultraviolet and Optical Light
  Distribution}

Independent of the offset distribution, an additional test of the
progenitor population is the spatial correlation (or lack thereof)
between short GRBs and the underlying light distribution of their
hosts.  In particular, for progenitors that track star formation
activity with a short delay (i.e., massive stars) we expect direct
spatial correlation with the underlying rest-frame UV light (as found
for long GRBs and core-collapse SNe;
\citealt{fls+06,kkp08,slt+10,kk12}).  On the other hand, an old
progenitor population without a kick mechanism will spatially track
rest-frame optical light (as is the case for Type Ia SNe;
\citealt{wwf+13}).  Finally, compact object mergers with significant
kicks will result in a distribution that is only weakly correlated
with the underlying rest-frame optical light.

\begin{figure}
\centerline{\psfig{file=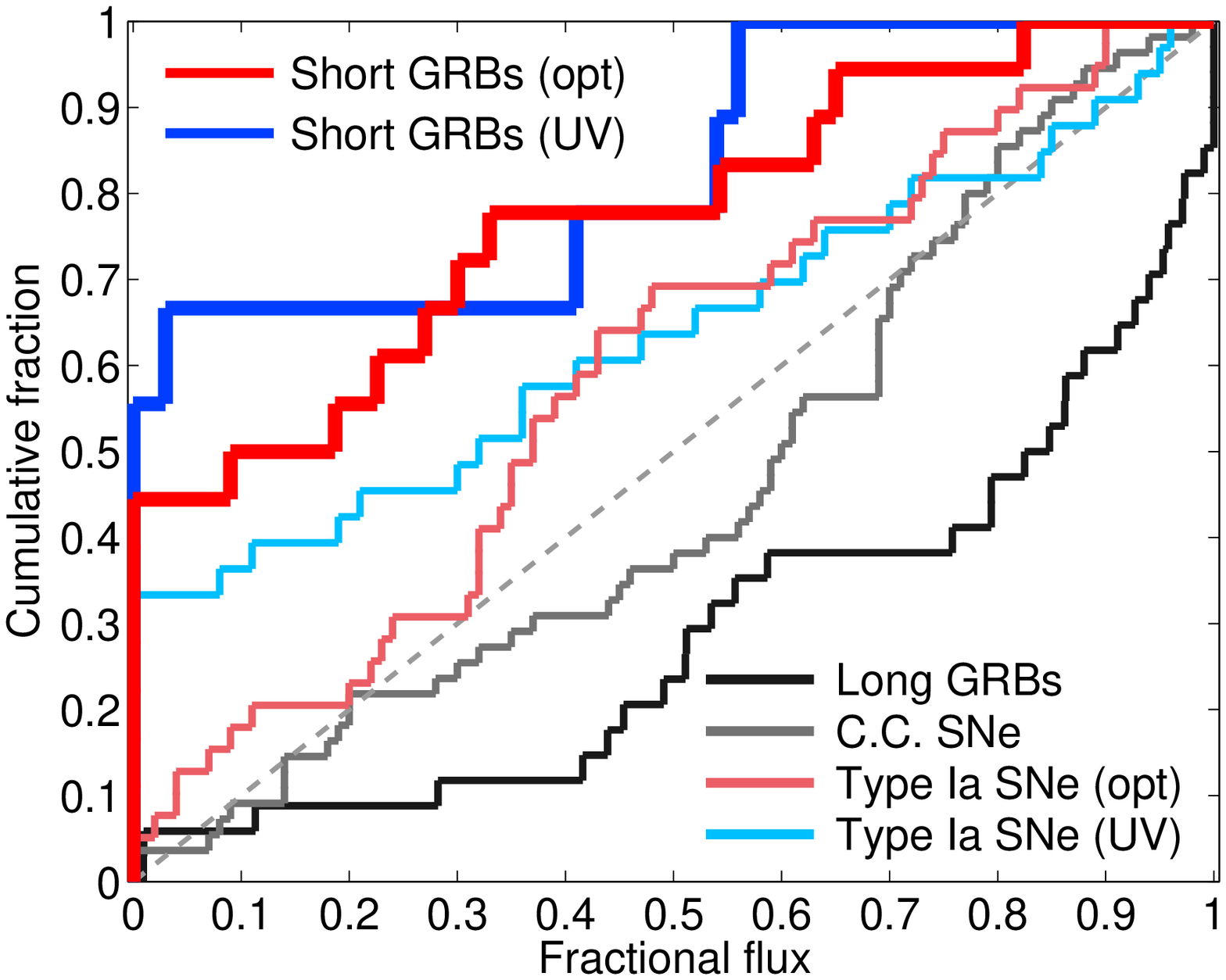,width=\textwidth}}
\caption{Cumulative distribution of the fractional flux at short GRB
  locations relative to the underlying light distributions of their
  hosts.  Shown are the distributions for rest-frame optical (red) and
  rest-frame UV (blue) from {\it HST} observations of short GRBs with
  sub-pixel positions \citep{fbf10,fb13}.  Also shown are the
  distributions for long GRBs (UV: black; \citealt{fls+06,slt+10}),
  Type Ia SNe (UV: light blue; optical: pink; \citealt{wwf+13}), and
  core-collapse SNe (UV: grey; \citealt{slt+10}).  The diagonal dashed
  line marks the expected distribution for a population that linearly
  tracks the underlying stellar light.  Long GRBs and core-collapse
  SNe strongly track UV light (star formation), while Type Ia SNe
  track optical light (stellar mass).  Short GRBs, on the other hand,
  exhibit a poor correlation with UV or optical light, indicating that
  the progenitors migrate from their birth-sites to the eventual
  explosion sites.  From \citet{fb13}.}
\label{fig:lightdist}
\end{figure}

\citet{fbf10} and \citet{fb13} determined the light fraction at the
sub-pixel locations of 20 short GRBs relative to the overall light
distribution of each host galaxy (following the methodology of
\citealt{fls+06}).  These measurements are available only for bursts
with sub-arcsecond positions and {\it HST} follow-up, but as I
discussed above, the requirement of an optical afterglow detection
might only bias the resulting distribution to locations with higher
ambient densities, and hence to brighter regions within the hosts.
The cumulative distributions of the light fraction at short GRB
locations relative to the underlying rest-frame UV and optical light
are shown in Figure~\ref{fig:lightdist}.  The results are striking;
namely, about half of the short GRBs are located in regions of their
hosts that are fainter in both the UV and optical than any other
location within these galaxies (i.e., they have a fractional flux
value of zero).  This essentially means that even with the superior
depth of {\it HST}, no stellar light is detected at the short GRB
locations.  This result remains true even if the short GRBs at
projected offsets of tens of kpc (the host-less events) are ignored
\citep{fb13}.

A K-S test relative to the null hypothesis that the progenitors track
the underlying UV or optical light results in $p$-values of 0.01 and
0.04, respectively \citep{fb13}.  Similarly, a comparison to long GRBs
in terms of UV light leads to $p\approx 2\times 10^{-3}$, while a
comparison to Type Ia SNe in terms of optical light indicates
$p\approx 0.02$.  This indicates that short GRBs occur in distinct
environments from other explosive transients whose progenitors are not
expected to experience kicks (massive stars, single- or
double-degenerate white dwarf systems), and are overall weakly
correlated with the underlying distribution of star formation and even
stellar mass in their hosts \citep{fb13}.  Given this clear
discrepancy, the implication is that short GRB explosion sites are not
representative of the progenitor birth-sites, and therefore require
significant migration (kicks).  This is a strong line of argument in
favor of compact object mergers.

\subsection{Kick Velocities}

The offset distribution, and the locations of short GRBs relative to
the stellar light of their hosts, are indicative of systemic kicks.
The required kick velocities can be inferred from a comparison of the
projected physical offsets and the host galaxy stellar population
ages, taking into account the host masses.  While projection effects
and uncertainty about the age of any specific short GRB progenitor
relative to the host mean stellar population age prevent the
extraction of a detailed distribution, it is still possible to
determine characteristic kick velocities.  Making the assumption
\citep{bpc+07} that the required kick velocities are roughly the
geometric mean between the minimum required velocity ($v_{\rm
  kick,min}=\delta R/\tau_*$) and the host velocity dispersion
($v_{\rm disp}\approx 120$ km s$^{-1}$ for late-type hosts and
$\approx 250$ km s$^{-1}$ for the more massive early-type hosts;
\citealt{bhm+05,xrz+08}), the resulting projected kick velocities span
$v_{\rm kick}\sim 20-140$ km s$^{-1}$ with a median value of about 60
km s$^{-1}$ \citep{fb13}.  This range is consistent with the kick
velocities derived for Galactic NS-NS binaries based on population
synthesis models, $v_{\rm kick}\approx 5-500$ km s$^{-1}$
\citep{fk97,fbb98,wlh06,wwk10}.  It is important to note that the
inferred distribution of kicks for short GRBs is consistent with some
progenitors having negligible kicks, and it does not strictly require
velocities that will unbind the progenitors from their hosts.

\subsection{Globular Clusters}

An alternative explanation for the broad offset distribution and weak
correlation with stellar light is that some short GRB progenitors
originate in globular clusters, due to dynamically-formed compact
object binaries, or to the tidal capture and collisions of compact
objects (e.g., \citealt{bv91,gpm06,lrv10,sdc+10}).  Estimates of the
event rate for these processes compared to primordial binaries is
$\simlt 10\%$ \citep{gpm06,lrv10}.  This is roughly similar to the
$\sim 10\%$ of short GRBs with inferred offsets of $\simgt 20$ kpc,
but it is not sufficient to explain the weak correlation with stellar
light, i.e., $\sim 50\%$ of short GRBs with fractional flux values of
zero.

A substantial fraction of short GRBs originating in globular clusters
will also lead to the prediction of a dominant population of
early-type hosts since the globular cluster specific frequency is
several times higher in early-type than in late-type galaxies
\citep{har91,bs06}.  This is in contrast to the observed low fraction
of early-type hosts in the short GRB sample, which is already
deficient relative to the cosmic stellar mass budget
(\S\ref{sec:hosts}).  In addition, formation of binaries in globular
clusters generally requires core-collapsed clusters, thereby leading
to a significant time delay relative to the cluster formation epoch
\citep{gpm06,lrv10,sdc+10}.  As a result, a dominant globular cluster
population will lead to systematically lower redshifts; for example,
\citet{lrv10} predict a median redshift of $z\approx 0.5$, and
essentially no short GRBs at $z\simgt 1.5$.  This result is not in
strong conflict with the current redshift distribution, although the
median redshift of short GRBs is probably somewhat higher than $z\sim
0.5$ (\S\ref{sec:hosts}).  Thus, while it is unlikely that most short
GRBs originate in globular clusters, it is possible that $\sim 10\%$
of the events originate in these environments.

\section{The Afterglows and Explosion Properties of Short GRBs}
\label{sec:afterglow}

Having explored the galactic and sub-galactic environments of short
GRBs, and the implications for the nature of their progenitors, I now
turn to an investigation of the short GRB explosion properties and
their parsec-scale circumburst environments.  The key parameters of
interest that can be extracted from the prompt and afterglow emission
are the energy scale in $\gamma$-rays ($E_\gamma$) and in the
blastwave powering the afterglow ($E_K$), the geometry of the outflow
(characterized by a jet half opening angle, $\theta_j$), and the
density of the ambient medium ($n$).  The standard formulation of the
afterglow synchrotron emission includes additional free parameters
related to the relativistic shock microphysics: the fraction of
post-shock energy in the magnetic fields ($\epsilon_B$) and in the
radiating relativistic electrons ($\epsilon_e$), which are assumed to
follow a power law distribution, $N(\gamma)\propto\gamma^{-p}$, above
a minimum Lorentz factor, $\gamma_m$ \citep{spn98}.  To determine the
values of the various parameters requires multi-wavelength,
multi-epoch afterglow observations spanning radio to X-rays.

The synchrotron emission spectrum resulting from the interaction of
the relativistic outflow with the circumburst medium is characterized
by three break frequencies, corresponding to self-absorption
($\nu_a$), the minimum Lorentz factor ($\nu_m$), and synchrotron
cooling ($\nu_c$), as well as an overall flux normalization
\citep{spn98}.  The instantaneous values of these four parameters and
the slope of the spectrum at $\nu>\nu_m$ can be converted to $E_{\rm
  K,iso}$, $n$, $\epsilon_e$, $\epsilon_B$, and $p$; here $E_{\rm
  K,iso}$ is the isotropic-equivalent blastwave kinetic energy.  In
addition, the time evolution of the spectrum can be used to extract
$\theta_j$ and $p$.  This simple formulation has been confirmed with
observations of multiple long GRBs (e.g., \citealt{pk02,yhs+03}),
although some deviations from the standard model are apparent at
$\simlt 10^3$ s, particularly in the X-ray band (e.g., initial steep
decline, flares, plateaus; \citealt{nkg+06,cmr+07}).  Using the
formulation of the afterglow model in \citet{gs02} I employ the
following relations to determine the locations of the break
frequencies (relevant for the dominant slow-cooling regime, with
$\nu_a<\nu_m<\nu_c$):
\begin{eqnarray} 
\nu_a=1.24\times 10^9\,{\rm Hz}\,\frac{(p-1)^{3/5}}{(3p+2)^{3/5}}
(1+z)^{-1} \bar{\epsilon_e}^{-1} \epsilon_B^{1/5} n_0^{3/5}
E_{52}^{1/5} \\
\nu_m=3.73\times 10^{15}\,{\rm Hz}\,(p-0.67) (1+z)^{1/2}
\bar{\epsilon_e}^{2} \epsilon_B^{1/2} E_{52}^{1/2} t_d^{-3/2} \\
\nu_c=6.37\times 10^{13}\,{\rm Hz}\,(p-0.46) e^{-1.16p} (1+z)^{-1/2}
\epsilon_B^{-3/2} n_0^{-1} E_{52}^{-1/2} t_d^{-1/2} \\
F_{\nu_a}=0.65\,{\rm mJy}\,\frac{(p-1)^{6/5}}{(3p-1)(3p+2)^{1/5}}
(1+z)^{1/2} \bar{\epsilon_e}^{-1} \epsilon_B^{2/5} n_0^{7/10}
E_{52}^{9/10} t_d^{1/2} d_{L,28}^{-2}
\label{eqn:ag}
\end{eqnarray} 
where $\bar{\epsilon_e}\equiv\epsilon_e(p-2)/(p-1)$, $E_{52}$ is the
isotropic-equivalent blastwave kinetic energy in units of $10^{52}$
erg, $t_d$ is the observer-frame time in days, and $d_{L,28}$ is the
luminosity distance in units of $10^{28}$ cm.  For fiducial parameters
relevant to short GRBs ($z\sim 0.5$, $E_{52}\sim E_{\rm
  \gamma,iso,52}\sim 0.1$, $n_0\sim 0.1$), and assuming the typical
microphysical parameters from long GRB afterglow data ($\epsilon_e\sim
0.1$, $\epsilon_B\sim 0.01$, $p\sim 2.4$; \citealt{pk02,yhs+03}), the
typical values of the synchrotron parameters are: $\nu_a\sim 1$ GHz,
$\nu_m\sim 200\,t_d^{-1.5}$ GHz, $\nu_c\sim 2\times
10^{17}\,t_d^{-0.5}$ Hz, and $F_{\nu_a}\sim 20$ $\mu$Jy.  These values
confirm that observations from radio to X-rays are required to probe
the full set of synchrotron parameters.  They also indicate that the
typical brightness level in the radio band ($\sim 1-10$ GHz) is quite
challenging for detection, even with the upgraded Karl G.~Jansky Very
Large Array.  Similarly, these parameters predict a faint optical
afterglow brightness level of $\sim 23$ AB mag at $\sim 7$ hr, the
mean timescale for existing searches (see also \citealt{pkn01}).

In reality, the afterglow data for short GRBs are much sparser than
for long GRBs, and the complete set of physical parameters for most
individual events cannot be determined uniquely.  For example, only
three short GRBs to date have been detected in the radio
\citep{bpc+05,sbk+06,fbm+13}, providing meaningful constraints on the
location of the self-absorption frequency.  Still, using basic
constraints, including expectations for the values of the
microphysical parameters, and a combination of detections and upper
limits in the radio, optical, and X-rays, it is possible to determine
the distributions of some key short GRB properties.

Of particular interest for our understanding of the central engine and
energy extraction mechanism is the beaming-corrected energy, which
requires the determination of jet opening angles (see
\S\ref{sec:jets}).  The jet collimation also influences the true short
GRB event rate, which in turn provides an additional constraint on the
progenitor model, and potentially useful information on the rate of
gravitational wave detections in the specific context of compact
object merger progenitors.  Finally, the circumburst density provides
insight on the progenitors, through constraints on their parsec-scale
environments.  Beyond the direct properties of short GRBs, the energy
scale and circumburst densities can also be used to shed light on the
expected brightness of electromagnetic counterparts to gravitational
wave sources, both on- and off-axis; I discuss this point in detail in
\S\ref{sec:gwem}.

To date there have been only a few studies of short GRB afterglows.
\citet{bpc+05} determined the properties of GRB\,050724 from radio,
optical/near-IR, and X-ray observations, while \citet{gbp+06} placed
lower bounds on its collimation.  \citet{sbk+06} determined the
explosion properties of GRB\,051221A, and along with \citet{bgc+06}
measured its opening angle.  \citet{fbm+13} presented the radio,
optical, and X-ray afterglow of GRB\,130603B and determined its
properties and jet collimation angle.  \citet{ber07} presented an
analysis of the $\gamma$-ray and X-ray observations of 16 short GRBs
that revealed a clear correlation between $E_{\rm \gamma,iso}$ and
$L_{\rm X,iso}$ (a rough proxy for $E_{\rm K,iso}$), as well as a
broad range of isotropic-equivalent energies in the prompt emission
and afterglows (see also \citealt{nak07,gbb+08}).  Similarly,
\citet{nfp09} investigated correlations between the $\gamma$-ray,
X-ray, and optical emission.

\subsection{X-ray Afterglow Emission}

\begin{figure}
\centerline{\psfig{figure=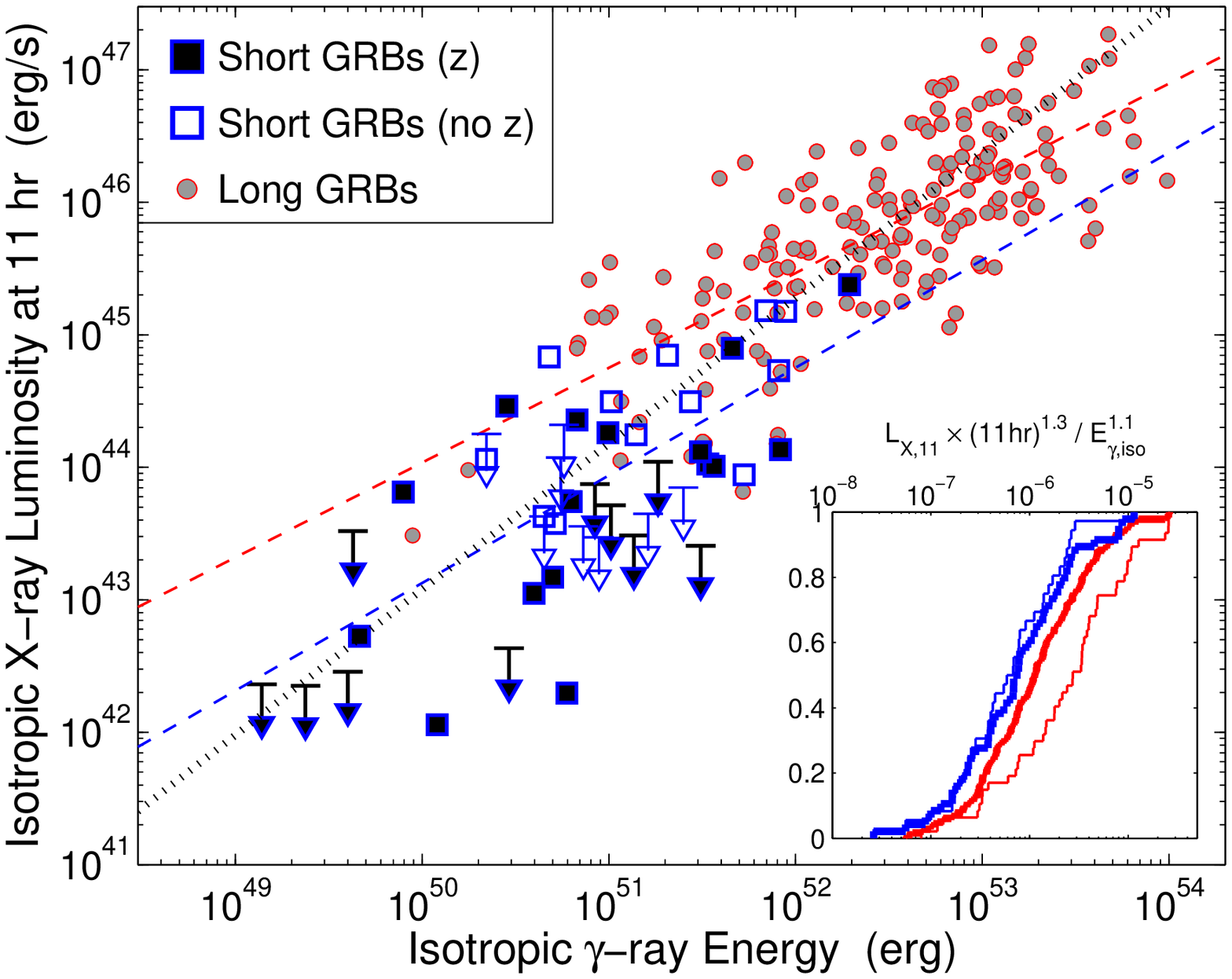,width=\textwidth}}
\caption{Isotropic-equivalent afterglow X-ray luminosity at a
  rest-frame time of 11 hr ($L_{\rm X,11}$) versus the
  isotropic-equivalent $\gamma$-ray energy ($E_{\rm \gamma,iso}$) for
  short GRBs (blue) and long GRBs (gray).  Open symbols for short GRBs
  indicate events without a known redshift, for which a fiducial value
  of $z=0.75$ is assumed.  The dashed blue and red lines are the
  best-fit power law relations to the trends for short and long GRBs,
  respectively, while the dotted black line is the expected
  correlation based on the afterglow synchrotron model with
  $\nu_X>\nu_c$ and $p=2.4$ ($L_{X,11}\propto E_{\rm K,iso}^{1.1}$).
  The inset shows the distribution of the ratio $L_{X,11} \times
  (11\,{\rm hr})^{1.3}/ E_{\rm\gamma,iso}^{1.1}$, for the full samples
  (thick lines) and for bursts in the region of $E_{\rm\gamma,iso}$
  overlap (thin lines).  The lower level of $L_{X,11}$ relative to
  $E_{\rm\gamma,iso}$ for short GRBs is evident from these various
  comparisons.}
\label{fig:lxeg}
\end{figure}

The most extensive afterglow data set for short GRBs is in the X-ray
band from \swift/XRT (Table~\ref{tab:afterglows}).  As shown in
\S\ref{sec:sample}, of the 50 short GRBs with X-ray detections, $28$
events exhibit detectable long-term X-ray emission beyond $\sim 10^3$
s, as expected for afterglow emission, while the remaining events
rapidly fade below the XRT detection threshold at $\simlt 10^3$ s, and
therefore yield only upper limits on the timescale dominated by the
afterglow.  The initial rapid decline phase is generally attributed to
high latitude emission associated with the prompt phase (e.g.,
\citealt{nkg+06}), and therefore cannot be reliably used to constrain
the afterglow properties.  In Figure~\ref{fig:lxeg}, I plot the
isotropic-equivalent X-ray luminosity of short and long GRBs at a
fiducial rest-frame time of 11 hr in a rest-frame band of $0.3-10$ keV
($L_{X,11}$) as a function of the isotropic-equivalent $\gamma$-ray
energy ($E_{\rm\gamma,iso}$).  I use $E_{\rm\gamma,iso}$ as a proxy
for the more directly relevant, but generally unknown $E_{\rm K,iso}$.
This is a reasonable approach for each GRB sample individually since
variations in the radiative $\gamma$-ray efficiency,
$\epsilon_\gamma\equiv E_{\rm\gamma,iso}/ (E_{\rm\gamma,iso}+E_{\rm
  K,iso}$, will increase the scatter, but should not introduce
systematic trends.  On the other hand, any systematic trends in
$\epsilon_\gamma$ between short and long GRBs may lead to systematic
effects when using $E_{\rm\gamma,iso}$ as a proxy for $E_{\rm K,iso}$.
I return to this point in \S\ref{sec:implications}.

As expected, both the short and long GRB samples exhibit a clear
correlation between $L_{X,11}$ and $E_{\rm\gamma,iso}$
(Figure~\ref{fig:lxeg}; \citealt{ber07,gbb+08,nfp09}).  For long GRBs
the best-fit power law relation is $L_{X,11}\approx 5.6\times
10^{44}\, E_{\rm\gamma,iso,51}^{0.72}$ erg s$^{-1}$, while for short
GRBs it is systematically lower, with $L_{X,11}\approx 8.5\times
10^{43}\, E_{\rm\gamma,iso,51}^{0.83}$ erg s$^{-1}$.  In the region of
overlap between the two GRB populations ($E_{\rm\gamma,iso}\approx
3\times 10^{50}-10^{52}$ erg), the short GRB isotropic X-ray afterglow
luminosity is on average about 7 times lower than for long GRBs.

Comparing the short GRBs with afterglow detections to those with only
a rapid decline phase and no X-ray detections beyond $\sim 10^3$ s, I
find no clear difference in the ratio $L_{X,11}/E_{\rm \gamma,iso}$.
In particular, short GRBs with X-ray afterglow detections span
$L_{X,11}/E_{\rm \gamma,iso}\approx 10^{-8}-10^{-6}$, with $40\%$ of
these events having $L_{X,11}/E_{\rm \gamma,iso}\simlt 10^{-7}$.  The
upper limits are generally $L_{X,11}/E_{\rm \gamma,iso} \simlt
10^{-7}$ (although some shallower limits exist as well).  Thus, it
appears that events without detected X-ray emission beyond $\sim 10^3$
s are generally due to inadequate searches, rather than to
intrinsically weaker X-ray afterglows.

In the context of the standard afterglow synchrotron model, the X-ray
band is expected to be located near or above the synchrotron cooling
frequency ($\nu_X>\nu_c$; see discussion following
Equation~\ref{eqn:ag}).  In this spectral regime the afterglow X-ray
flux is given by $F_{\nu,X}\propto E_{\rm K,iso}^{(p+2)/4}\,
\epsilon_e^{p-1}\, \epsilon_B^{(p-2)/4} \approx \epsilon_e\,E_{\rm
  K,iso}$ \citep{gs02}.  Assuming that $E_{\rm\gamma,iso}$ is a close
proxy for $E_{\rm K,iso}$, it is evident that the observed
correlations for short and long GRBs are flatter than the theoretical
expectation.  At the same time, the observed dispersion around this
relation is at least a factor of few, suggestive of dispersion in the
values of $p$, $\epsilon_e$, and $\epsilon_B$, as well as cases with
$\nu_X<\nu_c$ for which the density also influences the observed flux.

In Figure~\ref{fig:lxeg} I also plot the distribution of the ratio
$L_{X,11}\times (11\,{\rm hr})^{1.3}/E_{\rm \gamma,iso}^{1.1}$, which
is the expected relation for $p=2.4$ and $E_{\rm \gamma,iso}=E_{\rm
  K,iso}$ based on the afterglow model.  For the full samples of short
and long GRBs I find that the distributions of this ratio are similar,
although with a systematically lower value for short GRBs; a K-S test
yields a $p$-value of about 0.23, indicating that the two samples are
compatible with the null hypothesis of being drawn from the same
underlying distribution.  On the other hand, using only events in the
overlap region of $E_{\rm\gamma,iso}\approx 3\times 10^{50}-10^{52}$
erg yields a substantial difference (K-S test value of $p\approx
1.1\times 10^{-4}$).  I therefore conclude that short GRBs have
systematically weaker X-ray afterglow emission compared to long GRBs,
relative to both the observed correlation between $L_{X,11}$ and
$E_{\rm \gamma,iso}$, and the relation expected from afterglow theory.

\subsection{Optical Afterglow Emission}

I next turn to the optical afterglow emission, which has been detected
in 24 short GRBs, with upper limits for an additional 16 bursts
(\S\ref{sec:sample}).  Most detections and limits have been obtained
on a timescale of about 1 hr to 1 day after the burst, with a
magnitude range of about $21-26$ AB mag (Table~\ref{tab:afterglows}).
The median $r$-band brightness level at a fiducial time of 7 hr is
$\langle m_r(7\,{\rm hr})\rangle\approx 23.2$ AB mag, in excellent
agreement with the expectation for the fiducial set of parameters
discussed in relation to Equations~\ref{eqn:ag}, and with the
prediction of \citet{pkn01}.  For comparison, long GRB afterglows are
generally an order of magnitude brighter, with $\langle m_r(7\,{\rm
  hr})\rangle \approx 20.8$ AB mag (e.g., \citealt{kkz+11}), despite
their significantly larger redshifts (Figure~\ref{fig:z}).

\begin{figure}
\centerline{\psfig{figure=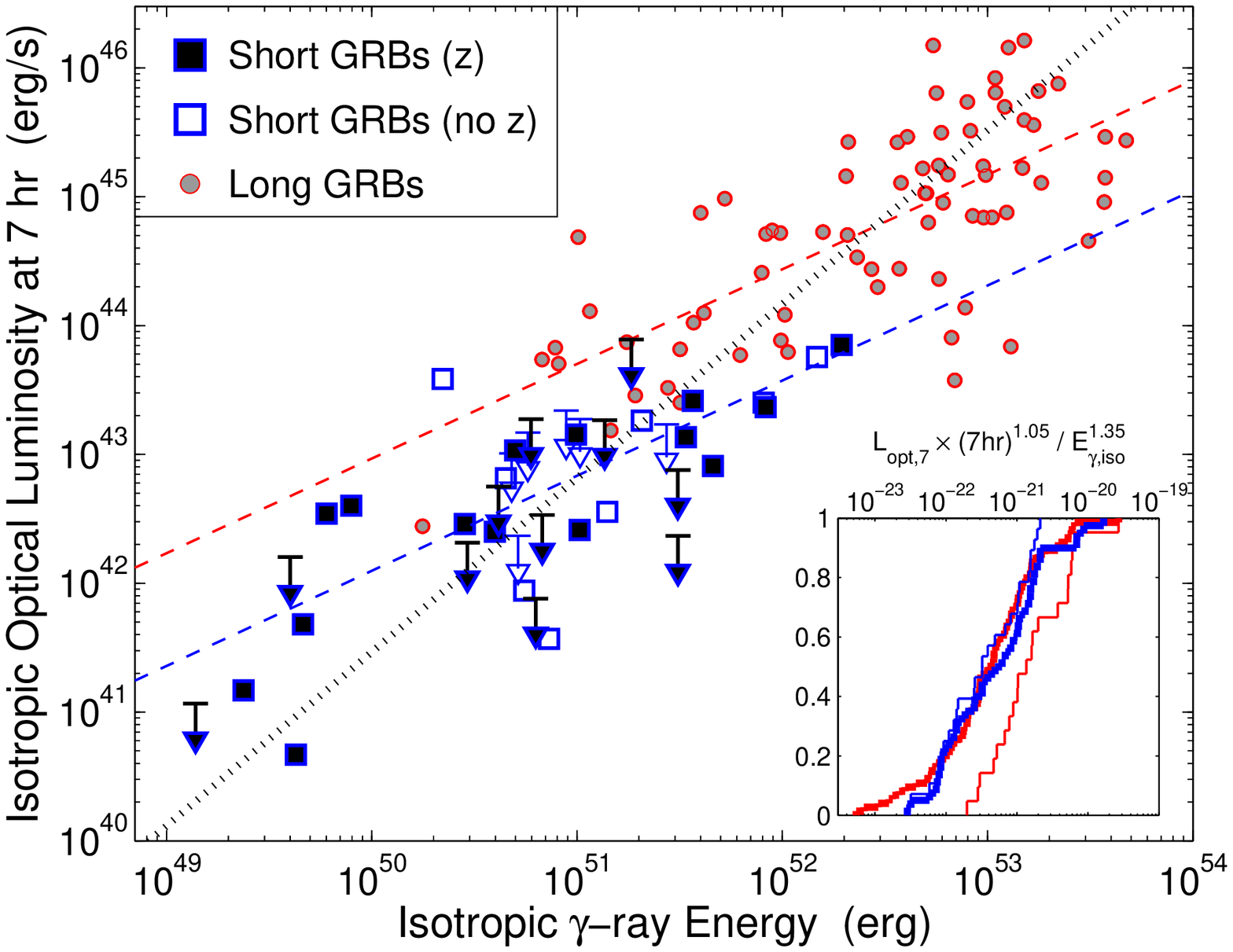,width=\textwidth}}
\caption{Same as Figure~\ref{fig:lxeg} but for the
  Isotropic-equivalent afterglow optical luminosity at a rest-frame
  time of 7 hr ($L_{\rm opt,7}$).  The dotted black line is the
  expected correlation based on the afterglow model for
  $\nu_m<\nu_{\rm opt}<\nu_c$ and $p=2.4$ ($L_{\rm opt,7}\propto
  E_{\rm K,iso}^{1.35}$).  The inset shows the distribution of the
  ratio $L_{\rm opt,7} \times (7\,{\rm hr})^{1.05}/
  E_{\rm\gamma,iso}^{1.35}$, for the full samples (thick lines) and
  for bursts in the region of $E_{\rm\gamma,iso}$ overlap (thin
  lines).  The lower level of $L_{\rm opt,7}$ relative to
  $E_{\rm\gamma,iso}$ for short GRBs is evident from these various
  comparisons.}
\label{fig:loeg}
\end{figure}

Converting to rest-frame properties, in Figure~\ref{fig:loeg} I plot
the isotropic-equivalent optical luminosity in the rest-frame $r$-band
at a fiducial rest-frame time of 7 hours ($L_{\rm opt,7}$) as a
function of $E_{\rm\gamma,iso}$.  As in the case of the isotropic
X-ray luminosity, there is a clear correlation between the optical
luminosity and $\gamma$-ray energy.  For short GRBs, the best-fit
power law relation is $L_{\rm opt,7}\approx 6.9\times 10^{42}\,
E_{\rm\gamma,iso,51}^{0.74}$ erg s$^{-1}$, while for long GRBs the
relation is systematically higher, $L_{\rm opt,7}\approx 5.0\times
10^{43}\,E_{\rm\gamma,iso,51}^{0.73}$ erg s$^{-1}$.  Thus, in the
range of $E_{\rm\gamma,iso}$ in which the short and long GRB
populations overlap, the optical afterglow luminosity of short GRBs is
on average about 7 times fainter than for long GRBs, just as in the
case of the X-ray luminosity.

The expected trend based on the afterglow synchrotron model (for
the case of $\nu_m<\nu_{\rm opt}<\nu_c$) is $F_{\rm\nu,opt}\propto E_{\rm
  K,iso}^{(p+3)/4}\, n_0^{1/2}\, \epsilon_e^{p-1}\,
\epsilon_B^{(p+1)/4}$.  For a typical value of $p=2.4$, this indicates
a steeper slope than inferred from the data, with $F_{\rm\nu,opt}
\propto E_{\rm K,iso}^{1.35}$.  This trend is shown in
Figure~\ref{fig:loeg} with a dotted line, and indeed seems to be a
reasonable representation of the full range of $E_{\rm\gamma,iso}$
spanned by short and long GRBs.  This is also borne out in the nearly
identical cumulative distributions of the ratio $L_{\rm opt,7}\times
(7\,{\rm hr})^{1.05}/E_{\rm \gamma, iso}^{1.35}$ for both short and
long GRBs (Figure~\ref{fig:loeg}).  However, despite the overall
similarity, the short GRBs have systematically lower values of this
ratio in the overlap range of $E_{\rm\gamma,iso}\approx 3\times
10^{50}-10^{52}$ erg.  A K-S test yields a $p$-value of about 0.016,
indicating that the two samples are not likely to be drawn from the
same underlying distribution.  Thus, despite the larger scatter in the
optical band compared to the X-ray band, it appears that short GRBs
have systematically weaker optical afterglow emission compared to long
GRBs, relative to both the observed correlation between $L_{\rm
  opt,7}$ and $E_{\rm \gamma,iso}$, and to the relation expected from
afterglow theory.

\subsubsection{X-ray / Optical Comparison}

\begin{figure}
\centerline{\psfig{figure=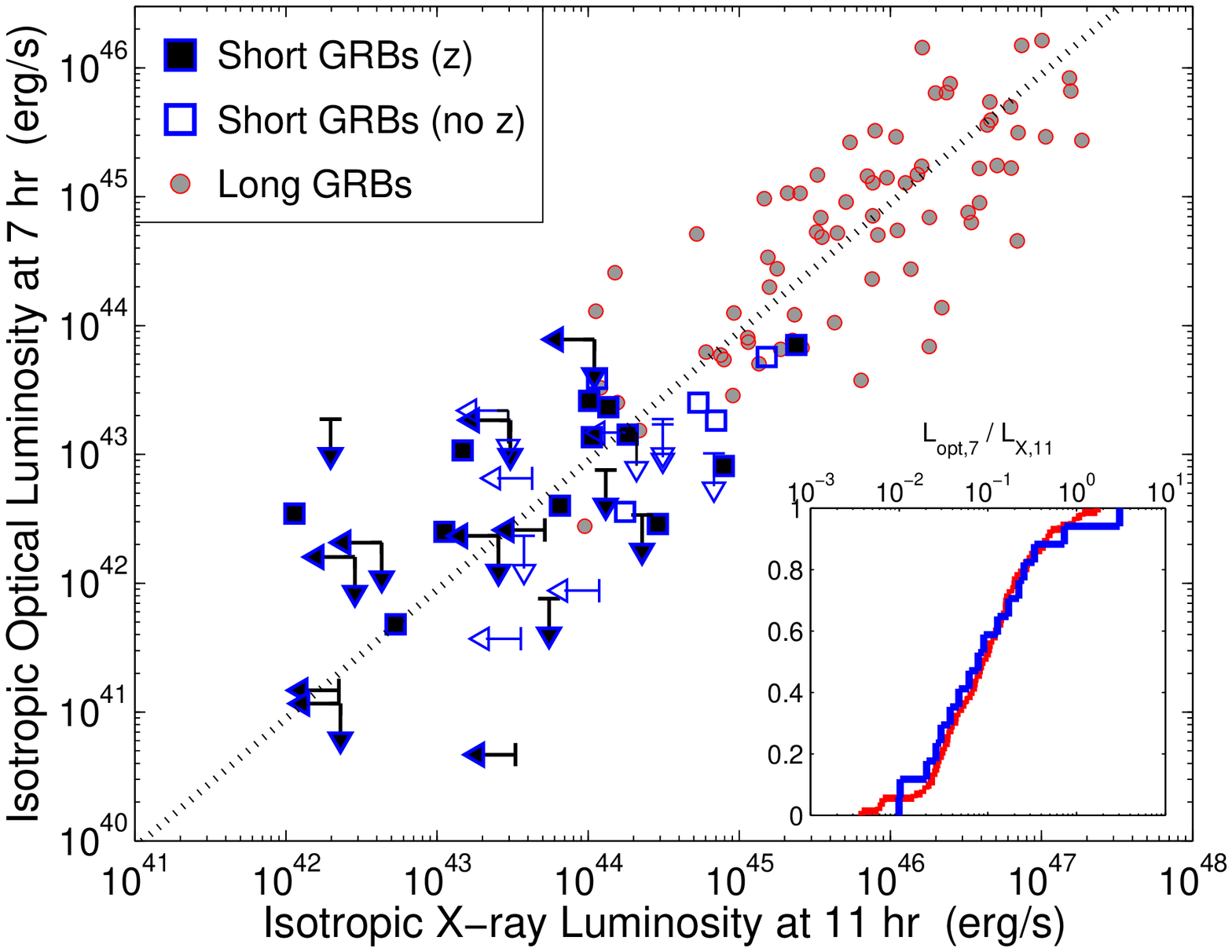,width=\textwidth}}
\caption{Isotropic-equivalent afterglow optical luminosity at a
  rest-frame time of 7 hr ($L_{\rm opt,7}$) versus
  isotropic-equivalent afterglow X-ray luminosity at a rest-frame time
  of 11 hr ($L_{\rm X,11}$).  Symbols are as in Figure~\ref{fig:lxeg}.
  The dotted black line marks a linear correlation, expected for
  $\nu_X\simlt \nu_c$.  The inset shows the distribution of the ratio
  $L_{\rm opt,7}/L_{X,11}$, indicating that both short and long GRBs
  exhibit a similar ratio, and that in general $L_{\rm opt,7}/
  L_{X,11}\simlt 1$, indicative of $\nu_X\simlt \nu_c$ for short
  GRBs.}
\label{fig:lolx} 
\end{figure}

The X-ray and optical luminosities of short GRB afterglows appear to
be fainter relative to $E_{\rm\gamma,iso}$ than for long GRBs, by a
similar factor of about 7.  This indicates that the ratio $L_{\rm
  opt,7}/ L_{X,11}$ should have a similar distribution for both GRB
samples.  In Figure~\ref{fig:lolx} I plot $L_{\rm opt,7}$ versus
$L_{X,11}$ for short and long GRBs, and indeed find that both
populations follow a similar trend, consistent with a linear
correlation (see also \citealt{nfp09}).  The median value for both the
short and long GRB samples is $\langle L_{\rm opt,7}/L_{X,11}\rangle
\approx 0.08$, and nearly all events have $L_{\rm opt,7}/L_{X,11}
\simlt 1$.  In the context of the synchrotron model, the expected
ratio for the case in which both the optical and X-ray bands are
located below $\nu_c$ is $L_{\rm opt,7}/L_{X,11}\approx 0.2$ (for
$p=2.4$).  On the other hand, for the case when $\nu_c$ is
intermediate between the optical and X-ray bands (e.g., $\nu_c\sim
10^{16}$ Hz) the expected ratio is $L_{\rm opt,7}/L_{X,11} \approx 1$,
higher than most observed values.  Thus, the comparison of optical and
X-ray luminosities suggests that generally $\nu_c$ is located near or
above the X-ray band.

Since the ratio of optical to X-ray luminosity for $\nu_c\sim\nu_X$
and for $\nu_c\gg\nu_X$ is essentially the same, and since in the
latter case both the X-ray and optical luminosity depend on density
(with $F_\nu\propto n^{0.5}$), the lower values of $L_{\rm opt,7}/
E_{\rm\gamma,iso}$ and $L_{X,11}/ E_{\rm\gamma,iso}$ for short GRBs
can be interpreted as being due to a lower density scale; this will
also naturally lead to $\nu_c\gg\nu_X$ since $\nu_c\propto n^{-1}$.
On the other hand, long GRBs generally have higher densities and hence
$\nu_c\sim\nu_X$.  To match the observed weaker optical and X-ray
afterglows of short GRBs, requires a density scale that is about 50
times lower than for long GRBs, corresponding to $n\sim 0.1$
cm$^{-3}$.

An alternative possibility is that the systematic offset in the trends
of $L_{\rm opt,7}$ and $L_{X,11}$ versus $E_{\rm\gamma,iso}$ for short
and long GRBs is due to systematic differences between the
$\gamma$-ray and blastwave kinetic energies.  In particular, if
$\epsilon_\gamma$ for short GRBs is systematically higher than for
long GRBs, this would lead to values of $E_{\rm\gamma,iso}$ that
over-predict $E_{\rm K,iso}$ and hence to lower ratios of $L_{\rm
  opt,7}/E_{\rm\gamma,iso}$ and $L_{X,11}/E_{\rm\gamma,iso}$ compared
to those for long GRBs.  Given the overall shift of the best-fit
relations shown in Figures~\ref{fig:lxeg} and \ref{fig:loeg}, this
scenario would require an efficiency that is about an order of
magnitude larger in short GRBs.  With a fiducial value of
$\epsilon_\gamma\sim 0.5$ for long GRBs (e.g., \citealt{pk02}), this
requires $\epsilon_\gamma\sim 0.95$ for short GRBs, an unlikely
scenario.

A possible variation of this scenario is that the blastwave kinetic
energies of long GRBs are systematically larger relative to $E_{\rm
  \gamma,iso}$ due to significant energy injection into the forward
shock following the prompt emission phase, which is absent in short
GRBs.  Indeed, the X-ray plateaus in long GRBs, spanning $\sim
10^3-10^4$ s, have been interpreted as evidence for significant energy
injection \citep{nkg+06}, but a smaller fraction of short GRBs exhibit
such plateaus \citep{mzb+13}.  Effectively, larger energy injection in
long GRBs will translate to a lower value of $\epsilon_\gamma$ than
for short GRBs.  An energy injection factor in long GRBs of about an
order of magnitude will bring the relations for short and long GRBs
into alignment.

\subsubsection{The Energy and Density Scale Extracted from Optical
  Observations}

\begin{figure}
\centerline{\psfig{figure=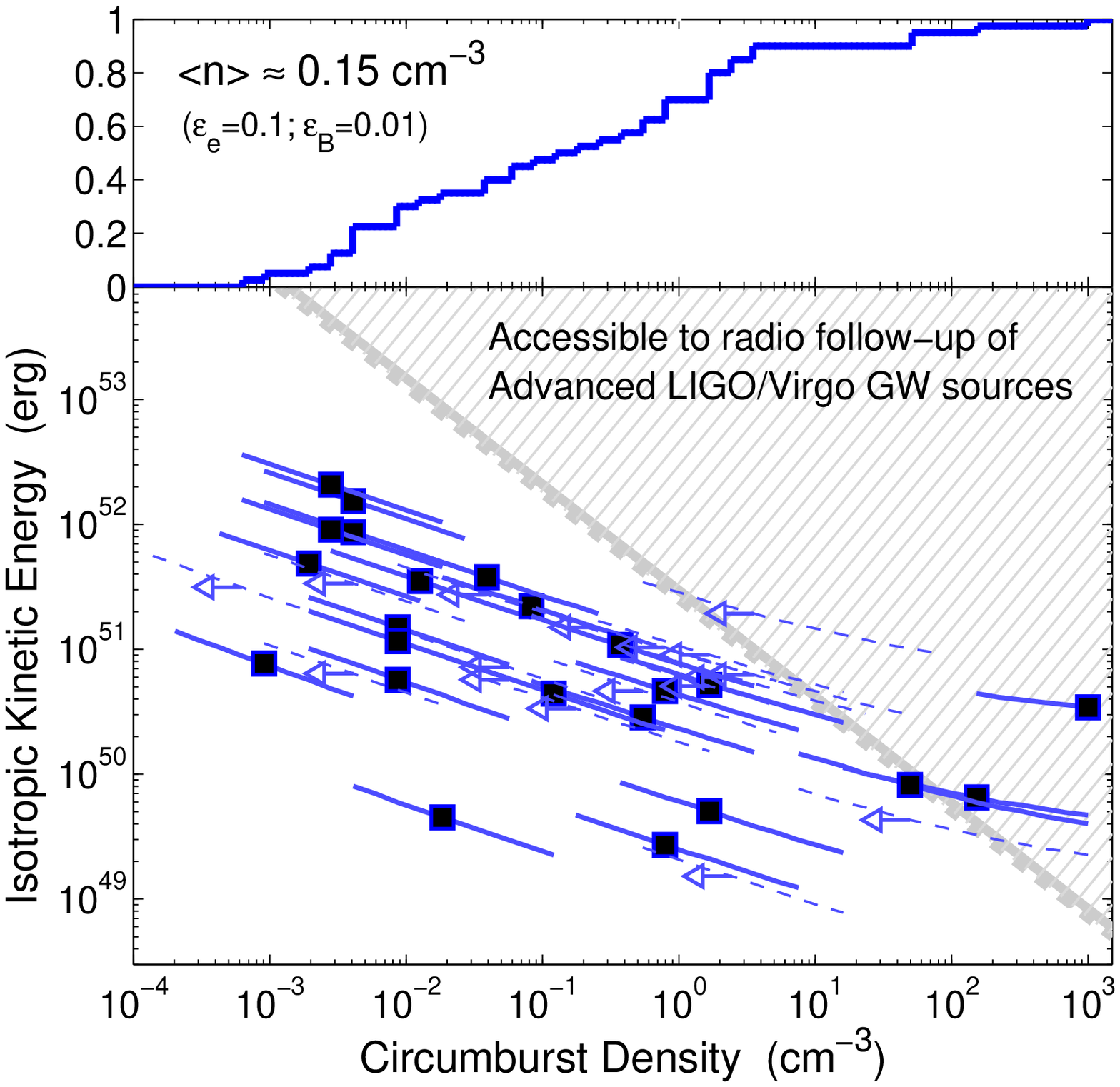,width=\textwidth}}
\caption{Bottom: Blastwave isotropic-equivalent kinetic energy
  ($E_{\rm K,iso}$) versus circumburst density ($n$) for individual
  short GRBs based on their optical afterglow emission (arrows
  indicate optical non-detections).  The lines span $E_{\rm K,iso}
  =(0.3-3) \times E_{\rm \gamma,iso}$.  The diagonal gray lines
  delineate the phase-space required for the detections of radio
  counterparts to Advanced LIGO/Virgo gravitational wave sources
  (hatched gray region; based on the models in \citealt{np11}).  The
  solid line is for an off-axis short GRB afterglow with fiducial
  parameters (\S\ref{sec:afterglow}), while the dashed line is for an
  isotropic mildly-relativistic outflow.  The short GRB results point
  to an energy-density scale that will generally prevent radio
  detections of Advanced LIGO/Virgo sources in searches that cover a
  full error region (\S\ref{sec:gwem}).  Top: Cumulative distribution
  of the circumburst densities (for $E_{\rm K,iso}=E_{\rm
    \gamma,iso}$).  The median circumburst density is $\simlt 0.15$
  cm$^{-3}$, at least an order of magnitude lower than for long GRBs,
  pointing to explosions in more tenuous environments.}
\label{fig:EKn}
\end{figure}

Independent of the relative behavior of short and long GRB afterglows,
each optical detection or upper limit can be converted to an allowed
range in the phase-space of $E_{\rm K,iso}$ and $n$.  For this purpose
I use fixed fiducial values of the microphysical parameters,
$\epsilon_e=0.1$, $\epsilon_B=0.01$, and $p=2.4$.  In addition, I
restrict the range of $E_{\rm K,iso}$ for each burst to $(0.3-3)\times
E_{\rm\gamma,iso}$ for a reasonable expected range of
$\epsilon_\gamma$.  The resulting distribution is shown in
Figure~\ref{fig:EKn}.  For the bulk of the short GRB population, the
inferred density range is $n\sim 10^{-3}-1$ cm$^{-3}$, with a median
for the full distribution of $\langle n\rangle\simlt 0.15$ cm$^{-3}$;
this is an upper limit due to the substantial fraction of optical
non-detections.  This result confirms our inference of a lower density
scale compared to long GRBs based on the ratios of optical and X-ray
luminosity to $E_{\rm \gamma,iso}$.

There are three bursts that may stand out from this distribution, with
inferred densities of $\sim 50-10^3$ cm$^{-3} $ (if $E_K\approx
E_\gamma$ and if their microphysical parameters are as listed above).
Two of these events (GRBs 070724A and 110112A) have low $\gamma$-ray
fluences of $F_\gamma\approx 3\times 10^{-8}$ erg cm$^{-2}$, which may
indicate that the \swift/BAT energy band severely underestimates the
value of $E_{\rm \gamma,iso}$, and hence their likely $E_{\rm K,iso}$
values.  With larger energies, the required densities will be lower
than the values inferred above.  On the other hand, the third event
(GRB\,050709) does not fall in this category.  Clearly, the inferred
distribution of circumburst densities is affected by dispersions in
the microphysical parameters and in the ratio of $E_{\rm \gamma,iso}$
to $E_{\rm K,iso}$, and may therefore be somewhat wider or narrower
than inferred here.  However, the median of the distribution is fairly
robust.  The relatively low density is not unexpected considering that
most short GRB offsets are larger than 1 $r_e$ and that nearly half of
all short GRBs reside in regions with no detectable stellar light
(\S\ref{sec:subgal}).

\subsection{Radio Afterglow Emission}

Radio afterglow observations have been undertaken for $28$ short GRBs,
and have so far led to only 3 detections (GRBs 050724, 051221A, and
130603B; \citealt{bpc+05,skn+06,fbm+13}); see
Table~\ref{tab:afterglows}.  With the exception of a few short GRBs
observed with the improved sensitivity of the JVLA starting in 2011,
most upper limits are in the range of $F_\nu\sim 60-150$ $\mu$Jy
($3\sigma$).  It is clear from a comparison to the fiducial
synchrotron model listed above (with $F_\nu\sim 40$ $\mu$Jy at $\sim
10$ GHz; Equation~\ref{eqn:ag}) that most of the existing limits are
too shallow compared to the expected flux density scale for $n\sim
0.1$ cm$^{-3}$.  On the other hand, since in the relevant regime
($\nu_a<\nu_{\rm radio}<\nu_m$) the flux density scales as
$F_\nu\propto n^{0.5}$, the non-detections generally indicate $n\simlt
{\rm few}$ cm$^{-3}$.  This result is supported by the three events
with detected radio afterglows \citep{bpc+05,sbk+06,fbm+13}, and
agrees with the more constraining results from the optical afterglow
data.  I also note that of the three bursts with potential high
density based on optical observations, I find that radio observations
of GRB\,050709 instead indicate $n\simlt 0.05$ cm$^{-3}$ (see also
\citealt{ffp+05}); the radio limits for GRBs 070724A and 110112A are
not constraining beyond the information available from the optical
detections.

\subsection{Jets and Implications for the Energy Scale and Event
  Rates}
\label{sec:jets}

The collimation of GRB outflows has direct implications for the true
energy scale and event rate, as well as for our understanding of the
energy extraction mechanism.  The signature of collimation in GRB
afterglows is the so-called jet break that occurs at the time ($t_j$)
when the bulk Lorentz factor of the outflow declines to
$\Gamma(t_j)\approx 1/\theta_j$ \citep{rho99,sph99}.  The break is due
to a combination of an edge effect, when the entire emitting surface
of the jet becomes visible, and sideways expansion of the outflow.
Together, these effects lead to a change in the evolution of the
synchrotron spectrum.  In particular, at frequencies above the
spectral peak ($\nu\simgt\nu_m$) the jet break is manifested as a
steepening in the decline rate from roughly $F_\nu\propto t^{-1}$ to
$F_\nu\propto t^{-p}$.  This is relevant for the optical and X-ray
bands.  In the radio band, which is located below $\nu_m$ for typical
jet break timescales, the evolution transitions from a mild rise to a
shallow decline, subsequently followed by $F_\nu\propto t^{-p}$ when
$\nu_m$ declines below the radio band.  Since the jet break is due to
relativistic and hydrodynamic effects, it is expected to be achromatic
from radio to X-rays.  The relation between the jet break time and
opening angle is given by \citep{sph99}:
\begin{equation}
  \theta_j=0.13\,\left(\frac{t_{j,d}}{1+z}\right)^{3/8}\left(\frac{n_0}
    {E_{52}}\right)^{1/8}
\end{equation}

In the case of long GRBs, there is ample evidence for jet collimation
based on the detection of jet breaks in the X-rays, optical, and/or
radio on a wide range of timescales
\citep{hbf+99,bfs01,fks+01,pk02,bkf03}.  The resulting jet opening
angles are mainly in the range $\theta_j\approx 3-10^\circ$, with some
events extending to about $20^\circ$ \citep{bfs01,fks+01,bkf03}.  The
typical beaming correction factor is therefore $f_b\equiv [1-{\rm
  cos}(\theta_j)] \approx 6\times 10^{-3}$.  This leads to a two
orders of magnitude downward correction in the energy scale, with a
resulting typical value of $E_\gamma\sim E_K\sim {\rm few}\times
10^{51}$ erg \citep{bfs01,fks+01,bkf03}, albeit with a spread of about
an order of magnitude \citep{skn+06,cfh+11}.  Similarly, the upward
correction to the long GRB rate is $f_b^{-1}\sim 10^2$.

The information on jet breaks in short GRBs is more sparse, and with
the exception of GRB\,130603B \citep{fbm+13}, the claimed breaks have
not been detected in multiple wave-bands that can confirm the expected
achromatic behavior.  The key challenge for jet break detections is
the faintness of short GRB afterglows, which typically fade below
detectable levels on a day timescale, translating to a weak constraint
of $\theta_j\simgt 3^\circ$ if no break is observed \citep{fbm+12}.

\begin{figure}
\centerline{\psfig{figure=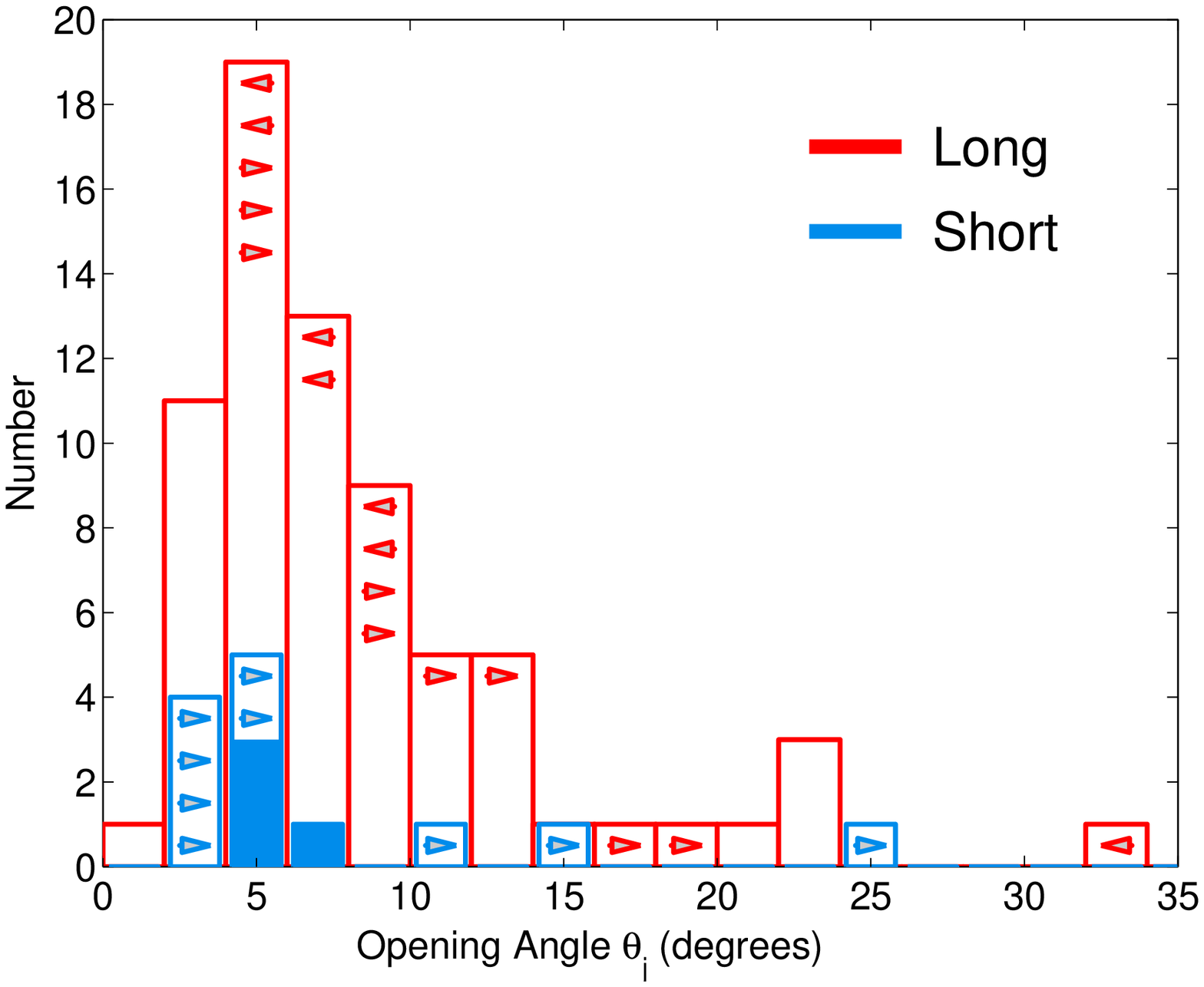,width=\textwidth}}
\caption{Distributions of jet opening angles for short (blue) and long
  (red) GRBs, based on breaks in their afterglow emission. Arrows mark
  lower or upper limits on the opening angles.  The observations are
  summarized in \S\ref{sec:jets}.  From \citet{fbm+13} and references
  therein.}
\label{fig:jets}
\end{figure}

Despite this observational challenge, there are a few credible
detections of jet breaks so far.  GRB\,051221A was the first short GRB
with evidence for collimation based on a break in its X-ray light
curve at about 5 d, with the expected steepening to a power law index
of $\alpha_X\simlt -2$ \citep{bgc+06,sbk+06}.  The inferred opening
angle is $\theta_j\approx 6-8^\circ$ \citep{sbk+06}, similar to the
opening angles of long GRBs.  GRB\,111020A similarly exhibited a break
in its X-ray light curve, at about 2 d, leading to an inferred opening
angle of $\theta_j\approx 3-8^\circ$ \citep{fbm+12}; the larger
uncertainty compared to GRB\,051221A is due to the unknown redshift of
this burst.  Most recently, GRB\,130603B exhibited a significant break
in its optical light curve at about 0.45 d, accompanied by a decline
in the radio band that matches the expectations of a post-jet break
behavior \citep{fbm+13}.  The inferred opening angle is
$\theta_j\approx 4-8^\circ$ \citep{fbm+13}.  A potential break was
also noted in the optical light curve of GRB\,090426 at about $0.4$ d,
with a steep decline typical of post jet break evolution
\citep{nkr+11}, and an inferred opening angle of $\theta_j\approx
4^\circ$.

In addition to these likely breaks, there are also several meaningful
lower limits on jet opening angles.  X-ray observations of GRB\,050724
with \swift/XRT and {\it Chandra} revealed no break to about $22$ d,
leading to an inferred lower limit of $\theta_j\simgt 20^\circ$
\citep{gbp+06}.  The optical afterglow of GRB\,050709 exhibited
potential steepening at about 10 d based on a single data point
\citep{ffp+05}, which if attributed to a jet break leads to
$\theta_j\approx 15^\circ$; however, this interpretation has been
disputed \citep{whj+06}, and here I conservatively use $\theta_j\simgt
15^\circ$ as a lower limit.  X-ray observations of GRB\,111117A with
\swift/XRT and {\it Chandra} revealed no break to about $3$ d, leading
to an inferred lower limit of $\theta_j\simgt 6^\circ$ \citep{mbf+12}.
Finally, the lack of a break in the X-ray afterglow of GRB\,120804A to
about 46 d, indicates an opening angle of $\theta_j\simgt 11^\circ$
\citep{bzl+13}.

There are also several claimed breaks at much earlier times, $\sim
0.5-2$ hr, with a post-break steep decline rate that is reminiscent of
jet break behavior (GRBs 061201, 090305, 090510).  If these are indeed
jet breaks, then the resulting opening angles are even narrower than
for most long GRBs, $\theta_j\sim 1-2^\circ$.  However, in the case of
GRB\,090305 the claimed break is based on a single optical data point
\citep{nkg+12}, while for GRB\,090510 there is no corresponding break
in the optical band despite simultaneous coverage \citep{nkk+12}.  In
addition to these putative breaks, several short GRBs exhibit no break
in their X-ray light curves to $\sim {\rm day}$, leading to typical
limits of $\simgt 3^\circ$ (GRBs 070714B, 070724A, 071227, 081226,
101219A; \citealt{fbm+12}).  The distribution of jet opening angles
for short GRBs, along with a comparison to long GRB jets is shown in
Figure~\ref{fig:jets}.

Using the most robust detections and constraints described above (3
detections and 4 lower limits), I find that the mean opening angle is
$\langle\theta_j\rangle\simgt 10^\circ$.  The resulting mean beaming
factor is $f_b\simgt 0.015$, indicating that the correction to the
observed short GRB rate is a factor of $\simlt 70$, or a rate of up to
$\sim 10^3$ Gpc$^{-3}$ yr$^{-1}$ \citep{ngf06,chp+12,fbm+12,fbm+13}.
In the context of compact object mergers, this value coincides with
the middle of the range based on estimates from the Galactic NS-NS
binary population and from population synthesis models
\citep{kkl+04,aaa+10}.  When converted to the rate within the expected
sensitivity volume of Advanced LIGO ($\sim 200$ Mpc), the resulting
value is $\sim 25$ yr$^{-1}$.  I note that if claimed values of
$\theta_j\sim 1-2^\circ$ are indeed correct, then it is possible that
the correction to the short GRB rate is up to $\sim 3\times 10^3$,
leading to an Advanced LIGO detection rate of $\sim 10^3$ yr$^{-1}$.
Thus, in the framework of compact object binary progenitors, the
discovery rate of NS-NS binaries with Advanced LIGO will provide
insight on short GRB beaming.

The inferred opening angles also determine the true energy scale of
short GRBs.  For the 3 events with likely jet breaks, the inferred
beaming-corrected $\gamma$-ray energies are $E_\gamma\approx
(0.5-5)\times 10^{49}$ erg.  The 4 events with lower limits on the
opening angles lead to a similar range for their minimum
beaming-corrected $\gamma$-ray energies (with the upper bounds
determined by the isotropic-equivalent values).  Since the short GRB
sample spans $E_{\rm \gamma,iso} \sim 10^{49}-10^{52}$ erg
(Figures~\ref{fig:lxeg} and \ref{fig:loeg}), it is likely that the
beaming-corrected energies are generally $E_\gamma\simlt 10^{50}$ erg
(using the typical beaming correction of $f_b\sim 0.015$).  It remains
to be seen, when a larger sample of opening angles becomes available,
whether the true energies span a narrow range, centered on
$E_\gamma\sim 10^{49}$ erg, about two orders of magnitude lower than
the energy scale of long GRBs.

\subsection{Implications for the Progenitors and Energy Source}
\label{sec:implications}

The comparison of short and long GRB afterglows, and the comparison of
short GRB afterglow data to the standard synchrotron model, reveal
several important trends.  First, the ratios $L_{X,11}/ E_{\rm
  \gamma,iso}$ and $L_{\rm opt,7} /E_{\rm \gamma,iso}$ are
systematically lower for short GRBs, by about a factor of 7.  On the
other hand, the ratio $L_{\rm opt,7}/L_{X,11}$ is similar for both GRB
samples.  One implication of these results is that the afterglows of
short and long GRBs are overall similar in their properties, and are
generated by the same mechanism, namely an external shock with the
circumburst environment that amplifies magnetic fields and accelerates
electrons to relativistic velocities.  However, short GRBs appear to
explode in lower density environments.  Using the optical afterglows,
I find a typical density scale of $\simlt 0.15$ cm$^{-3}$.  This is
supported by radio afterglow limits, which indicate $n\simlt 1$
cm$^{-3}$, and by the few events with radio to X-ray detections
\citep{bpc+05,sbk+06,fbm+13}.  The relatively low circumburst
densities are consistent with the larger offsets of short GRBs
compared to long GRBs, and with their weak spatial correlation with
the underlying distribution of stellar light in their hosts.  Although
it is less likely, the difference in the ratios of $L_{X,11}$ and
$L_{\rm opt,7}$ to $E_{\rm \gamma,iso}$ for short and long GRBs may
also reflect a systematic difference in the $\gamma$-ray efficiency,
either due to a higher radiative efficiency in the prompt emission of
short GRBs, or due to substantial energy injection in the early
afterglow phase of long GRBs.  In either case, a difference by about
an order of magnitude is required to reconcile the two samples.

Beyond the relative trends, it is clear from Figures~\ref{fig:lxeg}
and \ref{fig:loeg} that the distribution of $E_{\rm \gamma,iso}$ for
short GRBs is just as broad as for long GRBs, but with systematically
lower values of $\sim 10^{49}-10^{52}$ erg.  The beaming-corrected
energies in the several cases with likely jet opening angle
measurements or meaningful lower limits are $E_\gamma\sim
(0.5-5)\times 10^{49}$, about two orders of magnitude lower than for
long GRBs.  The opening angles are generally broader than for long
GRBs, with a range of $\theta_j\sim 5^\circ$ to $\simgt 20^\circ$ and
a median of $\langle\theta_j\rangle\simgt 10^\circ$.

It is instructive to compare these inferences to the results of
hydrodynamic and magnetohydrodynamic (MHD) simulations of compact
object mergers as they pertain to estimates of the geometry, energy
extraction mechanism, and the energy scale of relativistic outflows.
Several researchers carried out hydrodynamic simulations of jets
launched by the neutrino-antineutrino annihilation mechanism and found
jets with typical opening angles of $\theta_j\sim 5-20^\circ$, and
total isotropic-equivalent energies of $\sim 10^{51}$ erg (or typical
beaming-corrected energies of up to $\sim 10^{49}$ erg;
\citealt{jer+99,rrd03,srj04,ajm05,baj+07,dob+09}).  However, only
$\sim 10\%$ of this energy is available to produce the prompt
emission, leading a typical true energy of $E_\gamma\sim 10^{48}$ erg.
This value, and the estimated upper bound on $E_{\rm iso}$, are about
an order of magnitude lower than observed in short GRBs, although the
jet opening angle estimates agree with the observed values.  This
suggests that for at least some short GRBs, the neutrino-antineutrino
annihilation mechanism is not likely to provide sufficient energy to
power the observed emission.

Instead, recent general relativistic MHD simulations by \citet{rgb+11}
suggest the formation of a $\sim 10^{15}$ G magnetic field in both the
torus (toroidal) and along the rotation axis (poloidal) following a
compact object merger and the formation of a black hole.  The typical
opening angle of the poloidal component is $\sim 10-30^\circ$.  If an
outflow is then powered by the Blandford-Znajek mechanism
\citep{bz77}, the typical isotropic-equivalent energy scale is $\sim
{\rm few}\times 10^{51}\,(B/2\times 10^{15}\,{\rm G})^2$ erg (e.g.,
\citealt{lwb00,rrd03,ssk+11}); this assumes that $10\%$ of the
available energy is channeled into the relativistic outflow, and that
the typical opening angle is $\sim 10^\circ$ based on the existing
observations.  Thus, these simulations suggest that compact object
mergers can lead to conditions that give rise to opening angles and an
energy scale (via the Blandford-Znajek mechanism) that are
well-matched to the observations of short GRBs.

\section{``Kilonova'' Emission and the Short GRB\,130603B}
\label{sec:kilonova}

One of the key predictions of the compact object merger model is the
ejection of neutron-rich matter, either dynamically during the merger,
or by winds from an accretion disk \citep{ls74,elp+89}.  The typical
ejecta masses and velocities inferred from merger simulations are in
the range of $M_{\rm ej}\sim 10^{-3}-{\rm few}\times 10^{-2}$
M$_\odot$ and $v_{\rm ej}\sim 0.1-0.3c$
\citep{rlt+99,rdt00,rj01,ros05,efl+08,bgj13,pnr13,rpn13}; the ejecta
masses generally increase with larger asymmetry in the mass ratio
(i.e., they are larger for NS-BH mergers than for NS-NS mergers;
\citealt{hkk+13,kis13,rpn13}).  The resulting kinetic energy of the
outflow is therefore in the range $E_K\sim 10^{49}-3\times 10^{51}$
erg.  The ejecta will generate radio emission through interaction with
the ambient medium, with the ejecta velocity and kinetic energy, as
well as the ambient density, determining the peak brightness and
timescale of the radio emission (see \S\ref{sec:gwem}).

The rapid decompression of the neutron-rich matter from nuclear
densities will lead to the formation of heavy radioactive elements via
the $r$-process \citep{ls74,elp+89}.  The radioactive decay of these
elements spans a wide range of half-lives as the elements settle on
the region of $\beta$ stability, and provide a source of heating
\citep{lp98}.  Given the low ejecta mass and rapid expansion velocity,
the ejecta will become optically thin on a shorter timescale and with
a lower peak luminosity than supernova explosions.  This has led to
the terminology of ``mini-SN'', ``macronova'', or ``kilonova'' to
describe these counterparts \citep{lp98,mmd+10,gbj11,rkl+11,pnr13}; I
use the latter term here.  In addition to the dynamical properties of
the ejecta, the timescale, luminosity, and characteristic spectral
energy distribution of a kilonova also depend on the opacity of the
$r$-process material, since the peak of the light curve is expected
when the opacity becomes sufficiently low that photons can diffuse on
the expansion timescale (e.g., \citealt{arn82}).  The diffusion
timescale is given by (e.g., \citealt{mmd+10,bk13}):
\begin{equation}
t_d=\frac{B\kappa M_{\rm ej}}{cR},
\end{equation}
where $B\approx 0.07$ is a geometric factor, $\kappa$ is the opacity,
and $R=vt$ is the radius of the expanding ejecta.  In addition, the
peak luminosity is given by the radioactive heating rate ($Q$) at the
diffusion timescale:
\begin{equation}
L_p\approx \frac{Q_p}{t_d(R_p)},
\end{equation}
where $R_p$ is the radius of the ejecta at the diffusion timescale.
Finally, the characteristic blackbody temperature of the resulting
emission is given by:
\begin{equation}
T_p\approx \left(\frac{L_p}{4\pi R_p^2\sigma}\right)^{1/4}
\end{equation}
For an opacity of $\kappa\sim 0.1$ cm$^2$ g$^{-1}$ at $\sim 0.3-1$
$\mu$m, typical of the Fe-peak elements synthesized in SNe, the
resulting peak timescale is $t_p\sim 0.3$ d, with a luminosity of
$L_p\sim {\rm few}\times 10^{41}\,(f_{-6})$ erg s$^{-1}$ (where $f\sim
{\rm few}\times 10^{-6}$ is the fraction of ejecta rest-mass energy
that goes into radioactive heating; \citealt{lp98,mmd+10}), and the
photospheric temperature is $T_p\approx 10^4$ K, leading to a peak in
the UV/optical. 

However, recent investigations of the opacities due to $r$-process
elements lead to a different conclusion.  In a pioneering effort,
\citet{kbb13} and \citet{bk13} demonstrated that the opacities of
$r$-process matter, particularly the lanthanides, are about two orders
of magnitude higher than for Fe-peak elements, with $\kappa\sim
10-100$ cm$^2$ g$^{-1}$ at $\lambda\sim 0.3-3$ $\mu$m.  With these
opacity values, the expected peak timescale becomes longer by about an
order of magnitude, $t_p\sim {\rm few}$ days, the peak luminosity
becomes correspondingly lower, $L_p\sim {\rm few}\times 10^{40}$ erg
s$^{-1}$, and the characteristic temperatures shifts to $T_p\sim 3000$
K, or a peak in the near-IR.  Initial calculations of the $r$-process
opacities and the resulting light curves and spectral energy
distributions are provided in several recent papers
\citep{bk13,bgj13,gkr+13,th13}.

From an observational point of view, the effect of the larger
opacities is dramatic.  Instead of a transient that peaks in the blue
part of the optical spectrum on a timescale of $\sim {\rm day}$ and
with an absolute magnitude of about $-15$, the kilonova is expected to
peak with a similar absolute magnitude in the near-IR ($\sim 1.5$
$\mu$m) on a timescale of $\sim {\rm week}$.  In the optical bands, on
the other hand, the peak absolute magnitude and timescale are about
$-12$ mag and $\sim {\rm day}$.  This has significant implications for
the detectability of kilonovae in blind surveys or in optical
follow-up of gravitational wave sources (\S\ref{sec:gwem}).

Optical searches for kilonova emission have been carried out in
association with the short GRBs 050509B \citep{bpp+06}, 070724A
\citep{bcf+09,ktr+10}, 080503A \citep{pmg+09}, 080905A \citep{rwl+10},
and 100206A \citep{pmm+12}, but none have yielded detections.  These
searches were guided by the early predictions for bright UV/optical
emission with a $\sim{\rm day}$ timescale, but unfortunately, they do
not provide strong constraints in the context of the current opacity
estimates.

\begin{figure}
\centerline{\psfig{figure=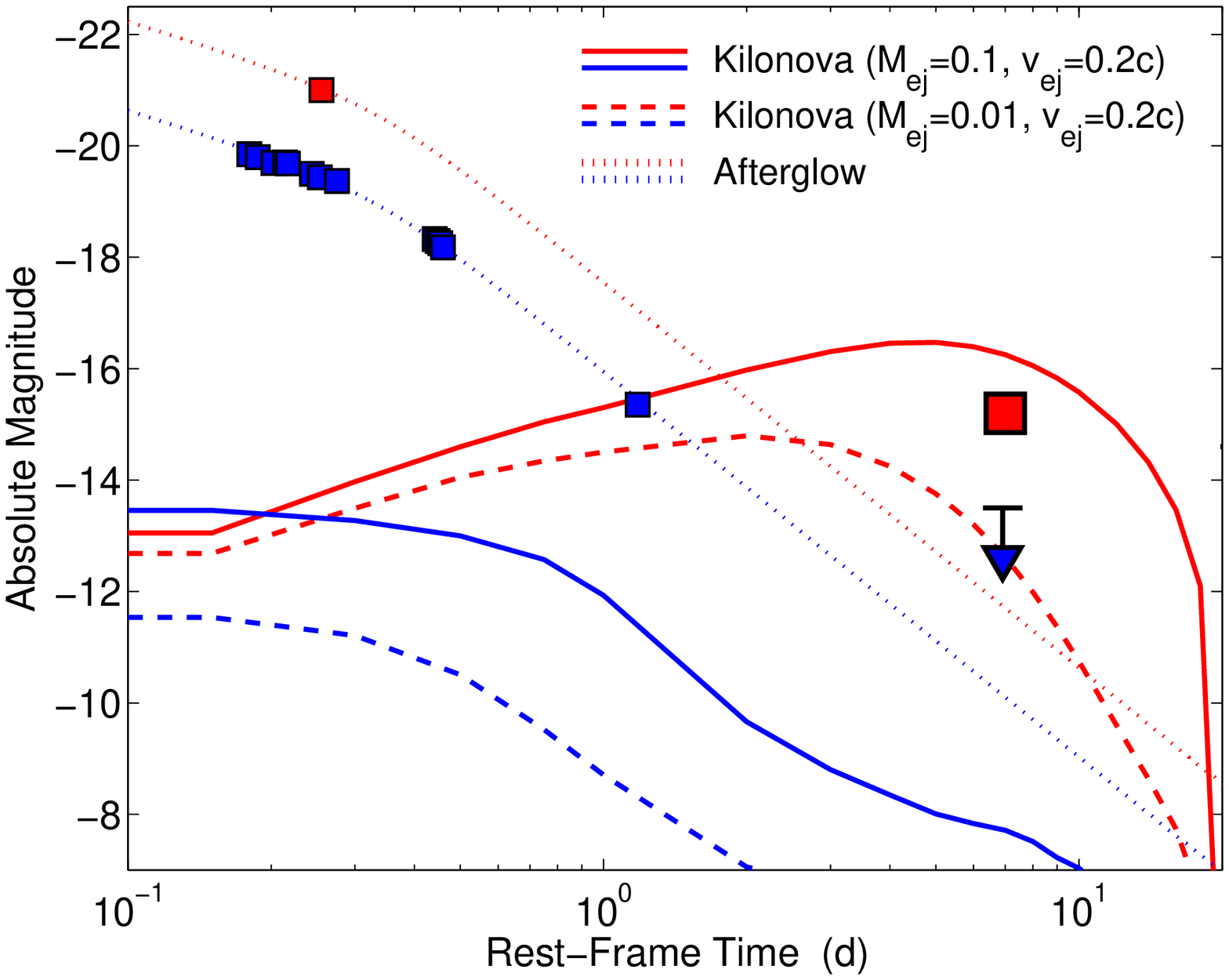,width=\textwidth}}
\caption{Optical (blue) and near-IR (red) observations of the short
  GRB\,130603b from ground-based telescopes ($\simlt 2$ d) and from
  {\it HST} \citep{bfc13,cpp+13,fbm+13,tlf+13}.  The dotted lines are
  a fit to the early afterglow evolution, which underestimates the
  near-IR flux detected in the {\it HST} observations.  The excess
  emission can be explained as emission from an $r$-process kilonova,
  resulting from a compact object binary merger.  Theoretical kilonova
  models from \citet{bk13} that take into account the opacities of
  $r$-process elements are shown as dashed and solid lines for ejecta
  masses of 0.01 and 0.1 M$_\odot$, respectively.  From
  \citet{bfc13}.}
\label{fig:130603b} 
\end{figure}

The first credible search focusing on the near-IR band was carried out
for GRB\,130603B at $z=0.356$ using a combination of ground-based
observations at $\simlt 2$ d and {\it HST} observations at about 9 and
30 d \citep{bfc13,tlf+13}; see Figure~\ref{fig:130603b}.  The
afterglow of this burst was relatively bright at early time, but faded
rapidly with $F_\nu\propto t^{-2.6}$ starting at about 0.45 d
(indicative of a jet break as discussed in \S\ref{sec:jets}).  Based
on the decline rate and spectral energy distribution, the afterglow
was predicted to have an exceedingly low flux at the time of the first
{\it HST} observation, $m_{\rm F606W}\approx 31$ mag and $m_{\rm
  F160W}\approx 29.3$ mag.  Instead, the {\it HST} observations
revealed a source with $m_{\rm F160W} \approx 25.8$ mag, but with no
corresponding counterpart in the optical data (to $m_{\rm
  F606W}\approx 27.7$ mag).  The excess emission in the near-IR by a
factor of about 25, coupled with the unusually red color, match the
predictions for a kilonova in the context of $r$-process opacities.
At the time of writing, this is one of the strongest lines of evidence
in favor of compact object binary progenitors.

The near-IR luminosity of the likely kilonova in GRB\,130603B provides
a rough estimate of the mass of $r$-process ejecta produced in this
event.  Utilizing the $r$-process opacity models, \citet{bfc13}
estimate $M_{\rm ej}\approx 0.03-0.08$ M$_\odot$ for ejecta velocities
of $v_{\rm ej}\approx 0.1-0.3c$.  Clearly, the theoretical models are
preliminary, and there are variations between the calculations carried
out by different groups \citep{bk13,bgj13,gkr+13,th13}, but the
results for GRB\,130603B suggest that at least some short GRB
progenitors may eject as much as $\sim {\rm few}\times 10^{-2}$
M$_\odot$ of $r$-process material.  Coupled with current estimates for
the rate of NS-NS mergers, this indicates that compact object binary
mergers may be the dominant site of $r$-process nucleosynthesis.

Detailed multi-epoch and multi-band follow-up observations of future
nearby short GRBs will establish the ubiquity of kilonova associations
and determine the range of ejecta masses.  In particular, near-IR
spectroscopy may reveal broad absorption features that will uniquely
establish a non-afterglow origin and will also provide a rough measure
of the typical ejecta velocities.  The key challenges for such
searches are the need for short GRBs at $z\simlt 0.3$, which occur at
a rate of $\simlt 1$ yr$^{-1}$; a fast-fading afterglow to avoid
significant contamination; and rapid follow-up with {\it HST} in the
optical and near-IR.  A key strategy to confirm a kilonova origin for
any late-time excess emission is the use of multiple near-IR and
optical filters to trace the expected sharp spectral cut-off at $\sim
1$ $\mu$m.

\section{Short GRBs as Electromagnetic Counterparts of Gravitational
  Wave Sources}
\label{sec:gwem}

Several lines of evidence, presented in the previous sections, point
to compact object mergers as the progenitors of short GRBs.  As a
result, existing observations of short GRBs and their afterglows can
uniquely shed light on the on- and off-axis electromagnetic (EM)
signatures that will accompany gravitational wave (GW) sources
detected by Advanced LIGO/Virgo.  The frequency range of these GW
detectors is designed to uncover the final inspiral and merger of
compact object binaries (NS-NS, NS-BH, BH-BH).  The expected EM
counterparts of Advanced LIGO/Virgo sources, and their detectability
with existing and upcoming facilities, are a topic of growing
interest, which has been explored in several recent papers (e.g.,
\citealt{mb12,pnr13}).  To assess the importance and detectability of
various EM counterparts it is important to note that the Advanced
LIGO/Virgo volume-averaged distance for neutron star binary mergers is
about 200 Mpc, while ideally-placed and oriented binaries (overhead
relative to the detector network and with a face-on binary
orientation) can be detected to about 450 Mpc \citep{aaa+10}.  The
face-on orientation may also be ideal for the detection of an on-axis
short GRB since the jet axis is expected to be oriented perpendicular
to the plane of the binary's orbit.  The typical localization region
for the Advanced LIGO/Virgo network (3 detectors) is $\sim 100$
deg$^2$, although this may improve with future detectors in Japan and
India \citep{fai09}.  The poor localization capability presents a
unique challenge for EM follow-up, and for counterpart discrimination
and identification, compared for example to studies of short GRBs from
\swift, which are localized to better than a few arcminutes radius.

\begin{figure}
\centerline{\psfig{figure=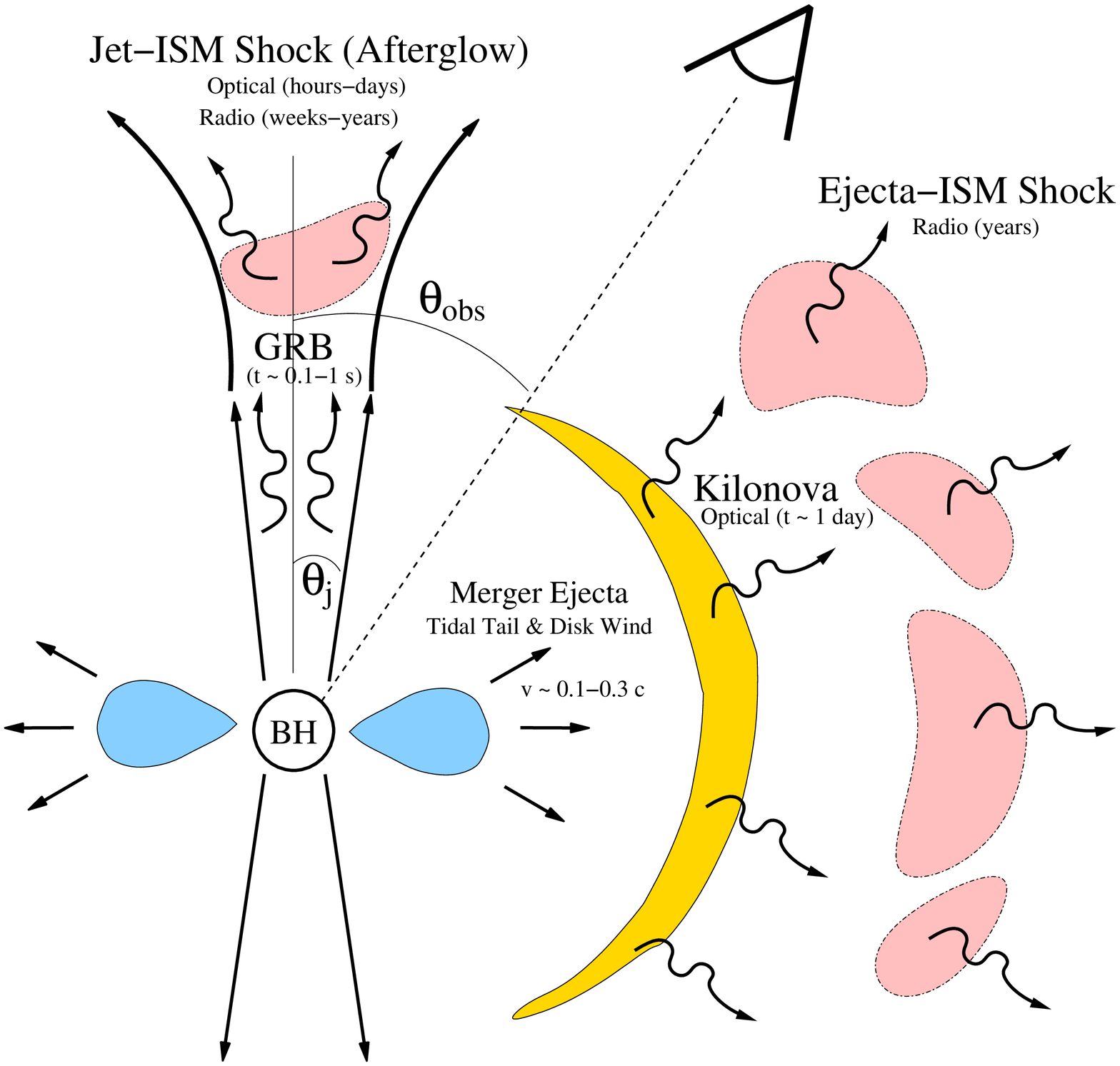,width=\textwidth}}
\caption{Potential electromagnetic counterparts of compact object
  binary mergers as a function of the observer viewing angle
  ($\theta_{\rm obs}$).  Rapid accretion of a centrifugally supported
  disk (blue) powers a collimated relativistic jet, which produces a
  short GRB.  Due to relativistic beaming, the $\gamma$-ray emission
  is restricted to observers with $\theta_{\rm obs}\simlt \theta_j$.
  Afterglow emission results from the interaction of the jet with the
  circumburst medium (pink).  Optical afterglow emission is detectable
  for observers with $\theta_{\rm obs} \simlt 2\theta_j$. Radio
  afterglow emission is observable from all viewing angles once the
  jet decelerates to mildly relativistic velocities on a timescale of
  months-years, and can also be produced on timescales of years from
  sub-relativistic ejecta.  Short-lived isotropic optical/near-IR
  emission lasting a few days (kilonova; yellow) can also accompany
  the merger, powered by the radioactive decay of $r$-process elements
  synthesized in the ejecta.  From \citet{mb12}.}
\label{fig:gwem} 
\end{figure}

Despite the obvious observational challenges, the information that can
be gleaned from EM counterparts is invaluable \citep{kp93,mb12}.  In
particular, the joint detection of gravitational waves and a short GRB
will definitively establish the compact object merger model, and may
also shed light on any differences in the short GRB population that
are due to mergers of NS-NS versus NS-BH binaries.  Even more broadly,
EM counterparts are essential for pinpointing the exact locations of
the mergers, thereby providing an association with host galaxies,
measurements of a distance scale, and a relation to specific stellar
populations that will shed light on the kick and merger timescale
distributions.  Similarly, the properties of the EM counterparts will
shed light on the behavior of matter following the merger, namely
accretion disks, jets, and dynamical or wind-driven outflows.  A
summary of potential counterparts is provided in
Figure~\ref{fig:gwem}.

\citet{mb12} examined a wide range of putative on- and off-axis
counterparts, and examined their potential impact for joint GW/EM
studies in relation to four critical virtues: (i) detectability with
present or upcoming telescope facilities using a reasonable allocation
of resources; (ii) the potential to accompany a high fraction of GW
detections; (iii) distinguishability from other astrophysical
transients that may contaminate the large search areas; and (iv)
allowing for a localization to arcsecond precision.  Other studies of
EM counterparts have also explored the issue of detectability and
background contamination, generally reaching similar conclusions,
although some studies employ different criteria for ``detectability''
(e.g., \citealt{np11,pnr13}).

\subsection{An On-Axis Short GRB Counterpart}

At least some Advanced LIGO/Virgo sources are expected to be
accompanied by on-axis short GRBs with detectable $\gamma$-ray
emission and a bright multi-wavelength afterglow.  Since existing
$\gamma$-ray satellites and large ground-based telescopes are capable
of detecting short GRBs and their afterglows to $z\sim 1$, such an
event at $\simlt 450$ Mpc ($z\simlt 0.1$) will be exceedingly bright.
In particular, using the measured X-ray and optical luminosities of
even the least energetic short GRBs (with $E_{\rm\gamma,iso}\sim
10^{49}$ erg; Figures~\ref{fig:lxeg} and \ref{fig:loeg}), I predict an
X-ray flux at $\sim 0.5$ d of $F_X\sim 5\times 10^{-14}$ erg s$^{-1}$
cm$^{-2}$ and an optical brightness of $r\sim 23$ mag.  These are
detectable levels assuming that the $\gamma$-ray emission can be
pinpointed to an area typical of \swift/BAT localizations.  For a
short GRB with a median energy scale, the expected X-ray flux is
$F_X\sim 5\times 10^{-12}$ erg s$^{-1}$ cm$^{-2}$ and the optical
brightness is $r\sim 20$ mag.  Such a bright optical counterpart may
be detectable even if the $\gamma$-ray position is as poor as the GW
localization, for example typical of {\it Fermi}/GBM positions.
However, such a large area cannot be effectively searched with
\swift/XRT, which has a small field-of-view of only 0.25 deg$^2$.

The temporally coincident detection of a short GRB and a GW source
will serve as the clearest smoking gun for the compact object merger
model.  It may also help to improve the positional accuracy of the GW
source (if detected by an instrument such as \swift/BAT), but even a
poor $\gamma$-ray localization will provide a convincing association
based on the temporal coincidence.  However, since the current
estimate of the beaming fraction is $f_b\sim 70$ (\S\ref{sec:jets}),
such joint detections will be rare.  The occurrence rate can be
estimated using the observed short GRB redshift distribution
\citep{mb12}.  In particular, there are currently no known short GRBs
within the Advanced LIGO/Virgo maximum detection distance for NS-NS
binaries of $z\approx 0.1$.  Extrapolating the observed redshift
distribution to $z\simlt 0.1$, and correcting from the \swift/BAT
field-of-view to roughly all-sky coverage (e.g., IPN, {\it Fermi}/GBM)
the expected coincidence rate is $\sim 0.3$ yr$^{-1}$ \citep{mb12}.
Thus, while joint GW and short GRB detections will play a critical
role in establishing the identity of the progenitors, and will provide
particularly bright counterparts, such detections are expected to be
rare.

Another profitable approach in the context of joint $\gamma$-ray and
GW detections is to carry out a systematic search for GW emission in
temporal coincidence with short GRBs \citep{aaa+12c,mb12,kmr13}.  The
key advantage is that a search around a narrow time window, and with a
position reconstruction that is consistent with the short GRB
location, will allow for a reduced significance threshold.  Such a
search was carried out during LIGO Science Run 6 and Virgo Science
Runs 2 and 3 \citep{aaa+12c}, but unsurprisingly it did not yield any
detections since the short GRB sample extends much beyond the reach of
these initial detectors (Figure~\ref{fig:z}).  On the other hand, the
improved sensitivity of Advanced LIGO/Virgo may reach the nearest
short GRBs.

\subsection{An Off-Axis Optical Afterglow Counterpart}

A coincident optical afterglow is more likely to be detected than a
$\gamma$-ray counterpart since even with an initial off-axis viewing
angle the jet will eventually decelerate and spread into the
observer's line of sight.  However, larger off-axis viewing angles
result in reduced peak brightness, due to the longer timescale for the
jet to spread into the line of sight.  Using the existing sample of
short GRB optical afterglows it is possible to predict their
appearance for off-axis observers.  \citet{mb12} utilized the off-axis
afterglow models of \citet{vzm10} with the parameters appropriate for
on-axis short GRB afterglows (\S\ref{sec:afterglow}) and found that
for an off-axis angle that is twice the jet opening angle
($\theta_{\rm obs}= 2\theta_j$) the peak brightness of the afterglows
at a typical distance of 200 Mpc is expected to be $\sim 23-25$ mag
with a peak time of $\sim 1-10$ d, barely within the reach of the
largest wide-field optical telescopes.  At even larger off-axis angles
the optical afterglow is essentially undetectable ($\sim 27-29$ mag
for $\theta_{\rm obs}=4\theta_j$).  

The fraction of GW events that will be oriented within an angle of
$2\theta_j$ is $f_{\rm opt}\approx 6.8\langle \theta_j\rangle^2\sim
0.1$ \citep{mb12}, where the pre-factor of 6.8 takes into account the
nearly face-on orientation of the binary, which leads to a larger
detection volume than for random orientations.  Thus, a few GW sources
per year may be accompanied by off-axis optical afterglow emission
with a peak brightness of $\sim 23-25$ mag and a peak timescale of
$\sim 1-10$ days.  Finding such counterparts will require deep and
rapid searches with wide-field optical telescopes, but this is within
the reach of existing and upcoming facilities.

\subsection{An Off-Axis Radio Afterglow Counterpart}

The underlying reason for the faintness of optical afterglows at large
off-axis viewing angles is that by the time the jet spreads into the
observer's line of sight (weeks to months to years after the event)
the outflow becomes only mildly relativistic and the peak of the
synchrotron spectrum is shifted to the radio GHz band.  In principle,
this means that radio afterglow emission can serve as an ideal
counterpart for all GW sources due to its detectability from any
off-axis angle \citep{np11}.  In reality, the detectability of such
radio emission depends on the kinetic energy of the blastwave and the
ambient density.  In addition to the initially collimated and
relativistic outflow, radio emission will also be produced by any
non-relativistic isotropic outflow as it interacts with the ambient
medium and decelerates \citep{np11}.  In the latter scenario the
initial velocity of the outflow is another critical parameter that
determines the detectability of the radio emission.  Such outflows are
expected in compact object mergers from tidally unbound debris or from
accretion disk winds
\citep{rlt+99,rdt00,rj01,ros05,efl+08,bgj13,pnr13,rpn13}.

For the case of an initially collimated and relativistic outflow it is
again useful to consider the observed properties of short GRB
afterglows to investigate the expected brightness and detectability of
an off-axis radio afterglow.  The timescale for the blastwave to
decelerate, isotropize, and reach peak brightness is $t_{\rm
  dec}\approx 30\,{\rm d}\, E_{K,49}^{1/3}\, n_0^{-1/3}\,
\beta_0^{-5/3}$, where $E_{K,49}$ is the kinetic energy in units of
$10^{49}$ erg and $\beta_0$ is the blastwave velocity in units of the
speed of light; for an initially collimated and relativistic jet
$\beta_0\approx 1$ \citep{np11}.  Thus, for the typical energy and
density scales inferred for short GRBs ($E_{K,49}\sim {\rm few}$ and
$n\sim 0.1$ cm$^{-3}$; \S\ref{sec:afterglow}), the resulting peak time
is $t_{\rm dec}\sim 100$ d.  The expected peak brightness (occurring
at $t=t_{\rm dec}$) is $F_{\rm\nu,p}\approx 0.15\,{\rm mJy}\,
E_{K,49}\, n_0^{0.8}\, \epsilon_{B,-2}^{0.8}\, \epsilon_{e,-1}^{1.2}\,
\beta_0^2\, \nu_{\rm GHz}^{-0.6}$ \citep{np11}, where I have scaled to
a distance of 200 Mpc and to the fiducial values of $\epsilon_e=0.1$
and $\epsilon_B=0.01$ used in \S\ref{sec:afterglow} to infer the short
GRB afterglow properties.  For the typical energy and density of short
GRBs, this leads to $F_{\rm\nu,p}\sim 0.1$ mJy.

For dynamical or wind-driven ejecta the typical mass scale is $\sim
0.01$ M$_\odot$ with a velocity of $\beta\sim 0.2$, corresponding to
$E_{K,49}\approx 40$ \citep{rpn13}.  Since the density of $\sim 0.1$
cm$^{-3}$ inferred from short GRB observations is still relevant for
this case, this indicates $t_{\rm dec}\approx 2$ years and
$F_{\rm\nu,p}\approx 0.03$ mJy.  For the most extreme outflow
parameters indicated by numerical simulations ($\sim 0.1$ M$_\odot$
and $\beta\sim 0.3$), the peak brightness is $F_{\rm\nu,p}\approx 1.5$
mJy, but it will only occur on a timescale of $t_{\rm dec}\approx 10$
years.  It is important to note that $t_{\rm dec}$ represents not only
the timescale to reach peak brightness, but also the typical timescale
for the source to display significant brightness variability.  From
the point of view of a counterpart search, a source that does not vary
significantly on a decade timescale, about a decade following a GW
trigger is unlikely to be identified as an associated transient with
any level of confidence.

The key question, therefore, is whether a radio transient with a
typical peak brightness of $\sim 0.03-0.1$ mJy and a peak/variability
timescale of $\sim {\rm year}$ can be detected and identified in a
$\sim 100$ deg$^2$ region with existing and upcoming facilities.  To
assess this question, I note that for the JVLA, with a 0.45 deg$^2$
field-of-view at 1 GHz, full coverage of such an area requires at
least 200 individual pointings.  With a synthesized beam of about
$5''$ diameter, there are about $10^8$ independent beams within this
region, indicating that the minimum required significance level for a
detection is $6\sigma$ (making the unrealistically optimistic
assumption that the noise in a 100 deg$^2$ region is purely Gaussian,
with no systematic effects).  To detect such a source as a transient
requires at least a factor of two change in brightness, which means
that the observations need to reach a typical $1\sigma$ noise level of
$\sim 3-10$ $\mu$Jy.  With the JVLA this requires a few hours of
integration per pointing, or about a month of continuous observing per
epoch.  This unrealistic follow-up scenario is further exacerbated by
the fact that at 1 GHz source confusion leads to a sensitivity floor
of $\sim 0.03-0.1$ mJy (depending on the JVLA configuration), which is
comparable to the level of the expected signal.  Thus, it is unlikely
that radio emission will be detectable with JVLA follow-up of a
typical Advanced LIGO/Virgo error region.

The Australian Square Kilometre Array Pathfinder (ASKAP) wide-field
interferometer is expected to image $\sim 10^2$ deg$^2$ in only a few
pointings, and is therefore better matched to typical GW error
regions.  However, ASKAP's limited angular resolution of about $10''$
will lead to source confusion at the level of the anticipated signal,
$\sim 0.05-0.1$ mJy \citep{jtb+08}.  Finally, lower frequency
instruments such as the Low-Frequency Array (LOFAR) and the Murchison
Widefield Array (MWA) have sufficiently wide fields to instantaneously
cover a full GW error region, but they are even more severely hampered
by source confusion due to their lower angular resolution and the
increased background at lower radio frequencies ($\sim 0.1-0.3$ GHz).
Moreover, at the these frequencies the typical timescale for the
outflow to reach peak brightness will be years to decades, instead of
months at GHz frequencies.

I therefore conclude that while radio counterparts are in principle
detectable from all viewing angles, existing facilities are not
well-matched to the faintness of the anticipated signals.  It is
possible that in the long term multi-detector GW networks will
localize some events to $\simlt 10$ deg$^2$ \citep{fai09}, making
radio searches more feasible.  In addition, since the radio signal is
delayed compared to counterparts at other wave-bands, a more
profitable approach may be to use radio observations to follow up
candidate counterparts from $\gamma$-ray, X-ray, or optical/near-IR
searches, potentially as a way of distinguishing a true counterpart
from unrelated sources (e.g., supernovae, AGN).

\subsection{A Kilonova Counterpart}

As I discussed in \S\ref{sec:kilonova}, NS-NS and NS-BH mergers are
also expected to be accompanied by isotropic optical/near-IR emission
produced by the radioactive decay of $r$-process elements in a
dynamical or accretion disk wind driven outflow
\citep{rlt+99,rdt00,rj01,ros05,efl+08,bgj13,pnr13,rpn13}.  To date,
the primary insight on the expected properties of such counterparts
comes from numerical simulations coupled with $r$-process opacity
calculations \citep{bk13,bgj13,gkr+13,th13}, and from the near-IR
counterpart associated with the short GRB\,130603B
(\S\ref{sec:kilonova}; \citealt{bfc13,tlf+13}).  In this context, the
peak of the kilonova spectral energy distribution is expected at $\sim
1.5-3$ $\mu$m (the near-IR $HK$ bands) with an apparent brightness of
$\sim 21.5-24$ AB mag, and a timescale of $\sim {\rm week}$.  In the
optical $izy$ bands ($0.7-1$ $\mu$m) the peak brightness is $\sim
22-24$ AB mag and the timescale is $\sim 1-3$ d (e.g.,
\citealt{bk13}).

While the kilonova spectral energy distribution is expected to peak in
the near-IR, deep wide-field searches in this band are not currently
feasible.  For example, ESO's Visible and Infrared Survey Telescope
for Astronomy (VISTA) has an instantaneous field-of-view of about 0.6
deg$^2$, requiring about 170 pointings to cover a typical GW error
region.  This limits the exposure time per pointing to $\sim 100$ s
for a single-night visit.  The $5\sigma$ point-source sensitivity in
such an exposure time is $H\approx 20.5$ AB mag, about a magnitude too
shallow even for a bright kilonova.  A better chance to detect the
near-IR emission from a kilonova counterpart may be presented by a
future space-based observatory, for example WFIRST.  The current
design reference document \citep{sgb+13} indicates an instantaneous
field-of-view of about 0.3 deg$^2$, requiring several hundred
pointings to cover a typical GW error region.  However, the
point-source sensitivity in a 1 min exposure is expected to be
$H\approx 25$ AB mag ($5\sigma$), allowing for the detection of a wide
range of kilonovae.

A more profitable approach, especially prior to the launch of a
facility such as WFIRST, is to use wide-field optical telescopes in
the reddest bands accessible to CCD detectors ($izy$).  In this
context, there are several existing and upcoming facilities capable of
mounting sensitive follow-up campaigns (Pan-STARRS, Blanco/DECam,
Subaru/HSC, LSST).  To detect the faint and fast kilonova signal with
sufficient significance requires at least nightly coverage of the full
GW error region.  For example, Pan-STARRS has a field-of-view of 7
deg$^2$ and can therefore cover a typical error region with about 15
min per pointing, achieving a $5\sigma$ point source sensitivity of
$iz\approx 23.5$ AB mag.  Similarly, with the DECam field-of-view of 3
deg$^2$ a typical GW error region can be covered to a depth of
$iz\approx 24.5$ AB mag ($5\sigma$) in a single night.  A similar
depth can be reached with Subaru/HSC which has a larger aperture, but
a smaller field of view.  Finally, LSST will be able to provide rapid
and efficient coverage to $iz\approx 24.5$ AB mag ($5\sigma$) in $\sim
0.5$ hr, or alternatively to $iz\approx 26$ AB mag in a single night.
These sensitivities are sufficient to detect kilonovae over a broad
range of expected ejecta masses, provided that a systematic campaign
lasting several days is undertaken.  Equally important, non-detections
with such searches will yield meaningful upper bounds on the ejecta
properties.

Unlike in $\gamma$-rays, and to some extent even in the radio band,
optical searches with the required area and depth to robustly detect a
kilonova will undoubtedly yield a wide range of unrelated variable and
transient sources.  To robustly identify or rule out a kilonova will
therefore require the ability to reject background and foreground
contaminants.  One possibility is to restrict the search area to
regions around galaxies within the expected Advanced LIGO/Virgo range
of 200 Mpc \citep{ns10,aaa+12a}.  To achieve this goal requires
all-sky spectroscopic information for galaxies down to at least $\sim
0.1$ L$^*$, which represent the hosts of short GRBs
(Figure~\ref{fig:sfr}; \citealt{ber09}).  Such spectroscopic
information can be obtained from a narrow-band H$\alpha$ imaging
survey, or using an H{\small I} emission line radio survey.

The potential of such surveys was investigated by \citet{mkb13} using
information from existing galaxy surveys, as well as guidance from the
distribution of short GRB host galaxy properties (e.g., stellar
masses, star formation rates).  The basic finding is that realistic
H$\alpha$ and H{\small I} surveys can uncover about $80-90\%$ of the
star formation within 200 Mpc, but only $\sim 30-50\%$ of the stellar
mass.  The high incompleteness for stellar mass is due to the fact
that about half of the stellar mass in the local universe is contained
in elliptical galaxies which produce weak or no H$\alpha$ emission and
contain little neutral hydrogen gas.  The level of completeness with
respect to the population of short GRB hosts is about 50\% for both
methods \citep{mkb13}.  Thus, using galaxy catalogs to reduce the
search area to locations around galaxies within 200 Mpc may result in
a factor of $2-3$ loss in the counterpart identification rate.  On the
other hand, this approach may reduce the background and foreground
contamination dramatically \citep{ns10,aaa+12a}.  For NS-BH mergers,
which can be detected to much larger distances, this method is likely
to be inefficient.

The mitigation of contaminating sources may also benefit from color
information.  In the existing kilonova models, the spectral energy
distribution is unusually red compared to most known transients.
Therefore, the combined use of $iz$ band photometry could
substantially reduce the contamination from AGN flares, supernovae,
supernova shock breakouts, or Galactic stellar flares which generally
peak in the optical/UV \citep{mb12}.  Finally, the shorter duration of
kilonova events, though still $\sim 1-3$ days, can help to distinguish
these sources from foreground stellar flares (generally $\sim {\rm
  hour}$ durations) and background supernovae (generally $\sim {\rm
  weeks}$ durations).  Clearly, the challenge of background/foreground
rejection should be investigated using observations that match the
anticipated signal in depth, timescale, and color.  These parameters
are not currently covered by on-going time-domain surveys such as
Pan-STARRS, PTF, or DES.

\subsection{Speculative Counterparts}

In addition to potential EM counterparts that are motivated by
existing short GRB observations, other speculative counterparts have
been theorized in recent years.  The validity of these theoretical
models, and the nature of any resulting EM radiation are difficult to
assess without observational evidence, while the proposed energy scale
for some counterparts is likely too low to be detectable with existing
facilities.  \citet{zm13} argue that compact object mergers lead to
the amplification of $\sim 10^{16}$ G magnetic fields on the merger
timescale, and speculate that if even $\sim 0.1\%$ of the available
energy is channeled into $\gamma$-rays, the resulting isotropic signal
will be detectable with \swift/BAT.  \citet{plp+13} consider the
interaction of the NS magnetospheres and show that it can power EM
emission with $\sim 10^{40}-10^{43}$ erg s$^{-1}$ prior to the merger.
However, the nature of the resulting radiation (e.g., radio,
$\gamma$-rays) is unclear.  Several researchers considered the
formation of a massive millisecond magnetar as a merger remnant, which
could power an isotropic counterpart with an optical/near-IR
luminosity approaching those of supernovae and a typical timescale of
hours to days, as well as bright X-ray emission on a timescale of
hours \citep{gdw+13,yzg13,zha13} .  \citet{lpl+12} consider the
collapse of a massive, rapidly-rotating magnetar remnant to a black
hole and find that the detached magnetosphere can power a bright EM
counterpart.  \citet{kis12} suggest that the shock produced by the
merger may accelerate the outer layer of the neutron star to
relativistic velocities, with an energy of $\sim 10^{47}$ erg, which
may produce an isotropic radio to X-ray afterglow through interaction
with the ambient medium.  Another proposed energy source for EM
radiation is the failure or cracking of the neutron star crust prior
to the merger, with a potential energy release of up to $\sim 10^{47}
$ erg, although the nature of any resulting radiation is unclear
\citep{paj+12,trh+12}.

\section{The Progenitors of Short GRBs and Future Directions}
\label{sec:conc}

The discovery of short GRB afterglows in 2005 was a watershed event
that led to the first identifications of host galaxies, to a
determination of the distance scale, and to a delineation of the burst
properties.  In the span of less than a decade, intense follow-up
efforts aimed at studying the afterglow emission and the host galaxy
population have led to several key findings that I summarized and
explored in this review:
\begin{itemize}

\item Short GRBs span a wide range of redshifts, at least $z\sim
  0.1-1.5$.  The redshift distribution appears to be affected by the
  sensitivity threshold of \swift/BAT, as evidenced by the similar
  redshift distributions in early- and late-type hosts, despite the
  anticipated longer time delay in the former.  The redshift
  distribution indicates typical progenitor delay times of $\simlt
  {\rm few}$ Gyr.

\item Short GRBs at $z\simlt 0.5$, for which sensitive SN searches
  have been carried out lack associations with Type Ic SNe similar to
  those that accompany long GRBs.  Nearly all of these limits are for
  short GRBs in star-forming galaxies, indicating that such
  environments are not indicative of an association with massive
  stars.

\item Short GRBs occur in both early- and late-type galaxies, with the
  former accounting for about $20\%$ of the sample.  The sub-dominant
  fraction of early-type hosts indicates that the short GRB rate is
  influenced by both stellar mass and star formation activity.  The
  inferred delay time distribution based on the host demographics is
  roughly $P(\tau)\propto \tau^{-1}$.

\item The distribution of host galaxy stellar masses follows the
  expectation from the field galaxy mass function for early-type
  hosts, but is lower than expected for late-type hosts.  This
  indicates that star formation activity results in an elevated short
  GRB rate per unit stellar mass.  Combined with the broad
  distribution of host galaxy stellar population ages, the inferred
  delay time distribution roughly follows $P(\tau)\propto \tau^{-1}$.

\item The host galaxies of short GRBs follow the general properties
  and trends of the field galaxy population (in terms of stellar
  masses, star formation rates, specific star formation rates,
  metallicities), but have systematically higher luminosities, larger
  stellar masses, older stellar population ages, higher metallicities,
  and lower star formation rates and specific star formation rates
  than the hosts of long GRBs.

\item Short GRBs have systematically larger radial offsets from their
  host galaxies than long GRBs, but match the predicted offset
  distribution of compact object binaries from population synthesis
  models that include natal kicks.  About $10\%$ of short GRBs have
  offsets of $\simgt 20$ kpc, extending beyond the typical visible
  extent of their host galaxies.  These bursts appear host-less in
  deep optical/near-IR imaging from {\it HST}, but exhibit nearby
  galaxies with a low probability of chance coincidence.  The short
  GRB offsets normalized by host galaxy size are similarly larger than
  those of long GRBs, core-collapse SNe, and Type Ia SNe, with only
  $20\%$ located at $\simlt 1$ $r_e$, and about $20\%$ located at
  $\simgt 5$ $r_e$.  These results are indicative of natal kicks, or
  an origin in globular clusters, both of which point to compact
  object binary mergers.  The inferred kick velocities are $\sim
  20-140$ km s$^{-1}$, in reasonable agreement with Galactic NS-NS
  binaries and population synthesis models.

\item Short GRBs exhibit a weak spatial correlation with the
  underlying distribution of UV and optical light in their host
  galaxies, with $\sim 50\%$ of all bursts located in the faintest
  regions of their hosts, often with no detectable underlying stellar
  light despite deep imaging with {\it HST}.  This distribution is
  distinct from the strong correlation of long GRBs and core-collapse
  SNe with UV light, and the correlation of Type Ia SNe with optical
  light.  This is indicative of natal kicks that displace the
  progenitors from their birth-sites to their evental explosion sites.

\item The afterglows of short GRBs are systematically less luminous
  than those of long GRBs.  This is partly a result of the lower
  isotropic-equivalent energy scale of short GRBs, but also reflects a
  lower density scale by about an order of magnitude.  The
  distribution of optical afterglow luminosities indicates a typical
  density scale of $\sim 0.1$ cm$^{-3}$, supported by radio limits
  that require a density scale of $\simlt 1$ cm$^{-3}$.

\item Short GRBs exhibit evidence for collimation based on breaks in
  their X-ray, optical, and radio light curves that closely match the
  expectations for jet breaks.  The inferred opening angles span $\sim
  5^\circ$ to $\simgt 20^\circ$, with a resulting mean beaming factor
  of about 70.  This leads to a beaming-corrected energy scale of
  $\sim (0.5-5)\times 10^{49}$ erg, and an event rate of $\sim 10^3$
  Gpc$^{-3}$ yr$^{-1}$.  The inferred rate within the Advanced LIGO
  sensitivity volume is $\sim 25$ yr$^{-1}$.

\item There is evidence for a kilonova association from late near-IR
  excess emission in the short GRB\,130603B, with an inferred
  $r$-process ejecta mass of $\sim 0.05$ M$_\odot$.  If confirmed with
  future kilonova detections, this result provides strong evidence for
  compact object binary progenitors, indicates that compact object
  mergers are the dominant site for $r$-process nucleosynthesis, and
  demonstrates that Advanced LIGO/Virgo gravitational wave sources
  will be accompanied by detectable optical/near-IR emission.

\item In the context of compact object binary progenitors, short GRBs
  provide unique insight on the expected electromagnetic counterparts
  of Advanced LIGO/Virgo gravitational wave sources, both on- and
  off-axis.  Coincident detections of gravitational waves and on-axis
  short GRBs are expected to be rare, $\sim 0.3$ yr$^{-1}$.  Off-axis
  optical afterglow emission is expected to be detectable for $\sim
  10\%$ of gravitational wave sources.  Radio emission from an
  off-axis jet or from an isotropic mildly-relativistic outflow is in
  principle detectable for all viewing angles, but the predicted
  fluxes are too low to be detected with existing or upcoming
  facilities.  Finally, kilonova emission can be detected within the
  Advanced LIGO/Virgo volume using large wide-field optical telescopes
  in the reddest optical bands.

\end{itemize}

Within the range of proposed progenitor systems for short GRBs, the
binary compact object merger scenario provides the best agreement with
the broad range of observations.  A dominant channel of young
magnetars conflicts with the occurrence of short GRBs in early-type
galaxies, with the large offsets, and with the lack of spatial
correlation with host UV light.  A dominant channel of delayed
magnetars, formed through white dwarf accretion-induced collapse or
white dwarf binary mergers, can explain the host demographics, but
does not naturally lead to the observed offset distribution or to the
weak spatial correlation with optical light.  Finally, models in which
short GRBs are produced by massive stars in a similar fashion to long
GRBs are ruled out by the lack of SN associations, the occurrence of
short GRBs in early-type galaxies, the clear differences in the host
galaxy population, the lack of spatial association with UV light, and
the large differences in energy and circumburst density scales.  It is
indeed quite remarkable that the underlying theoretical motivations
that led to the compact object binary merger model prior to the
identification of short GRBs as a distinct class (e.g.,
\citealt{elp+89,npp92}) have been borne out by the extensive
observational effort over the past decade.

At the same time, there are clear challenges for the compact object
binary merger model, which may require a more careful assessment of
the physical conditions during the merger, or perhaps a minor
contribution from a distinct progenitor channel.  For example, the
$\sim 100$-sec extended emission observed in about 15\% of short GRBs
does not naturally fit in the basic merger framework, although several
theoretical explanations exist for this emission component
\citep{mrz05,ros07,mqt08,maq+10,bmt+12}.  Similarly, there are
potential clues in the commonalities and differences between the
prompt emission properties of long and short GRBs (with and without
extended emission) that may shed light on the nature of the
progenitors and/or the engine.  In addition, it remains to be seen if
NS-BH mergers contribute to the short GRB population, and if so, what
observational signatures distinguish them from NS-NS binary mergers.
The answer may have to await gravitational wave detections that can
shed light on the relative fraction of occurrence of NS-NS and NS-BH
binaries.

Continued studies of short GRBs from \swift, {\it Fermi}, and future
$\gamma$-ray missions remain a top priority.  The existing sample
already provides deep insight into the nature of the progenitors and
the explosion properties, but larger samples, combined with individual
high-quality events (e.g., GRB\,130603B) are essential for addressing
some of the open questions listed above.  An operational $\gamma$-ray
satellite (preferably with all-sky coverage) in the Advanced
LIGO/Virgo era is also critical for establishing a direct connection
between compact object mergers and short GRBs.

\vspace{0.3in}

\noindent {\bf Acknowledgements}: I thank my primary collaborators in
the study of short GRBs: Wen-fai Fong, Ryan Chornock, Raffaella
Margutti, Brian Metzger, Ashley Zauderer, and Derek Fox.  I also
acknowledge helpful discussions and occasional collaborations with
Alicia Soderberg, Ramesh Narayan, Josh Grindlay, Christopher Fryer,
Enrico Ramirez-Ruiz, Eliot Quataert, Lars Bildsten, Daniel Kasen,
Camille Leibler, S.~Bradley Cenko, Ehud Nakar, Re'em Sari, Nial
Tanvir, Andrew Levan, Krzysztof Belczynski, Duncan Brown, Luis Lehner,
Daniel Holz, Alessandra Bounano, David Kaplan, Neil Gehrels, Jochen
Greiner, Ryan Foley, Dale Frail, Avishay Gal-Yam, Antonino Cucchiara,
Shrinivas Kulkarni, and Paul Price.  I also thank the participants of
the ``Chirps, Mergers and Explosions: The Final Moments of Coalescing
Compact Binaries'' workshop at the Kavli Institute for Theoretical
Physics, Santa Barbara for stimulating discussions about short GRBs
and gravitational wave astrophysics.  Finally, I acknowledge support
for some of this work from the National Science Foundation under Grant
AST-1107973, and from several NASA {\it Swift} and {\it Fermi} grants.

\clearpage
{\small
\begin{longtable}{lccccccccc}
\caption{Prompt Emission and Afterglow Properties of Short GRBs} \label{tab:afterglows} \\\hline
GRB & Satellite & $T_{90}$ & $F_\gamma$       & $z$ & $F_{\rm X,11}$                  & $t_{\rm opt}$ & $F_{\rm \nu,opt}$ & $t_{\rm radio}$ & $F_{\rm\nu,radio}$ \\
       &              & (s)          & (erg cm$^{-2}$) &        & (erg cm$^{-2}$ s$^{-1}$) & (hr)             & ($\mu$Jy)        & (hr)                & ($\mu$Jy)          \\
\hline\hline
\endfirsthead
\multicolumn{10}{c}{{\tablename\ \thetable{} -- Continued from previous page}} \\\hline
GRB & Satellite & $T_{90}$ & $F_\gamma$       & $z$ & $F_{\rm X,11}$                  & $t_{\rm opt}$ & $F_{\rm \nu,opt}$ & $t_{\rm radio}$ & $F_{\rm\nu,radio}$ \\
       &              & (s)          & (erg cm$^{-2}$) &        & (erg cm$^{-2}$ s$^{-1}$) & (hr)             & ($\mu$Jy)        & (hr)                & ($\mu$Jy)          \\
\hline\hline
\endhead
\hline\multicolumn{10}{r}{{Continued on next page}} \\ \hline
\endfoot
\hline
\endlastfoot
050202    & \swift\ & $0.27$ & $3.0\times 10^{-8}$ & ...           & ...$^a$     & ...   & ...   & ...    & ...   \\
050509B  & \swift\ & $0.04$ & $2.3\times 10^{-8}$ & $0.225$ & $<1.95\times 10^{-14}$  & $2.1 $  & $<0.8$ & ...         & ...            \\  
050709    & HETE-2 & $0.07$ & $4.0\times 10^{-7}$ & $0.161$ & $1.92\times 10^{-14} $ & $34.0$ & $2.3 $ & $1.57$ & $<65$    \\   
050724    & \swift\  & $3.0$/EE  & $6.3\times 10^{-7}$ & $0.257$ & $9.55\times 10^{-14} $ & $12.0$ & $8.4 $ & $1.69$ & $465$    \\        
050813    & \swift\ & $0.6 $  & $1.2\times 10^{-7}$ & ...           & $<2.12\times 10^{-14}$  & $12.8$ & $<1.9$ & $1.64$ & $<55$  \\
050906    & \swift\ & $0.26$ & $6.0\times 10^{-9}$ & ...           & ...$^b$     & ...   & ...   & ...   & ...       \\
050925    & \swift\ & $0.07$ & $7.6\times 10^{-8}$ & ...            & ...$^b$   & ...   & ...   & ...  & ...        \\
051210    & \swift\ & $1.3 $  & $8.3\times 10^{-8}$ & $1.3$     & $<2.70\times 10^{-14}$  & $19.2$ & $<1.6$ & ...         & ...           \\
051105A  & \swift\ & $0.09$ & $2.2\times 10^{-8}$ & ...            & ...$^b$    & ...   & ...   & ...    & ...     \\
051221A  & \swift\  & $1.4 $  & $1.2\times 10^{-6}$ & $0.546$ &$1.08\times 10^{-12} $ & $3.1 $ & $5.8 $  & $0.91$ & $173$    \\ \hline        
060121    & HETE-2 & $2.0 $  & $4.7\times 10^{-6}$ & ...           & $1.09\times10^{-12} $ & $7.4 $ & $8.8 $   & ...         & ...           \\       
060313    & \swift\  & $0.7 $  & $1.1\times 10^{-6}$ & ...           &$3.85\times 10^{-13} $ & $2.8 $ & $10.8$  & $2.12$ & $<110$ \\         
060502B  & \swift\ & $0.09$ & $4.0\times 10^{-8}$ & $0.287$ & $<1.47\times 10^{-14}$  & $16.8$ & $<0.7$ & ...         & ...         \\
060801    & \swift\ & $0.5 $  & $8.1\times 10^{-8}$ & $1.130$ & $<9.80\times 10^{-15}$ & $12.4$ & $<0.8$ & $0.49$ & $<105$ \\
061006    & \swift\  & $0.4$/EE  & $1.4\times 10^{-6}$ & $0.438$ & $2.27\times 10^{-13} $ & $14.9$ & $2.9 $ & ...         & ...           \\        
061201    & \swift\  & $0.8 $  & $3.3\times 10^{-7}$ & $0.111$ & $1.92\times 10^{-13} $ & $8.6 $ & $2.9 $  & ...         & ...           \\        
061210    & \swift\ & $0.2$/EE  & $3.0\times 10^{-7}$ & $0.409$ & $1.36\times 10^{-13} $   & $2.1 $  & $<1.4$ & $1.90$ & $<102$ \\    
061217    & \swift\ & $0.2 $  & $4.6\times 10^{-8}$ & $0.827$ & ...      & $2.8 $  & $<2.0$ & ...         & ...          \\ \hline
070209    & \swift\ & $0.09$ & $2.2\times 10^{-8}$ & ...            & ...$^b$    & ...   & ...   & ...    & ...      \\
070406    & \swift\ & $1.20$ & $3.6\times 10^{-8}$ & ...            & ...$^a$    & ...   & ...   & ...   & ...       \\
070429B  & \swift\ & $0.5 $  & $6.3\times 10^{-8}$ & $0.902$ & $1.13\times 10^{-13} $   & $4.8 $  & $<0.6$ & ...         & ...           \\   
070707    & INTEGRAL & $1.1 $ & $1.4\times 10^{-6}$ & ...         & $5.04\times 10^{-13} $ & $11.0$ & $1.9 $ & ...         & ...            \\        
070714B  & \swift\  & $2.0 $  & $7.2\times 10^{-7}$ & $0.923$ & $6.30\times 10^{-14} $ & $23.6$ & $0.7 $ & $15.5$ & $<139$  \\
070724A  & \swift\  & $0.4 $  & $3.0\times 10^{-8}$ & $0.457$ & $1.28\times 10^{-13} $ & $3.3 $ & $3.8 $ & $1.06$ & $<250$   \\         
070729    & \swift\ & $0.9 $  & $1.0\times 10^{-7}$ & $0.8$     & $<4.71\times 10^{-14}$ & $0   $    & $<0  $ & ...         & ...          \\   
070809    & \swift\  & $1.3 $  & $1.0\times 10^{-7}$ & $0.473$ & $5.30\times 10^{-13} $ & $11.2$ & $0.7 $ & ...         & ...           \\         
070810B  & \swift\ & $0.08$ & $1.2\times 10^{-8}$ & ...            & ...$^b$    & ...   & ...   & ...   & ...     \\
070923    & \swift\ & $0.05$ & $5.0\times 10^{-8}$ & ...            & ...$^a$    & ...   & ...   & ...   & ...      \\
071112B  & \swift\ & $0.30$ & $4.8\times 10^{-8}$ & ...            & $<2.00\times 10^{-13}$  & ...   & ...   & ...   & ...   \\
071227    & \swift\  & $1.8 $  & $2.2\times 10^{-7}$ & $0.381$ & $3.20\times 10^{-14} $ & $7.0 $ & $1.6 $  & ...         & ...         \\   \hline     
080121    & \swift\ & $0.7 $ & $3.0\times 10^{-8}$ & ...             & ...$^a$     & ...   & ...   & ...    & ...     \\
080426    & \swift\ & $1.7 $  & $3.7\times 10^{-7}$ & ...           & $2.24\times 10^{-13} $  & $7.5 $   & $<2.6$ & ...         & ...          \\
080503    & \swift\  & $0.3$/EE  & $6.1\times 10^{-8}$ & ...           &$<3.05\times 10^{-14}$ & $25.9$ & $0.3 $ & $3.05$ & $<55$    \\           
080702A  & \swift\ & $0.5 $ & $3.6\times 10^{-8}$ & ...             & $<1.06\times 10^{-13}$    & ...   & ...   & $0.83$ & $<155$ \\
080905A  & \swift\  & $1.0 $  & $1.4\times 10^{-7}$ & $0.122$ & $<6.70\times 10^{-14}$ & $8.5 $ & $0.8 $ & ...         & ...           \\         
080919    & \swift\ & $0.6 $ & $7.2\times 10^{-8}$ & ...             & $<2.12\times 10^{-14}$ & ...$^d$   & ... & ...   & ...       \\
081024    & \swift\ & $1.8 $ & $1.2\times 10^{-7}$ & ...              & $<4.56\times 10^{-14}$ & ...$^d$   & ...  & ...   & ...     \\
081226A  & \swift\  & $0.4 $  & $9.9\times 10^{-8}$ & ...           & $<2.57\times 10^{-14}$ & $0.37$ & $1.3 $ & ...         & ...          \\  \hline  
090305    & \swift\  & $0.4 $  & $7.5\times 10^{-8}$ & ...           & $<8.50\times 10^{-14}$ & $0.56$ & $2.0 $ & ...         & ...           \\         
090417A  & \swift\ & $0.07$ & $1.9\times 10^{-8}$ & ...            & ...$^a$    & ...   & ...   & ...    & ...      \\
090426    & \swift\ & $1.2 $ & $1.8\times 10^{-7}$ & $2.609$   & $2.63\times 10^{-13} $ & $0.35$ & $14.7$  & ...   & ...     \\
090510    & \swift\ & $0.3 $  & $3.4\times 10^{-7}$ & $0.903$ &$5.04\times 10^{-14} $   & $9.0 $ & $2.3 $ & $1.46$ & $<145$ \\   
090515    & \swift\  & $0.04$ & $2.1\times 10^{-8}$ & $0.403$ & $<8.43\times 10^{-14}$ & $1.9 $ & $0.1 $ & $0.87$ & $<60$  \\         
090621B  & \swift\ & $0.14$ & $7.0\times 10^{-8}$ & ...           & $2.70\times10^{-14} $   & $0.9 $   & $<3.3$ & $0.61$ & $<54$  \\   
090715    & \swift\ & $0.5$/EE & $9.7\times 10^{-7}$ & ...           & ...$^a$ & ...   & ...    & ...  & ...     \\
090815C  & \swift\ & $0.6 $ & $4.4\times 10^{-8}$ & ...             & $<7.00\times 10^{-14}$ & ...   & ...   & ...    & ...       \\
090916    & \swift\ & $0.3$/EE & $9.5\times 10^{-7}$ & ...            & $<3.00\times 10^{-14}$ & ...   & ...    & ...   & ...      \\
091109B  & \swift\  & $0.3 $  & $1.9\times 10^{-7}$ & ...           & $1.26\times 10^{-13} $   & $6.0 $ & $0.7 $ & ...         & ...           \\  \hline      
100117A  & \swift\  & $0.3 $  & $9.3\times 10^{-8}$ & $0.915$ & $<2.50\times 10^{-14}$ & $8.4 $ & $0.2 $ & ...         & ...           \\         
100206A  & \swift\ & $0.1 $  & $1.4\times 10^{-7}$ & $0.408$ & $<1.07\times 10^{-14}$ & $15.7$ & $<0.5$ & ...         & ...          \\    
100625A  & \swift\ & $0.3$/EE & $2.3\times 10^{-7}$ & $0.453$  & $3.95\times 10^{-15}$ & $12.7$ & $<4.4$ & ...         & ...           \\   
100628A  & \swift\ & $0.04$ & $2.5\times 10^{-8}$ & ...           & ...$^b$    & ...   & ...   & ...    & ...     \\
100702A  & \swift\ & $0.16$/EE & $1.2\times 10^{-7}$ & ...            & $<6.63\times 10^{-14}$ & ...$^d$   & ...  & ...   & ...     \\
101129A  & \swift\ & $0.35$ & $9.0\times 10^{-8}$ & ...            & ...$^a$   & ...   & ...   & ...    & ...    \\
101219A  & \swift\ & $0.6 $ & $4.6\times 10^{-7}$ & $0.718$  & $<2.00\times 10^{-14}$ & $1.0 $ & $<3.2$ & ...         & ...           \\    
101224A  & \swift\ & $0.2 $ & $5.8\times 10^{-7}$ & ...             & ...$^c$     & ...   & ...   & ...    & ...       \\ \hline
110112A  & \swift\  & $0.5 $  & $3.0\times 10^{-8}$ & ...           & $8.22\times 10^{-14} $  & $15.4$ & $2.8 $ & $1.90$ & $<75$    \\        
110420B  & \swift\ & $0.08$ & $5.3\times 10^{-8}$ & ...             & $<1.40\times 10^{-13}$ & ...   & ...   & $0.46$ & $<45$   \\
111020A  & \swift\ & $0.4 $ & $6.5\times 10^{-8}$ & ...            & $4.87\times10^{-13} $   & $17.7$ & $<0.6$ & $0.7$ & $<39$      \\ 
111117A  & \swift\ & $0.5 $ & $1.4\times 10^{-7}$ & $1.3$      & $3.21\times 10^{-14} $  & $13.2$ & $<0.2$ & $0.50$ & $<18$   \\    
111121A  & \swift\ & $0.45$/EE & $2.2\times 10^{-6}$ & ...           & $4.80\times 10^{-13}$   & ...$^d$     & ...          & $0.80$ & $<78$   \\
111126A  & \swift\ & $0.8 $ & $7.0\times 10^{-8}$ & ...             & ...$^a$     & ...   & ...   & ...    & ...       \\ \hline
120229A  & \swift\ & $0.22$ & $4.1\times 10^{-8}$ & ...           & ...$^a$    & ...   & ...   & ...  & ...         \\
120305A  & \swift\ & $0.10$ & $2.0\times 10^{-7}$ & ...           & $2.40\times10^{-14} $ & ...   & ...   & $2.25$ & $<18$   \\
120403A  & \swift\ & $1.25$ & $1.0\times 10^{-7}$ & ...            & ...$^a$   & ...   & ...   & ...    & ...     \\
120521A  & \swift\ & $0.45$ & $7.8\times 10^{-8}$ & ...           & $<1.50\times 10^{-13}$ & $17.7$ & $<0.9$ & ...         & ...           \\
120630A  & \swift\ & $0.6 $ & $6.1\times 10^{-8}$ & ...              & ...$^c$     & ...   & ...   & ...    & ...     \\
120804A  & \swift\  & $0.81$ & $8.8\times 10^{-7}$ & $1.3$     & $5.86\times 10^{-13} $  & $5.5 $  & $5.4 $ & $1.95$ & $<18$   \\   
121226A  & \swift\ & $1.0 $ & $1.4\times 10^{-7}$ & ...            & $2.24\times 10^{-13} $  & $11.1$ & $<1.9$ & ...         & ...            \\ \hline
130603B  & \swift\  & $0.18$ & $6.3\times 10^{-7}$ & $0.356$  & $6.00\times 10^{-13} $  & $8.2 $ & $8.6 $ & $0.37$ & $119$   \\    
\hline
\hline
\caption{Prompt emission and afterglow properties of the short GRB
  sample spanning January 2005 to January 2013, with the addition of
  GRB\,130603B.  The columns are: (i) GRB name; (ii) discovery
  satellite; (iii) duration (EE: extended emission); (iv) fluence in the $15-150$ keV band for
  \swift\ events; (v) redshift; (vi) observed X-ray flux at 11 hr
  post-burst; (vii) time of first optical detection; (viii) flux
  density in the optical $r$-band; (ix) time of radio observation for
  non-detections, or of peak radio brightness for detections; and (x)
  radio flux density (at 5 or 8.5 GHz).\\
  $^a$ No rapid \swift/XRT follow-up due to observing constraints.\\
  $^b$ No \swift/XRT detection despite rapid follow-up.\\
  $^c$ No \swift/XRT observations at $\simgt 10^3$ s.\\
  $^d$ Sight-line with significant Milky Way extinction. }
\end{longtable}
}

\clearpage
{\small
\begin{longtable}{lccccccc}
\caption{Properties of Short GRB Host Galaxies} \label{tab:hosts} \\\hline
GRB & $z$ & Type & $L_B$   & SFR                           & $\tau_*$ & ${\rm log}(M_*)$ & 12+log(O/H) \\
       &        &         & ($L^*$) & (M$_\odot$ yr$^{-1}$) & (Gyr)       & (M$_\odot$)          &                      \\
\hline\hline
\endfirsthead
\multicolumn{8}{c}{{\tablename\ \thetable{} -- Continued from previous page}} \\\hline
GRB & $z$ & Type & $L_B$   & SFR                           & $\tau_*$ & ${\rm log}(M_*)$ & 12+log(O/H) \\
       &        &         & ($L^*$) & (M$_\odot$ yr$^{-1}$) & (Gyr)       & (M$_\odot$)          &                      \\
\hline\hline
\endhead
\hline\multicolumn{8}{r}{{Continued on next page}} \\ \hline
\endfoot
\hline
\endlastfoot
\noalign{\smallskip}
\multicolumn{8}{c}{{Sub-arcsecond Localizations}}  \\
\noalign{\smallskip}
\hline
\noalign{\smallskip}
050709    & $0.161$  & L   & $0.1$ & $0.15$ & $0.26$ & $8.8 $ & $8.5$ \\
050724A  & $0.257$  & E   & $1.0$ & $<0.1$ & $0.94$ & $10.8$ & $...$ \\
051221A  & $0.546$  & L   & $0.3$ & $0.95$ & $0.17$ & $9.4 $ & $8.8$ \\
060121    & $<4.1$   & ?   & $...$ & $... $ & $... $ & $... $ & $...$ \\
060313    & $<1.7$   & ?   & $...$ & $... $ & $... $ & $... $ & $...$ \\
061006    & $0.438$  & L   & $0.1$ & $0.24$ & $0.24$ & $9.0 $ & $8.6$ \\
061201    & $0.111$  & H/L & $0.1$ & $0.14$ & $... $ & $... $ & $...$ \\
070429B  & $0.902$  & L   & $0.6$ & $1.1 $ & $0.46$ & $10.4$ & $...$ \\
070707    & $<3.6$   & ?   & $...$ & $... $ & $... $ & $... $ & $...$ \\
070714B  & $0.922$  & L   & $0.1$ & $0.44$ & $0.22$ & $9.4 $ & $...$ \\ 
070724A  & $0.457$  & L   & $1.4$ & $2.5 $ & $0.30$ & $10.1$ & $8.9$ \\
070809    & $0.473$  & H/E & $1.9$ & $<0.1$ & $3.10$ & $11.4$ & $...$ \\
071227    & $0.381$  & L   & $1.2$ & $0.6 $ & $0.49$ & $10.4$ & $8.5$ \\
080503    & $<4.2$   & H/? & $...$ & $... $ & $... $ & $... $ & $...$ \\
080905A  & $0.122$  & L   & $...$ & $... $ & $... $ & $... $ & $8.7$ \\
081226A  & $<4.1$   & ?   & $...$ & $... $ & $... $ & $... $ & $...$ \\
090305    & $<4.1$   & H/? & $...$ & $... $ & $... $ & $... $ & $...$ \\
090426A  & $2.609$  & L   & $...$ & $... $ & $... $ & $... $ & $...$ \\
090510    & $0.903$  & L   & $0.3$ & $0.3 $ & $0.14$ & $9.7 $ & $...$ \\
090515    & $0.403$  & H/E & $1.0$ & $0.1 $ & $4.35$ & $11.2$ & $...$ \\
091109B  & $<4.4$   & ?   & $...$ & $... $ & $... $ & $... $ & $...$ \\
100117A  & $0.915$ & E   & $0.5$ & $<0.2$ & $0.79$ & $10.3$ & $...$ \\
110112A  & $<5.3$   & H/? & $...$ & $... $ & $... $ & $... $ & $...$ \\
111020A  & ...            & ?     & $...$ & $... $ & $... $ & $... $ & $...$ \\
111117A  & $1.2$      & L   & $0.6$ & $6.0 $ & $0.09$ & $9.6 $ & $...$ \\
120804A  & $1.3$      & L   & $0.8$ & $8.0 $ & $0.13$ & $10.8$ & $...$ \\
130603B  & $0.356$  & L   & $1.0$ & $1.7 $ & ...       & $9.7$   & $8.7$ \\
\noalign{\smallskip}
\hline
\noalign{\smallskip}
\multicolumn{8}{c}{{{\it Swift}/XRT Positions Only}} \\
\noalign{\smallskip}
\hline
\noalign{\smallskip}
050509B & $0.225$  & E   & $5.0$ & $<0.15$ & $3.18$ & $11.6$ & $...$ \\
050813  & ...             & E   & $...$ & $...  $ & $... $ & $... $ & $...$ \\
051210  & $>1.4$    & ?    & $...$   & $...  $     & $... $    & $... $    & $...$ \\
060502B & $0.287$  & E   & $1.6$ & $0.8  $   & $1.3 $  & $11.8$ & $...$ \\
060801   & $1.130$  & L   & $0.6$ & $6.1  $   & $0.03$ & $9.1 $ & $...$ \\
061210   & $0.409$  & L   & $0.9$ & $1.2  $   & $0.38$ & $9.6 $ & $8.8$ \\
061217   & $0.827$  & L   & $0.4$ & $2.5  $   & $0.03$ & $9.1 $ & $...$ \\
070729   & $0.8$      & E   & $1.0$ & $<1.5 $ & $0.98$ & $10.6$ & $...$ \\ 
080123   & $0.495$  & L   & $1.2$ & $...  $     & $0.31$ & $10.1$ & $...$ \\  
100206A & $0.407$  & L   & $1.0$ & $30   $   & $0.10$ & $10.8$ & $9.2$ \\
100625A & $0.452$  & E   & $0.2$ & $0.3  $   & $0.79$ & $10.3$ & $...$ \\
101219A & $0.718$  & L   & $1.3$ & $16   $   & $0.03$ & $9.2 $ & $...$ \\
\hline
\hline
\caption{Properties of short GRB host galaxies.  The columns are: (i)
  GRB name; (ii) redshift; (iii) galaxy type (E: early-type; L:
  late-type; ?: unknown; H: host-less); (iv) optical $B$-band
  luminosity normalized by $L_B^*$ at the appropriate redshift; (v)
  star formation rate; (vi) dominant stellar population age; (vii)
  logarithm of the stellar mass; and (viii) metallicity.  Data are
  from \citet{ber09}, \citet{lb10}, \citet{ber10}, \citet{fbc+13} and
  references therein.}
\end{longtable}
}


\begin{thebibliography}{}
\expandafter\ifx\csname natexlab\endcsname\relax\def\natexlab#1{#1}\fi

\bibitem[{{Abadie} et~al.(2012{\natexlab{a}}){Abadie}, {Abbott}, {Abbott},
  {Abbott}, {Abernathy} et~al.}]{aaa+12c}
{Abadie} J, {Abbott} BP, {Abbott} R, {Abbott} TD, {Abernathy} M, et~al.
  2012{\natexlab{a}}.
\newblock \textit{\apj} 760:12

\bibitem[{{Abadie} et~al.(2010){Abadie}, {Abbott}, {Abbott}, {Abernathy},
  {Accadia} et~al.}]{aaa+10}
{Abadie} J, {Abbott} BP, {Abbott} R, {Abernathy} M, {Accadia} T, et~al. 2010.
\newblock \textit{Classical and Quantum Gravity} 27:173001

\bibitem[{{Abadie} et~al.(2012{\natexlab{b}}){Abadie}, {Abbott}, {Abbott},
  {Abbott}, {Abernathy} et~al.}]{aaa+12b}
{Abadie} J, {Abbott} BP, {Abbott} TD, {Abbott} R, {Abernathy} M, et~al.
  2012{\natexlab{b}}.
\newblock \textit{\apj} 755:2

\bibitem[{{Abbott} et~al.(2008){Abbott}, {Abbott}, {Adhikari}, {Agresti},
  {Ajith} et~al.}]{aaa+08a}
{Abbott} B, {Abbott} R, {Adhikari} R, {Agresti} J, {Ajith} P, et~al. 2008.
\newblock \textit{\apj} 681:1419--1430

\bibitem[{{Abdo} et~al.(2009){Abdo}, {Ackermann}, {Ajello}, {Asano}, {Atwood}
  et~al.}]{aaa+09a}
{Abdo} AA, {Ackermann} M, {Ajello} M, {Asano} K, {Atwood} WB, et~al. 2009.
\newblock \textit{\nat} 462:331--334

\bibitem[{{Abdo} et~al.(2010){Abdo}, {Ackermann}, {Ajello}, {Asano}, {Atwood}
  et~al.}]{aaa+10a}
{Abdo} AA, {Ackermann} M, {Ajello} M, {Asano} K, {Atwood} WB, et~al. 2010.
\newblock \textit{\apj} 712:558--564

\bibitem[{{Abramovici} et~al.(1992){Abramovici}, {Althouse}, {Drever},
  {Gursel}, {Kawamura} et~al.}]{aad+92}
{Abramovici} A, {Althouse} WE, {Drever} RWP, {Gursel} Y, {Kawamura} S, et~al.
  1992.
\newblock \textit{Science} 256:325--333

\bibitem[{{Accadia} et~al.(2011){Accadia}, {Acernese}, {Antonucci}, {Astone},
  {Ballardin} et~al.}]{aaa+11a}
{Accadia} T, {Acernese} F, {Antonucci} F, {Astone} P, {Ballardin} G, et~al.
  2011.
\newblock \textit{Classical and Quantum Gravity} 28:114002

\bibitem[{{Ackermann} et~al.(2010){Ackermann}, {Asano}, {Atwood}, {Axelsson},
  {Baldini} et~al.}]{aaa10b}
{Ackermann} M, {Asano} K, {Atwood} WB, {Axelsson} M, {Baldini} L, et~al. 2010.
\newblock \textit{\apj} 716:1178--1190

\bibitem[{{Aloy}, {Janka} \& {M{\"u}ller}(2005)}]{ajm05}
{Aloy} MA, {Janka} HT, {M{\"u}ller} E. 2005.
\newblock \textit{\aap} 436:273--311

\bibitem[{{Amati} et~al.(2002){Amati}, {Frontera}, {Tavani}, {in't Zand},
  {Antonelli} et~al.}]{aft+02}
{Amati} L, {Frontera} F, {Tavani} M, {in't Zand} JJM, {Antonelli} A, et~al.
  2002.
\newblock \textit{\aap} 390:81--89

\bibitem[{{Antonelli} et~al.(2009){Antonelli}, {D'Avanzo}, {Perna}, {Amati},
  {Covino} et~al.}]{adp+09}
{Antonelli} LA, {D'Avanzo} P, {Perna} R, {Amati} L, {Covino} S, et~al. 2009.
\newblock \textit{\aap} 507:L45--L48

\bibitem[{{Antoniadis} et~al.(2013){Antoniadis}, {Freire}, {Wex}, {Tauris},
  {Lynch} et~al.}]{afw+13}
{Antoniadis} J, {Freire} PCC, {Wex} N, {Tauris} TM, {Lynch} RS, et~al. 2013.
\newblock \textit{Science} 340:448

\bibitem[{{Arnett}(1982)}]{arn82}
{Arnett} WD. 1982.
\newblock \textit{\apj} 253:785--797

\bibitem[{{Asano}, {Guiriec} \& {M{\'e}sz{\'a}ros}(2009)}]{agm09}
{Asano} K, {Guiriec} S, {M{\'e}sz{\'a}ros} P. 2009.
\newblock \textit{\apjl} 705:L191--L194

\bibitem[{{Band} et~al.(1993){Band}, {Matteson}, {Ford}, {Schaefer}, {Palmer}
  et~al.}]{bmf+93}
{Band} D, {Matteson} J, {Ford} L, {Schaefer} B, {Palmer} D, et~al. 1993.
\newblock \textit{\apj} 413:281--292

\bibitem[{{Baring} \& {Harding}(1997)}]{bh97}
{Baring} MG, {Harding} AK. 1997.
\newblock \textit{\apj} 491:663

\bibitem[{{Barnes} \& {Kasen}(2013)}]{bk13}
{Barnes} J, {Kasen} D. 2013.
\newblock \textit{\apj} 775:18-26

\bibitem[{{Barthelmy} et~al.(2005){Barthelmy}, {Chincarini}, {Burrows},
  {Gehrels}, {Covino} et~al.}]{bcb+05}
{Barthelmy} SD, {Chincarini} G, {Burrows} DN, {Gehrels} N, {Covino} S, et~al.
  2005.
\newblock \textit{\nat} 438:994--996

\bibitem[{{Battaglia} et~al.(2005){Battaglia}, {Helmi}, {Morrison}, {Harding},
  {Olszewski} et~al.}]{bhm+05}
{Battaglia} G, {Helmi} A, {Morrison} H, {Harding} P, {Olszewski} EW, et~al.
  2005.
\newblock \textit{\mnras} 364:433--442

\bibitem[{{Bauswein}, {Goriely} \& {Janka}(2013)}]{bgj13}
{Bauswein} A, {Goriely} S, {Janka} HT. 2013.
\newblock \textit{\apj} 773:78

\bibitem[{{Beckwith} et~al.(2006){Beckwith}, {Stiavelli}, {Koekemoer},
  {Caldwell}, {Ferguson} et~al.}]{bsk+06}
{Beckwith} SVW, {Stiavelli} M, {Koekemoer} AM, {Caldwell} JAR, {Ferguson} HC,
  et~al. 2006.
\newblock \textit{\aj} 132:1729--1755

\bibitem[{{Belczynski} et~al.(2010){Belczynski}, {Holz}, {Fryer}, {Berger},
  {Hartmann} \& {O'Shea}}]{bhf+10}
{Belczynski} K, {Holz} DE, {Fryer} CL, {Berger} E, {Hartmann} DH, {O'Shea} B.
  2010.
\newblock \textit{\apj} 708:117--126

\bibitem[{{Belczynski} et~al.(2006){Belczynski}, {Perna}, {Bulik}, {Kalogera},
  {Ivanova} \& {Lamb}}]{bpb+06}
{Belczynski} K, {Perna} R, {Bulik} T, {Kalogera} V, {Ivanova} N, {Lamb} DQ.
  2006.
\newblock \textit{\apj} 648:1110--1116

\bibitem[{{Bell} et~al.(2003){Bell}, {McIntosh}, {Katz} \& {Weinberg}}]{bmk+03}
{Bell} EF, {McIntosh} DH, {Katz} N, {Weinberg} MD. 2003.
\newblock \textit{\apjs} 149:289--312

\bibitem[{{Berger}(2007)}]{ber07}
{Berger} E. 2007.
\newblock \textit{\apj} 670:1254--1259

\bibitem[{{Berger}(2009)}]{ber09}
{Berger} E. 2009.
\newblock \textit{\apj} 690:231--237

\bibitem[{{Berger}(2010)}]{ber10}
{Berger} E. 2010.
\newblock \textit{\apj} 722:1946--1961

\bibitem[{{Berger}(2011)}]{ber11}
{Berger} E. 2011.
\newblock \textit{\nar} 55:1--22

\bibitem[{{Berger} et~al.(2009){Berger}, {Cenko}, {Fox} \&
  {Cucchiara}}]{bcf+09}
{Berger} E, {Cenko} SB, {Fox} DB, {Cucchiara} A. 2009.
\newblock \textit{\apj} 704:877--882

\bibitem[{{Berger} et~al.(2011){Berger}, {Chornock}, {Holmes}, {Foley},
  {Cucchiara} et~al.}]{bch+11}
{Berger} E, {Chornock} R, {Holmes} TR, {Foley} RJ, {Cucchiara} A, et~al. 2011.
\newblock \textit{\apj} 743:204

\bibitem[{{Berger}, {Fong} \& {Chornock}(2013)}]{bfc13}
{Berger} E, {Fong} W, {Chornock} R. 2013.
\newblock \textit{\apjl} 774:L23--L27

\bibitem[{{Berger} et~al.(2007{\natexlab{a}}){Berger}, {Fox}, {Price}, {Nakar},
  {Gal-Yam} et~al.}]{bfp+07}
{Berger} E, {Fox} DB, {Price} PA, {Nakar} E, {Gal-Yam} A, et~al.
  2007{\natexlab{a}}.
\newblock \textit{\apj} 664:1000--1010

\bibitem[{{Berger} et~al.(2005{\natexlab{a}}){Berger}, {Kulkarni}, {Fox},
  {Soderberg}, {Harrison} et~al.}]{bkf+05}
{Berger} E, {Kulkarni} SR, {Fox} DB, {Soderberg} AM, {Harrison} FA, et~al.
  2005{\natexlab{a}}.
\newblock \textit{\apj} 634:501--508

\bibitem[{{Berger}, {Kulkarni} \& {Frail}(2003)}]{bkf03}
{Berger} E, {Kulkarni} SR, {Frail} DA. 2003.
\newblock \textit{\apj} 590:379--385

\bibitem[{{Berger} et~al.(2005{\natexlab{b}}){Berger}, {Price}, {Cenko},
  {Gal-Yam}, {Soderberg} et~al.}]{bpc+05}
{Berger} E, {Price} PA, {Cenko} SB, {Gal-Yam} A, {Soderberg} AM, et~al.
  2005{\natexlab{b}}.
\newblock \textit{\nat} 438:988--990

\bibitem[{{Berger} et~al.(2007{\natexlab{b}}){Berger}, {Shin}, {Mulchaey} \&
  {Jeltema}}]{bsm+07}
{Berger} E, {Shin} MS, {Mulchaey} JS, {Jeltema} TE. 2007{\natexlab{b}}.
\newblock \textit{\apj} 660:496--503

\bibitem[{{Berger} et~al.(2013){Berger}, {Zauderer}, {Levan}, {Margutti},
  {Laskar} et~al.}]{bzl+13}
{Berger} E, {Zauderer} BA, {Levan} A, {Margutti} R, {Laskar} T, et~al. 2013.
\newblock \textit{\apj} 765:121

\bibitem[{{Bhattacharya} \& {van den Heuvel}(1991)}]{bv91}
{Bhattacharya} D, {van den Heuvel} EPJ. 1991.
\newblock \textit{\physrep} 203:1--124

\bibitem[{{Birkl} et~al.(2007){Birkl}, {Aloy}, {Janka} \&
  {M{\"u}ller}}]{baj+07}
{Birkl} R, {Aloy} MA, {Janka} HT, {M{\"u}ller} E. 2007.
\newblock \textit{\aap} 463:51--67

\bibitem[{{Blandford} \& {Znajek}(1977)}]{bz77}
{Blandford} RD, {Znajek} RL. 1977.
\newblock \textit{\mnras} 179:433--456

\bibitem[{{Bloom}, {Butler} \& {Perley}(2008)}]{bbp08}
{Bloom} JS, {Butler} NR, {Perley} DA. 2008.
\newblock In \textit{American Institute of Physics Conference Series}, eds.
  M~{Galassi}, D~{Palmer}, E~{Fenimore}, vol. 1000 of \textit{American
  Institute of Physics Conference Series}

\bibitem[{{Bloom} et~al.(1998){Bloom}, {Djorgovski}, {Kulkarni} \&
  {Frail}}]{bdk+98}
{Bloom} JS, {Djorgovski} SG, {Kulkarni} SR, {Frail} DA. 1998.
\newblock \textit{\apjl} 507:L25--L28

\bibitem[{{Bloom}, {Frail} \& {Sari}(2001)}]{bfs01}
{Bloom} JS, {Frail} DA, {Sari} R. 2001.
\newblock \textit{\aj} 121:2879--2888

\bibitem[{{Bloom}, {Kulkarni} \& {Djorgovski}(2002)}]{bkd02}
{Bloom} JS, {Kulkarni} SR, {Djorgovski} SG. 2002.
\newblock \textit{\aj} 123:1111--1148

\bibitem[{{Bloom} et~al.(1999){Bloom}, {Kulkarni}, {Djorgovski},
  {Eichelberger}, {C{\^o}t{\'e}} et~al.}]{bkd+99}
{Bloom} JS, {Kulkarni} SR, {Djorgovski} SG, {Eichelberger} AC, {C{\^o}t{\'e}}
  P, et~al. 1999.
\newblock \textit{\nat} 401:453--456

\bibitem[{{Bloom} et~al.(2002){Bloom}, {Kulkarni}, {Price}, {Reichart},
  {Galama} et~al.}]{bkp+02}
{Bloom} JS, {Kulkarni} SR, {Price} PA, {Reichart} D, {Galama} TJ, et~al. 2002.
\newblock \textit{\apjl} 572:L45--L49

\bibitem[{{Bloom} et~al.(2007){Bloom}, {Perley}, {Chen}, {Butler}, {Prochaska}
  et~al.}]{bpc+07}
{Bloom} JS, {Perley} DA, {Chen} HW, {Butler} N, {Prochaska} JX, et~al. 2007.
\newblock \textit{\apj} 654:878--884

\bibitem[{{Bloom} et~al.(2006){Bloom}, {Prochaska}, {Pooley}, {Blake}, {Foley}
  et~al.}]{bpp+06}
{Bloom} JS, {Prochaska} JX, {Pooley} D, {Blake} CH, {Foley} RJ, et~al. 2006.
\newblock \textit{\apj} 638:354--368

\bibitem[{{Bloom}, {Sigurdsson} \& {Pols}(1999)}]{bsp99}
{Bloom} JS, {Sigurdsson} S, {Pols} OR. 1999.
\newblock \textit{\mnras} 305:763--769

\bibitem[{{Brodie} \& {Strader}(2006)}]{bs06}
{Brodie} JP, {Strader} J. 2006.
\newblock \textit{\araa} 44:193--267

\bibitem[{{Bromberg} et~al.(2013){Bromberg}, {Nakar}, {Piran} \&
  {Sari}}]{bnp+13}
{Bromberg} O, {Nakar} E, {Piran} T, {Sari} R. 2013.
\newblock \textit{\apj} 764:179

\bibitem[{{Bucciantini} et~al.(2012){Bucciantini}, {Metzger}, {Thompson} \&
  {Quataert}}]{bmt+12}
{Bucciantini} N, {Metzger} BD, {Thompson} TA, {Quataert} E. 2012.
\newblock \textit{\mnras} 419:1537--1545

\bibitem[{{Burgay} et~al.(2003){Burgay}, {D'Amico}, {Possenti}, {Manchester},
  {Lyne} et~al.}]{bdp+03}
{Burgay} M, {D'Amico} N, {Possenti} A, {Manchester} RN, {Lyne} AG, et~al. 2003.
\newblock \textit{\nat} 426:531--533

\bibitem[{{Burrows} et~al.(2006){Burrows}, {Grupe}, {Capalbi}, {Panaitescu},
  {Patel} et~al.}]{bgc+06}
{Burrows} DN, {Grupe} D, {Capalbi} M, {Panaitescu} A, {Patel} SK, et~al. 2006.
\newblock \textit{\apj} 653:468--473

\bibitem[{{Burrows} et~al.(2005){Burrows}, {Romano}, {Falcone}, {Kobayashi},
  {Zhang} et~al.}]{brf+05}
{Burrows} DN, {Romano} P, {Falcone} A, {Kobayashi} S, {Zhang} B, et~al. 2005.
\newblock \textit{Science} 309:1833--1835

\bibitem[{{Campana} et~al.(2006){Campana}, {Tagliaferri}, {Lazzati},
  {Chincarini}, {Covino} et~al.}]{ctl+06}
{Campana} S, {Tagliaferri} G, {Lazzati} D, {Chincarini} G, {Covino} S, et~al.
  2006.
\newblock \textit{\aap} 454:113--117

\bibitem[{{Caputi} et~al.(2007){Caputi}, {Lagache}, {Yan}, {Dole}, {Bavouzet}
  et~al.}]{cly+07}
{Caputi} KI, {Lagache} G, {Yan} L, {Dole} H, {Bavouzet} N, et~al. 2007.
\newblock \textit{\apj} 660:97--116

\bibitem[{{Caputi} et~al.(2006){Caputi}, {McLure}, {Dunlop}, {Cirasuolo} \&
  {Schael}}]{cmd+06}
{Caputi} KI, {McLure} RJ, {Dunlop} JS, {Cirasuolo} M, {Schael} AM. 2006.
\newblock \textit{\mnras} 366:609--623

\bibitem[{{Castro-Tirado} et~al.(2005){Castro-Tirado}, {de Ugarte Postigo},
  {Gorosabel}, {Fathkullin}, {Sokolov} et~al.}]{cdg+05}
{Castro-Tirado} AJ, {de Ugarte Postigo} A, {Gorosabel} J, {Fathkullin} T,
  {Sokolov} V, et~al. 2005.
\newblock \textit{\aap} 439:L15--L18

\bibitem[{{Cenko} et~al.(2011){Cenko}, {Frail}, {Harrison}, {Haislip},
  {Reichart} et~al.}]{cfh+11}
{Cenko} SB, {Frail} DA, {Harrison} FA, {Haislip} JB, {Reichart} DE, et~al.
  2011.
\newblock \textit{\apj} 732:29

\bibitem[{{Champion} et~al.(2004){Champion}, {Lorimer}, {McLaughlin}, {Cordes},
  {Arzoumanian} et~al.}]{clm+04}
{Champion} DJ, {Lorimer} DR, {McLaughlin} MA, {Cordes} JM, {Arzoumanian} Z,
  et~al. 2004.
\newblock \textit{\mnras} 350:L61--L65

\bibitem[{{Chandra} \& {Frail}(2012)}]{cf12}
{Chandra} P, {Frail} DA. 2012.
\newblock \textit{\apj} 746:156

\bibitem[{{Chevalier} \& {Li}(2000)}]{cl00}
{Chevalier} RA, {Li} ZY. 2000.
\newblock \textit{\apj} 536:195--212

\bibitem[{{Chincarini} et~al.(2010){Chincarini}, {Mao}, {Margutti},
  {Bernardini}, {Guidorzi} et~al.}]{cmm+10}
{Chincarini} G, {Mao} J, {Margutti} R, {Bernardini} MG, {Guidorzi} C, et~al.
  2010.
\newblock \textit{\mnras} 406:2113--2148

\bibitem[{{Chincarini} et~al.(2007){Chincarini}, {Moretti}, {Romano},
  {Falcone}, {Morris} et~al.}]{cmr+07}
{Chincarini} G, {Moretti} A, {Romano} P, {Falcone} AD, {Morris} D, et~al. 2007.
\newblock \textit{\apj} 671:1903--1920

\bibitem[{{Christensen}, {Hjorth} \& {Gorosabel}(2004)}]{chg04}
{Christensen} L, {Hjorth} J, {Gorosabel} J. 2004.
\newblock \textit{\aap} 425:913--926

\bibitem[{{Church} et~al.(2011){Church}, {Levan}, {Davies} \&
  {Tanvir}}]{cld+11}
{Church} RP, {Levan} AJ, {Davies} MB, {Tanvir} N. 2011.
\newblock \textit{\mnras} 413:2004--2014

\bibitem[{{Connaughton}(2002)}]{con02}
{Connaughton} V. 2002.
\newblock \textit{\apj} 567:1028--1036

\bibitem[{{Costa} et~al.(1997){Costa}, {Frontera}, {Heise}, {Feroci}, {in't
  Zand} et~al.}]{cfh+97}
{Costa} E, {Frontera} F, {Heise} J, {Feroci} M, {in't Zand} J, et~al. 1997.
\newblock \textit{\nat} 387:783--785

\bibitem[{{Coward} et~al.(2012){Coward}, {Howell}, {Piran}, {Stratta},
  {Branchesi} et~al.}]{chp+12}
{Coward} DM, {Howell} EJ, {Piran} T, {Stratta} G, {Branchesi} M, et~al. 2012.
\newblock \textit{\mnras} 425:2668--2673

\bibitem[{{Cucchiara} et~al.(2011){Cucchiara}, {Levan}, {Fox}, {Tanvir},
  {Ukwatta} et~al.}]{clf+11}
{Cucchiara} A, {Levan} AJ, {Fox} DB, {Tanvir} NR, {Ukwatta} TN, et~al. 2011.
\newblock \textit{\apj} 736:7

\bibitem[{{Cucchiara} et~al.(2013){Cucchiara}, {Prochaska}, {Perley}, {Cenko},
  {Werk} et~al.}]{cpp+13}
{Cucchiara} A, {Prochaska} JX, {Perley} DA, {Cenko} SB, {Werk} J, et~al. 2013.
\newblock \textit{ArXiv}:1306.2028

\bibitem[{{Dahle} et~al.(2013){Dahle}, {Sarazin}, {Lopez}, {Kouveliotou},
  {Patel} et~al.}]{dsl+13}
{Dahle} H, {Sarazin} CL, {Lopez} LA, {Kouveliotou} C, {Patel} SK, et~al. 2013.
\newblock \textit{\apj} 772:23

\bibitem[{{D'Avanzo} et~al.(2009){D'Avanzo}, {Malesani}, {Covino},
  {Piranomonte}, {Grazian} et~al.}]{dmc+09}
{D'Avanzo} P, {Malesani} D, {Covino} S, {Piranomonte} S, {Grazian} A, et~al.
  2009.
\newblock \textit{\aap} 498:711--721

\bibitem[{{de Ugarte Postigo} et~al.(2006){de Ugarte Postigo}, {Castro-Tirado},
  {Guziy}, {Gorosabel}, {J{\'o}hannesson} et~al.}]{dcg+06}
{de Ugarte Postigo} A, {Castro-Tirado} AJ, {Guziy} S, {Gorosabel} J,
  {J{\'o}hannesson} G, et~al. 2006.
\newblock \textit{\apjl} 648:L83--L87

\bibitem[{{de Ugarte Postigo} et~al.(2013){de Ugarte Postigo}, {Thoene},
  {Rowlinson}, {Garcia-Benito}, {Levan} et~al.}]{dtr+13}
{de Ugarte Postigo} A, {Thoene} CC, {Rowlinson} A, {Garcia-Benito} R, {Levan}
  AJ, et~al. 2013.
\newblock \textit{ArXiv}:1308.2984

\bibitem[{{Della Valle} et~al.(2006){Della Valle}, {Chincarini}, {Panagia},
  {Tagliaferri}, {Malesani} et~al.}]{dcp+06}
{Della Valle} M, {Chincarini} G, {Panagia} N, {Tagliaferri} G, {Malesani} D,
  et~al. 2006.
\newblock \textit{\nat} 444:1050--1052

\bibitem[{{Demorest} et~al.(2010){Demorest}, {Pennucci}, {Ransom}, {Roberts} \&
  {Hessels}}]{dpr+10}
{Demorest} PB, {Pennucci} T, {Ransom} SM, {Roberts} MSE, {Hessels} JWT. 2010.
\newblock \textit{\nat} 467:1081--1083

\bibitem[{{Dessart} et~al.(2009){Dessart}, {Ott}, {Burrows}, {Rosswog} \&
  {Livne}}]{dob+09}
{Dessart} L, {Ott} CD, {Burrows} A, {Rosswog} S, {Livne} E. 2009.
\newblock \textit{\apj} 690:1681--1705

\bibitem[{{Dezalay} et~al.(1992){Dezalay}, {Barat}, {Talon}, {Syunyaev},
  {Terekhov} \& {Kuznetsov}}]{dbt+92}
{Dezalay} JP, {Barat} C, {Talon} R, {Syunyaev} R, {Terekhov} O, {Kuznetsov} A.
  1992.
\newblock In \textit{American Institute of Physics Conference Series}, eds.
  WS~{Paciesas}, GJ~{Fishman}, vol. 265 of \textit{American Institute of
  Physics Conference Series}

\bibitem[{{Dhawan} et~al.(2007){Dhawan}, {Mirabel}, {Rib{\'o}} \&
  {Rodrigues}}]{dmr+07}
{Dhawan} V, {Mirabel} IF, {Rib{\'o}} M, {Rodrigues} I. 2007.
\newblock \textit{\apj} 668:430--434

\bibitem[{{Djorgovski} et~al.(1998){Djorgovski}, {Kulkarni}, {Bloom},
  {Goodrich}, {Frail} et~al.}]{dkb+98}
{Djorgovski} SG, {Kulkarni} SR, {Bloom} JS, {Goodrich} R, {Frail} DA, et~al.
  1998.
\newblock \textit{\apjl} 508:L17--L20

\bibitem[{{Drout} et~al.(2011){Drout}, {Soderberg}, {Gal-Yam}, {Cenko}, {Fox}
  et~al.}]{dsg+11}
{Drout} MR, {Soderberg} AM, {Gal-Yam} A, {Cenko} SB, {Fox} DB, et~al. 2011.
\newblock \textit{\apj} 741:97

\bibitem[{{Duncan} \& {Thompson}(1992)}]{dt92}
{Duncan} RC, {Thompson} C. 1992.
\newblock \textit{\apjl} 392:L9--L13

\bibitem[{{Eichler} et~al.(1989){Eichler}, {Livio}, {Piran} \&
  {Schramm}}]{elp+89}
{Eichler} D, {Livio} M, {Piran} T, {Schramm} DN. 1989.
\newblock \textit{\nat} 340:126--128

\bibitem[{{Eke} et~al.(2005){Eke}, {Baugh}, {Cole}, {Frenk}, {King} \&
  {Peacock}}]{ebc+05}
{Eke} VR, {Baugh} CM, {Cole} S, {Frenk} CS, {King} HM, {Peacock} JA. 2005.
\newblock \textit{\mnras} 362:1233--1246

\bibitem[{{Etienne} et~al.(2008){Etienne}, {Faber}, {Liu}, {Shapiro},
  {Taniguchi} \& {Baumgarte}}]{efl+08}
{Etienne} ZB, {Faber} JA, {Liu} YT, {Shapiro} SL, {Taniguchi} K, {Baumgarte}
  TW. 2008.
\newblock \textit{\prd} 77:084002

\bibitem[{{Fairhurst}(2009)}]{fai09}
{Fairhurst} S. 2009.
\newblock \textit{New Journal of Physics} 11:123006

\bibitem[{{Falcone} et~al.(2007){Falcone}, {Morris}, {Racusin}, {Chincarini},
  {Moretti} et~al.}]{fmr+07}
{Falcone} AD, {Morris} D, {Racusin} J, {Chincarini} G, {Moretti} A, et~al.
  2007.
\newblock \textit{\apj} 671:1921--1938

\bibitem[{{Fenimore}, {Epstein} \& {Ho}(1993)}]{feh93}
{Fenimore} EE, {Epstein} RI, {Ho} C. 1993.
\newblock \textit{\aaps} 97:59--62

\bibitem[{{Filippenko}(1997)}]{fil97}
{Filippenko} AV. 1997.
\newblock \textit{\araa} 35:309--355

\bibitem[{{Fishman} \& {Meegan}(1995)}]{fm95}
{Fishman} GJ, {Meegan} CA. 1995.
\newblock \textit{\araa} 33:415--458

\bibitem[{{Fong} et~al.(2013){Fong}, {Berger}, {Chornock}, {Margutti}, {Levan}
  et~al.}]{fbc+13}
{Fong} W, {Berger} E, {Chornock} R, {Margutti} R, {Levan} AJ, et~al. 2013.
\newblock \textit{\apj} 769:56

\bibitem[{{Fong} et~al.(2011){Fong}, {Berger}, {Chornock}, {Tanvir}, {Levan}
  et~al.}]{fbc+11}
{Fong} W, {Berger} E, {Chornock} R, {Tanvir} NR, {Levan} AJ, et~al. 2011.
\newblock \textit{\apj} 730:26

\bibitem[{{Fong}, {Berger} \& {Fox}(2010)}]{fbf10}
{Fong} W, {Berger} E, {Fox} DB. 2010.
\newblock \textit{\apj} 708:9--25

\bibitem[{{Fong} et~al.(2012){Fong}, {Berger}, {Margutti}, {Zauderer}, {Troja}
  et~al.}]{fbm+12}
{Fong} W, {Berger} E, {Margutti} R, {Zauderer} BA, {Troja} E, et~al. 2012.
\newblock \textit{\apj} 756:189

\bibitem[{{Fong} et~al.(2013){Fong}, {Berger}, {Metzger}, {Margutti}, {Chornock}
  et~al.}]{fbm+13}
{Fong} W, {Berger} E, {Metzger} BD, {Margutti} R, {Chornock} R, et~al. 2013.
\newblock \textit{arXiv}: 1309.7479

\bibitem[{{Fong} \& {Berger}(2013)}]{fb13}
{Fong} Wf, {Berger} E. 2013.
\newblock \textit{\apj} 776:18

\bibitem[{{Fox} et~al.(2005){Fox}, {Frail}, {Price}, {Kulkarni}, {Berger}
  et~al.}]{ffp+05}
{Fox} DB, {Frail} DA, {Price} PA, {Kulkarni} SR, {Berger} E, et~al. 2005.
\newblock \textit{\nat} 437:845--850

\bibitem[{{Frail} et~al.(1997){Frail}, {Kulkarni}, {Nicastro}, {Feroci} \&
  {Taylor}}]{fkn+97}
{Frail} DA, {Kulkarni} SR, {Nicastro} L, {Feroci} M, {Taylor} GB. 1997.
\newblock \textit{\nat} 389:261--263

\bibitem[{{Frail} et~al.(2001){Frail}, {Kulkarni}, {Sari}, {Djorgovski},
  {Bloom} et~al.}]{fks+01}
{Frail} DA, {Kulkarni} SR, {Sari} R, {Djorgovski} SG, {Bloom} JS, et~al. 2001.
\newblock \textit{\apjl} 562:L55--L58

\bibitem[{{Frederiks} et~al.(2004){Frederiks}, {Aptekar}, {Golenetskii},
  {Il'Inskii}, {Mazets} et~al.}]{fag+04}
{Frederiks} DD, {Aptekar} RL, {Golenetskii} SV, {Il'Inskii} VN, {Mazets} EP,
  et~al. 2004.
\newblock In \textit{Gamma-Ray Bursts in the Afterglow Era}, eds. M~{Feroci},
  F~{Frontera}, N~{Masetti}, L~{Piro}, vol. 312 of \textit{Astronomical Society
  of the Pacific Conference Series}

\bibitem[{{Fruchter} et~al.(2006){Fruchter}, {Levan}, {Strolger}, {Vreeswijk},
  {Thorsett} et~al.}]{fls+06}
{Fruchter} AS, {Levan} AJ, {Strolger} L, {Vreeswijk} PM, {Thorsett} SE, et~al.
  2006.
\newblock \textit{\nat} 441:463--468

\bibitem[{{Fryer}, {Burrows} \& {Benz}(1998)}]{fbb98}
{Fryer} C, {Burrows} A, {Benz} W. 1998.
\newblock \textit{\apj} 496:333

\bibitem[{{Fryer} \& {Kalogera}(1997)}]{fk97}
{Fryer} C, {Kalogera} V. 1997.
\newblock \textit{\apj} 489:244

\bibitem[{{Fryer}, {Woosley} \& {Hartmann}(1999)}]{fwh99}
{Fryer} CL, {Woosley} SE, {Hartmann} DH. 1999.
\newblock \textit{\apj} 526:152--177

\bibitem[{{Fukugita}, {Hogan} \& {Peebles}(1998)}]{fhp98}
{Fukugita} M, {Hogan} CJ, {Peebles} PJE. 1998.
\newblock \textit{\apj} 503:518--530

\bibitem[{{Fynbo} et~al.(2006){Fynbo}, {Watson}, {Th{\"o}ne}, {Sollerman},
  {Bloom} et~al.}]{fwt+06}
{Fynbo} JPU, {Watson} D, {Th{\"o}ne} CC, {Sollerman} J, {Bloom} JS, et~al.
  2006.
\newblock \textit{\nat} 444:1047--1049

\bibitem[{{Gal-Yam} et~al.(2006){Gal-Yam}, {Fox}, {Price}, {Ofek}, {Davis}
  et~al.}]{gfp+06}
{Gal-Yam} A, {Fox} DB, {Price} PA, {Ofek} EO, {Davis} MR, et~al. 2006.
\newblock \textit{\nat} 444:1053--1055

\bibitem[{{Gal-Yam} et~al.(2008){Gal-Yam}, {Nakar}, {Ofek}, {Cenko}, {Kulkarni}
  et~al.}]{gno+08}
{Gal-Yam} A, {Nakar} E, {Ofek} EO, {Cenko} SB, {Kulkarni} SR, et~al. 2008.
\newblock \textit{\apj} 686:408--416

\bibitem[{{Galama} et~al.(1998){Galama}, {Vreeswijk}, {van Paradijs},
  {Kouveliotou}, {Augusteijn} et~al.}]{gvv+98}
{Galama} TJ, {Vreeswijk} PM, {van Paradijs} J, {Kouveliotou} C, {Augusteijn} T,
  et~al. 1998.
\newblock \textit{\nat} 395:670--672

\bibitem[{{Galbany} et~al.(2012){Galbany}, {Miquel}, {{\"O}stman}, {Brown},
  {Cinabro} et~al.}]{gmo+12}
{Galbany} L, {Miquel} R, {{\"O}stman} L, {Brown} PJ, {Cinabro} D, et~al. 2012.
\newblock \textit{\apj} 755:125

\bibitem[{{Gao} et~al.(2013){Gao}, {Ding}, {Wu}, {Zhang} \& {Dai}}]{gdw+13}
{Gao} H, {Ding} X, {Wu} XF, {Zhang} B, {Dai} ZG. 2013.
\newblock \textit{\apj} 771:86

\bibitem[{{Gehrels} et~al.(2008){Gehrels}, {Barthelmy}, {Burrows}, {Cannizzo},
  {Chincarini} et~al.}]{gbb+08}
{Gehrels} N, {Barthelmy} SD, {Burrows} DN, {Cannizzo} JK, {Chincarini} G,
  et~al. 2008.
\newblock \textit{\apj} 689:1161--1172

\bibitem[{{Gehrels} et~al.(2004){Gehrels}, {Chincarini}, {Giommi}, {Mason},
  {Nousek} et~al.}]{gcg+04}
{Gehrels} N, {Chincarini} G, {Giommi} P, {Mason} KO, {Nousek} JA, et~al. 2004.
\newblock \textit{\apj} 611:1005--1020

\bibitem[{{Gehrels} et~al.(2006){Gehrels}, {Norris}, {Barthelmy}, {Granot},
  {Kaneko} et~al.}]{gnb+06}
{Gehrels} N, {Norris} JP, {Barthelmy} SD, {Granot} J, {Kaneko} Y, et~al. 2006.
\newblock \textit{\nat} 444:1044--1046

\bibitem[{{Gehrels}, {Ramirez-Ruiz} \& {Fox}(2009)}]{grf09}
{Gehrels} N, {Ramirez-Ruiz} E, {Fox} DB. 2009.
\newblock \textit{\araa} 47:567--617

\bibitem[{{Gehrels} et~al.(2005){Gehrels}, {Sarazin}, {O'Brien}, {Zhang},
  {Barbier} et~al.}]{gso+05}
{Gehrels} N, {Sarazin} CL, {O'Brien} PT, {Zhang} B, {Barbier} L, et~al. 2005.
\newblock \textit{\nat} 437:851--854

\bibitem[{{Ghirlanda}, {Ghisellini} \& {Celotti}(2004)}]{ggc04}
{Ghirlanda} G, {Ghisellini} G, {Celotti} A. 2004.
\newblock \textit{\aap} 422:L55--L58

\bibitem[{{Ghirlanda}, {Ghisellini} \& {Nava}(2010)}]{ggn10}
{Ghirlanda} G, {Ghisellini} G, {Nava} L. 2010.
\newblock \textit{\aap} 510:L7

\bibitem[{{Ghirlanda}, {Ghisellini} \& {Nava}(2011)}]{ggn11}
{Ghirlanda} G, {Ghisellini} G, {Nava} L. 2011.
\newblock \textit{\mnras} 418:L109--L113

\bibitem[{{Ghirlanda} et~al.(2006){Ghirlanda}, {Magliocchetti}, {Ghisellini} \&
  {Guzzo}}]{gmg+06}
{Ghirlanda} G, {Magliocchetti} M, {Ghisellini} G, {Guzzo} L. 2006.
\newblock \textit{\mnras} 368:L20--L24

\bibitem[{{Ghirlanda} et~al.(2009){Ghirlanda}, {Nava}, {Ghisellini}, {Celotti}
  \& {Firmani}}]{gng+09}
{Ghirlanda} G, {Nava} L, {Ghisellini} G, {Celotti} A, {Firmani} C. 2009.
\newblock \textit{\aap} 496:585--595

\bibitem[{{Giannios}(2006)}]{gia06}
{Giannios} D. 2006.
\newblock \textit{\aap} 455:L5--L8

\bibitem[{{Goodman}(1986)}]{goo86}
{Goodman} J. 1986.
\newblock \textit{\apjl} 308:L47--L50

\bibitem[{{Goriely}, {Bauswein} \& {Janka}(2011)}]{gbj11}
{Goriely} S, {Bauswein} A, {Janka} HT. 2011.
\newblock \textit{\apjl} 738:L32

\bibitem[{{Granot} \& {Sari}(2002)}]{gs02}
{Granot} J, {Sari} R. 2002.
\newblock \textit{\apj} 568:820--829

\bibitem[{{Grindlay}, {Portegies Zwart} \& {McMillan}(2006)}]{gpm06}
{Grindlay} J, {Portegies Zwart} S, {McMillan} S. 2006.
\newblock \textit{Nature Physics} 2:116--119

\bibitem[{{Grossman} et~al.(2013){Grossman}, {Korobkin}, {Rosswog} \&
  {Piran}}]{gkr+13}
{Grossman} D, {Korobkin} O, {Rosswog} S, {Piran} T. 2013.
\newblock \textit{ArXiv}:1307.2943

\bibitem[{{Grupe} et~al.(2006){Grupe}, {Burrows}, {Patel}, {Kouveliotou},
  {Zhang} et~al.}]{gbp+06}
{Grupe} D, {Burrows} DN, {Patel} SK, {Kouveliotou} C, {Zhang} B, et~al. 2006.
\newblock \textit{\apj} 653:462--467

\bibitem[{{Guetta} \& {Piran}(2005)}]{gp05}
{Guetta} D, {Piran} T. 2005.
\newblock \textit{\aap} 435:421--426

\bibitem[{{Hakkila} \& {Preece}(2011)}]{hp11}
{Hakkila} J, {Preece} RD. 2011.
\newblock \textit{\apj} 740:104

\bibitem[{{Hansen} \& {Lyutikov}(2001)}]{hl01}
{Hansen} BMS, {Lyutikov} M. 2001.
\newblock \textit{\mnras} 322:695--701

\bibitem[{{Harris}(1991)}]{har91}
{Harris} WE. 1991.
\newblock \textit{\araa} 29:543--579

\bibitem[{{Harrison} et~al.(1999){Harrison}, {Bloom}, {Frail}, {Sari},
  {Kulkarni} et~al.}]{hbf+99}
{Harrison} FA, {Bloom} JS, {Frail} DA, {Sari} R, {Kulkarni} SR, et~al. 1999.
\newblock \textit{\apjl} 523:L121--L124

\bibitem[{{Harry} \& {LIGO Scientific Collaboration}(2010)}]{har10}
{Harry} GM, {LIGO Scientific Collaboration}. 2010.
\newblock \textit{Classical and Quantum Gravity} 27:084006

\bibitem[{{Hjorth} \& {Bloom}(2012)}]{hb12}
{Hjorth} J, {Bloom} JS. 2012.
\newblock \textit{{The Gamma-Ray Burst - Supernova Connection}}.
\newblock  169--190

\bibitem[{{Hjorth} et~al.(2005{\natexlab{a}}){Hjorth}, {Sollerman},
  {Gorosabel}, {Granot}, {Klose} et~al.}]{hsg+05}
{Hjorth} J, {Sollerman} J, {Gorosabel} J, {Granot} J, {Klose} S, et~al.
  2005{\natexlab{a}}.
\newblock \textit{\apjl} 630:L117--L120

\bibitem[{{Hjorth} et~al.(2003){Hjorth}, {Sollerman}, {M{\o}ller}, {Fynbo},
  {Woosley} et~al.}]{hsm+03}
{Hjorth} J, {Sollerman} J, {M{\o}ller} P, {Fynbo} JPU, {Woosley} SE, et~al.
  2003.
\newblock \textit{\nat} 423:847--850

\bibitem[{{Hjorth} et~al.(2005{\natexlab{b}}){Hjorth}, {Watson}, {Fynbo},
  {Price}, {Jensen} et~al.}]{hwf+05}
{Hjorth} J, {Watson} D, {Fynbo} JPU, {Price} PA, {Jensen} BL, et~al.
  2005{\natexlab{b}}.
\newblock \textit{\nat} 437:859--861

\bibitem[{{Hogg} et~al.(1997){Hogg}, {Pahre}, {McCarthy}, {Cohen}, {Blandford}
  et~al.}]{hpm+97}
{Hogg} DW, {Pahre} MA, {McCarthy} JK, {Cohen} JG, {Blandford} R, et~al. 1997.
\newblock \textit{\mnras} 288:404--410

\bibitem[{{Hotokezaka} et~al.(2013){Hotokezaka}, {Kiuchi}, {Kyutoku}, {Okawa},
  {Sekiguchi} et~al.}]{hkk+13}
{Hotokezaka} K, {Kiuchi} K, {Kyutoku} K, {Okawa} H, {Sekiguchi} Yi, et~al.
  2013.
\newblock \textit{\prd} 87:024001

\bibitem[{{Hulse} \& {Taylor}(1975)}]{ht75}
{Hulse} RA, {Taylor} JH. 1975.
\newblock \textit{\apjl} 195:L51--L53

\bibitem[{{Hurley} et~al.(2002){Hurley}, {Berger}, {Castro-Tirado}, {Castro
  Cer{\'o}n}, {Cline} et~al.}]{hbc+02}
{Hurley} K, {Berger} E, {Castro-Tirado} A, {Castro Cer{\'o}n} JM, {Cline} T,
  et~al. 2002.
\newblock \textit{\apj} 567:447--453

\bibitem[{{Hurley} et~al.(2010){Hurley}, {Rowlinson}, {Bellm}, {Perley},
  {Mitrofanov} et~al.}]{hrb+10}
{Hurley} K, {Rowlinson} A, {Bellm} E, {Perley} D, {Mitrofanov} IG, et~al. 2010.
\newblock \textit{\mnras} 403:342--352

\bibitem[{{Ilbert} et~al.(2010){Ilbert}, {Salvato}, {Le Floc'h}, {Aussel},
  {Capak} et~al.}]{isl+10}
{Ilbert} O, {Salvato} M, {Le Floc'h} E, {Aussel} H, {Capak} P, et~al. 2010.
\newblock \textit{\apj} 709:644--663

\bibitem[{{Jakobsson} et~al.(2006){Jakobsson}, {Levan}, {Fynbo}, {Priddey},
  {Hjorth} et~al.}]{jlf+06}
{Jakobsson} P, {Levan} A, {Fynbo} JPU, {Priddey} R, {Hjorth} J, et~al. 2006.
\newblock \textit{\aap} 447:897--903

\bibitem[{{Janka} et~al.(1999){Janka}, {Eberl}, {Ruffert} \& {Fryer}}]{jer+99}
{Janka} HT, {Eberl} T, {Ruffert} M, {Fryer} CL. 1999.
\newblock \textit{\apjl} 527:L39--L42

\bibitem[{{Johnston} et~al.(2008){Johnston}, {Taylor}, {Bailes}, {Bartel},
  {Baugh} et~al.}]{jtb+08}
{Johnston} S, {Taylor} R, {Bailes} M, {Bartel} N, {Baugh} C, et~al. 2008.
\newblock \textit{Experimental Astronomy} 22:151--273

\bibitem[{{Kalogera} et~al.(2004){Kalogera}, {Kim}, {Lorimer}, {Burgay},
  {D'Amico} et~al.}]{kkl+04}
{Kalogera} V, {Kim} C, {Lorimer} DR, {Burgay} M, {D'Amico} N, et~al. 2004.
\newblock \textit{\apjl} 601:L179--L182

\bibitem[{{Kaneko} et~al.(2006){Kaneko}, {Preece}, {Briggs}, {Paciesas},
  {Meegan} \& {Band}}]{kpb+06}
{Kaneko} Y, {Preece} RD, {Briggs} MS, {Paciesas} WS, {Meegan} CA, {Band} DL.
  2006.
\newblock \textit{\apjs} 166:298--340

\bibitem[{{Kann} et~al.(2011){Kann}, {Klose}, {Zhang}, {Covino}, {Butler}
  et~al.}]{kkz+11}
{Kann} DA, {Klose} S, {Zhang} B, {Covino} S, {Butler} NR, et~al. 2011.
\newblock \textit{\apj} 734:96

\bibitem[{{Kasen}, {Badnell} \& {Barnes}(2013)}]{kbb13}
{Kasen} D, {Badnell} NR, {Barnes} J. 2013.
\newblock \textit{\apj} 774:25

\bibitem[{{Kehoe} et~al.(2001){Kehoe}, {Akerlof}, {Balsano}, {Barthelmy},
  {Bloch} et~al.}]{kab+01}
{Kehoe} R, {Akerlof} C, {Balsano} R, {Barthelmy} S, {Bloch} J, et~al. 2001.
\newblock \textit{\apjl} 554:L159--L162

\bibitem[{{Kelley}, {Mandel} \& {Ramirez-Ruiz}(2013)}]{kmr13}
{Kelley} LZ, {Mandel} I, {Ramirez-Ruiz} E. 2013.
\newblock \textit{\prd} 87:123004

\bibitem[{{Kelly} \& {Kirshner}(2012)}]{kk12}
{Kelly} PL, {Kirshner} RP. 2012.
\newblock \textit{\apj} 759:107

\bibitem[{{Kelly}, {Kirshner} \& {Pahre}(2008)}]{kkp08}
{Kelly} PL, {Kirshner} RP, {Pahre} M. 2008.
\newblock \textit{\apj} 687:1201--1207

\bibitem[{{Klebesadel}, {Strong} \& {Olson}(1973)}]{kso73}
{Klebesadel} RW, {Strong} IB, {Olson} RA. 1973.
\newblock \textit{\apjl} 182:L85

\bibitem[{{Kobulnicky} \& {Kewley}(2004)}]{kk04}
{Kobulnicky} HA, {Kewley} LJ. 2004.
\newblock \textit{\apj} 617:240--261

\bibitem[{{Kocevski} et~al.(2010){Kocevski}, {Th{\"o}ne}, {Ramirez-Ruiz},
  {Bloom}, {Granot} et~al.}]{ktr+10}
{Kocevski} D, {Th{\"o}ne} CC, {Ramirez-Ruiz} E, {Bloom} JS, {Granot} J, et~al.
  2010.
\newblock \textit{\mnras} 404:963--974

\bibitem[{{Kochanek} \& {Piran}(1993)}]{kp93}
{Kochanek} CS, {Piran} T. 1993.
\newblock \textit{\apjl} 417:L17

\bibitem[{{Kouveliotou} et~al.(1998){Kouveliotou}, {Dieters}, {Strohmayer},
  {van Paradijs}, {Fishman} et~al.}]{kds+98}
{Kouveliotou} C, {Dieters} S, {Strohmayer} T, {van Paradijs} J, {Fishman} GJ,
  et~al. 1998.
\newblock \textit{\nat} 393:235--237

\bibitem[{{Kouveliotou} et~al.(1993){Kouveliotou}, {Meegan}, {Fishman}, {Bhat},
  {Briggs} et~al.}]{kmf+93}
{Kouveliotou} C, {Meegan} CA, {Fishman} GJ, {Bhat} NP, {Briggs} MS, et~al.
  1993.
\newblock \textit{\apjl} 413:L101--L104

\bibitem[{{Kouveliotou} et~al.(1987){Kouveliotou}, {Norris}, {Cline}, {Dennis},
  {Desai} et~al.}]{knc+87}
{Kouveliotou} C, {Norris} JP, {Cline} TL, {Dennis} BR, {Desai} UD, et~al. 1987.
\newblock \textit{\apjl} 322:L21--L25

\bibitem[{{Krolik} \& {Pier}(1991)}]{kp91}
{Krolik} JH, {Pier} EA. 1991.
\newblock \textit{\apj} 373:277--284

\bibitem[{{Kulkarni} et~al.(1999){Kulkarni}, {Djorgovski}, {Odewahn}, {Bloom},
  {Gal} et~al.}]{kdo+99}
{Kulkarni} SR, {Djorgovski} SG, {Odewahn} SC, {Bloom} JS, {Gal} RR, et~al.
  1999.
\newblock \textit{\nat} 398:389--394

\bibitem[{{Kumar} \& {Barniol Duran}(2010)}]{kb10}
{Kumar} P, {Barniol Duran} R. 2010.
\newblock \textit{\mnras} 409:226--236

\bibitem[{{Kyutoku}, {Ioka} \& {Shibata}(2012)}]{kis12}
{Kyutoku} K, {Ioka} K, {Shibata} M. 2012.
\newblock \textit{ArXiv}:1209.5747

\bibitem[{{Kyutoku}, {Ioka} \& {Shibata}(2013)}]{kis13}
{Kyutoku} K, {Ioka} K, {Shibata} M. 2013.
\newblock \textit{ArXiv}:1305.6309

\bibitem[{{Lamb}(1995)}]{lam95}
{Lamb} DQ. 1995.
\newblock \textit{\pasp} 107:1152

\bibitem[{{Laros} et~al.(1987){Laros}, {Fenimore}, {Klebesadel}, {Atteia},
  {Boer} et~al.}]{lfk+87}
{Laros} JG, {Fenimore} EE, {Klebesadel} RW, {Atteia} JL, {Boer} M, et~al. 1987.
\newblock \textit{\apjl} 320:L111--L115

\bibitem[{{Lattimer} et~al.(1977){Lattimer}, {Mackie}, {Ravenhall} \&
  {Schramm}}]{lmr+77}
{Lattimer} JM, {Mackie} F, {Ravenhall} DG, {Schramm} DN. 1977.
\newblock \textit{\apj} 213:225--233

\bibitem[{{Lattimer} \& {Schramm}(1974)}]{ls74}
{Lattimer} JM, {Schramm} DN. 1974.
\newblock \textit{\apjl} 192:L145--L147

\bibitem[{{Lattimer} \& {Schramm}(1976)}]{ls76}
{Lattimer} JM, {Schramm} DN. 1976.
\newblock \textit{\apj} 210:549--567

\bibitem[{{Lazzati}(2005)}]{laz05}
{Lazzati} D. 2005.
\newblock \textit{\mnras} 357:722--731

\bibitem[{{Lazzati} \& {Begelman}(2005)}]{lb05}
{Lazzati} D, {Begelman} MC. 2005.
\newblock \textit{\apj} 629:903--907

\bibitem[{{Lazzati}, {Morsony} \& {Begelman}(2010)}]{lmb10}
{Lazzati} D, {Morsony} BJ, {Begelman} MC. 2010.
\newblock \textit{\apj} 717:239--244

\bibitem[{{Lazzati}, {Ramirez-Ruiz} \& {Ghisellini}(2001)}]{lrg01}
{Lazzati} D, {Ramirez-Ruiz} E, {Ghisellini} G. 2001.
\newblock \textit{\aap} 379:L39--L43

\bibitem[{{Le Floc'h} et~al.(2005){Le Floc'h}, {Papovich}, {Dole}, {Bell},
  {Lagache} et~al.}]{lpd+05}
{Le Floc'h} E, {Papovich} C, {Dole} H, {Bell} EF, {Lagache} G, et~al. 2005.
\newblock \textit{\apj} 632:169--190

\bibitem[{{Lee}, {Wijers} \& {Brown}(2000)}]{lwb00}
{Lee} HK, {Wijers} RAMJ, {Brown} GE. 2000.
\newblock \textit{\physrep} 325:83--114

\bibitem[{{Lee} \& {Ramirez-Ruiz}(2007)}]{lr07}
{Lee} WH, {Ramirez-Ruiz} E. 2007.
\newblock \textit{New Journal of Physics} 9:17

\bibitem[{{Lee}, {Ramirez-Ruiz} \& {L{\'o}pez-C{\'a}mara}(2009)}]{lrl09}
{Lee} WH, {Ramirez-Ruiz} E, {L{\'o}pez-C{\'a}mara} D. 2009.
\newblock \textit{\apjl} 699:L93--L96

\bibitem[{{Lee}, {Ramirez-Ruiz} \& {van de Ven}(2010)}]{lrv10}
{Lee} WH, {Ramirez-Ruiz} E, {van de Ven} G. 2010.
\newblock \textit{\apj} 720:953--975

\bibitem[{{Lehner} et~al.(2012){Lehner}, {Palenzuela}, {Liebling}, {Thompson}
  \& {Hanna}}]{lpl+12}
{Lehner} L, {Palenzuela} C, {Liebling} SL, {Thompson} C, {Hanna} C. 2012.
\newblock \textit{\prd} 86:104035

\bibitem[{{Leibler} \& {Berger}(2010)}]{lb10}
{Leibler} CN, {Berger} E. 2010.
\newblock \textit{\apj} 725:1202--1214

\bibitem[{{Levan} et~al.(2006{\natexlab{a}}){Levan}, {Tanvir}, {Fruchter},
  {Rol}, {Fynbo} et~al.}]{ltf+06}
{Levan} AJ, {Tanvir} NR, {Fruchter} AS, {Rol} E, {Fynbo} JPU, et~al.
  2006{\natexlab{a}}.
\newblock \textit{\apjl} 648:L9--L12

\bibitem[{{Levan} et~al.(2006{\natexlab{b}}){Levan}, {Wynn}, {Chapman},
  {Davies}, {King} et~al.}]{lwc+06}
{Levan} AJ, {Wynn} GA, {Chapman} R, {Davies} MB, {King} AR, et~al.
  2006{\natexlab{b}}.
\newblock \textit{\mnras} 368:L1--L5

\bibitem[{{Levesque} et~al.(2010{\natexlab{a}}){Levesque}, {Bloom}, {Butler},
  {Perley}, {Cenko} et~al.}]{lbb+10}
{Levesque} EM, {Bloom} JS, {Butler} NR, {Perley} DA, {Cenko} SB, et~al.
  2010{\natexlab{a}}.
\newblock \textit{\mnras} 401:963--972

\bibitem[{{Levesque} et~al.(2010{\natexlab{b}}){Levesque}, {Kewley}, {Berger}
  \& {Zahid}}]{lkb+10}
{Levesque} EM, {Kewley} LJ, {Berger} E, {Zahid} HJ. 2010{\natexlab{b}}.
\newblock \textit{\aj} 140:1557--1566

\bibitem[{{Li} \& {Paczy{\'n}ski}(1998)}]{lp98}
{Li} L, {Paczy{\'n}ski} B. 1998.
\newblock \textit{\apjl} 507:L59--L62

\bibitem[{{Li} et~al.(2011){Li}, {Chornock}, {Leaman}, {Filippenko},
  {Poznanski} et~al.}]{lcl+11}
{Li} W, {Chornock} R, {Leaman} J, {Filippenko} AV, {Poznanski} D, et~al. 2011.
\newblock \textit{\mnras} 412:1473--1507

\bibitem[{{LIGO Scientific Collaboration} et~al.(2012){LIGO Scientific
  Collaboration}, {Virgo Collaboration}, {Abadie}, {Abbott}, {Abbott}
  et~al.}]{aaa+12a}
{LIGO Scientific Collaboration}, {Virgo Collaboration}, {Abadie} J, {Abbott}
  BP, {Abbott} R, et~al. 2012.
\newblock \textit{\aap} 539:A124

\bibitem[{{L{\"u}} et~al.(2010){L{\"u}}, {Liang}, {Zhang} \& {Zhang}}]{llz+10}
{L{\"u}} HJ, {Liang} EW, {Zhang} BB, {Zhang} B. 2010.
\newblock \textit{\apj} 725:1965--1970

\bibitem[{{MacFadyen}, {Ramirez-Ruiz} \& {Zhang}(2005)}]{mrz05}
{MacFadyen} AI, {Ramirez-Ruiz} E, {Zhang} W. 2005.
\newblock \textit{astro-ph}/0510192

\bibitem[{{MacFadyen} \& {Woosley}(1999)}]{mw99}
{MacFadyen} AI, {Woosley} SE. 1999.
\newblock \textit{\apj} 524:262--289

\bibitem[{{Malesani} et~al.(2007){Malesani}, {Covino}, {D'Avanzo}, {D'Elia},
  {Fugazza} et~al.}]{mcd+07}
{Malesani} D, {Covino} S, {D'Avanzo} P, {D'Elia} V, {Fugazza} D, et~al. 2007.
\newblock \textit{\aap} 473:77--84

\bibitem[{{Mannucci} et~al.(2010){Mannucci}, {Cresci}, {Maiolino}, {Marconi} \&
  {Gnerucci}}]{mcm+10}
{Mannucci} F, {Cresci} G, {Maiolino} R, {Marconi} A, {Gnerucci} A. 2010.
\newblock \textit{\mnras} 408:2115--2127

\bibitem[{{Mannucci}, {Della Valle} \& {Panagia}(2006)}]{mdp06}
{Mannucci} F, {Della Valle} M, {Panagia} N. 2006.
\newblock \textit{\mnras} 370:773--783

\bibitem[{{Mannucci} et~al.(2005){Mannucci}, {Della Valle}, {Panagia},
  {Cappellaro}, {Cresci} et~al.}]{mdp+05}
{Mannucci} F, {Della Valle} M, {Panagia} N, {Cappellaro} E, {Cresci} G, et~al.
  2005.
\newblock \textit{\aap} 433:807--814

\bibitem[{{Maoz} \& {Badenes}(2010)}]{mb10}
{Maoz} D, {Badenes} C. 2010.
\newblock \textit{\mnras} 407:1314--1327

\bibitem[{{Maoz}, {Mannucci} \& {Brandt}(2012)}]{mmb12}
{Maoz} D, {Mannucci} F, {Brandt} TD. 2012.
\newblock \textit{\mnras} 426:3282--3294

\bibitem[{{Maoz} et~al.(2011){Maoz}, {Mannucci}, {Li}, {Filippenko}, {Della
  Valle} \& {Panagia}}]{mml+11}
{Maoz} D, {Mannucci} F, {Li} W, {Filippenko} AV, {Della Valle} M, {Panagia} N.
  2011.
\newblock \textit{\mnras} 412:1508--1521

\bibitem[{{Margutti} et~al.(2012){Margutti}, {Berger}, {Fong}, {Zauderer},
  {Cenko} et~al.}]{mbf+12}
{Margutti} R, {Berger} E, {Fong} W, {Zauderer} BA, {Cenko} SB, et~al. 2012.
\newblock \textit{\apj} 756:63

\bibitem[{{Margutti} et~al.(2011){Margutti}, {Chincarini}, {Granot},
  {Guidorzi}, {Berger} et~al.}]{mcg+11}
{Margutti} R, {Chincarini} G, {Granot} J, {Guidorzi} C, {Berger} E, et~al.
  2011.
\newblock \textit{\mnras} 417:2144--2160

\bibitem[{{Margutti} et~al.(2010){Margutti}, {Guidorzi}, {Chincarini},
  {Bernardini}, {Genet} et~al.}]{mgc+10}
{Margutti} R, {Guidorzi} C, {Chincarini} G, {Bernardini} MG, {Genet} F, et~al.
  2010.
\newblock \textit{\mnras} 406:2149--2167

\bibitem[{{Margutti} et~al.(2013){Margutti}, {Zaninoni}, {Bernardini},
  {Chincarini}, {Pasotti} et~al.}]{mzb+13}
{Margutti} R, {Zaninoni} E, {Bernardini} MG, {Chincarini} G, {Pasotti} F,
  et~al. 2013.
\newblock \textit{\mnras} 428:729--742

\bibitem[{{McGlynn} et~al.(2008){McGlynn}, {Foley}, {McBreen}, {Hanlon},
  {O'Connor} et~al.}]{mfm+08}
{McGlynn} S, {Foley} S, {McBreen} S, {Hanlon} L, {O'Connor} R, et~al. 2008.
\newblock \textit{\aap} 486:405--410

\bibitem[{{Meegan} et~al.(1992){Meegan}, {Fishman}, {Wilson}, {Horack}, {Brock}
  et~al.}]{mfw+92}
{Meegan} CA, {Fishman} GJ, {Wilson} RB, {Horack} JM, {Brock} MN, et~al. 1992.
\newblock \textit{\nat} 355:143--145

\bibitem[{{M{\'e}sz{\'a}ros}(2002)}]{mes02}
{M{\'e}sz{\'a}ros} P. 2002.
\newblock \textit{\araa} 40:137--169

\bibitem[{{Meszaros} \& {Rees}(1993)}]{mr93}
{Meszaros} P, {Rees} MJ. 1993.
\newblock \textit{\apjl} 418:L59

\bibitem[{{Meszaros} \& {Rees}(1997)}]{mr97}
{Meszaros} P, {Rees} MJ. 1997.
\newblock \textit{\apj} 476:232

\bibitem[{{Metzger} et~al.(2010{\natexlab{a}}){Metzger}, {Arcones}, {Quataert}
  \& {Mart{\'{\i}}nez-Pinedo}}]{maq+10}
{Metzger} BD, {Arcones} A, {Quataert} E, {Mart{\'{\i}}nez-Pinedo} G.
  2010{\natexlab{a}}.
\newblock \textit{\mnras} 402:2771--2777

\bibitem[{{Metzger} \& {Berger}(2012)}]{mb12}
{Metzger} BD, {Berger} E. 2012.
\newblock \textit{\apj} 746:48

\bibitem[{{Metzger}, {Kaplan} \& {Berger}(2013)}]{mkb13}
{Metzger} BD, {Kaplan} DL, {Berger} E. 2013.
\newblock \textit{\apj} 764:149

\bibitem[{{Metzger} et~al.(2010{\natexlab{b}}){Metzger},
  {Mart{\'{\i}}nez-Pinedo}, {Darbha}, {Quataert}, {Arcones} et~al.}]{mmd+10}
{Metzger} BD, {Mart{\'{\i}}nez-Pinedo} G, {Darbha} S, {Quataert} E, {Arcones}
  A, et~al. 2010{\natexlab{b}}.
\newblock \textit{\mnras} 406:2650--2662

\bibitem[{{Metzger}, {Quataert} \& {Thompson}(2008)}]{mqt08}
{Metzger} BD, {Quataert} E, {Thompson} TA. 2008.
\newblock \textit{\mnras} 385:1455--1460

\bibitem[{{Metzger} et~al.(1997){Metzger}, {Djorgovski}, {Kulkarni}, {Steidel},
  {Adelberger} et~al.}]{mdk+97}
{Metzger} MR, {Djorgovski} SG, {Kulkarni} SR, {Steidel} CC, {Adelberger} KL,
  et~al. 1997.
\newblock \textit{\nat} 387:878--880

\bibitem[{{Modjaz} et~al.(2008){Modjaz}, {Kewley}, {Kirshner}, {Stanek},
  {Challis} et~al.}]{mkk+08}
{Modjaz} M, {Kewley} L, {Kirshner} RP, {Stanek} KZ, {Challis} P, et~al. 2008.
\newblock \textit{\aj} 135:1136--1150

\bibitem[{{Nakar}(2007)}]{nak07}
{Nakar} E. 2007.
\newblock \textit{\physrep} 442:166--236

\bibitem[{{Nakar}, {Gal-Yam} \& {Fox}(2006)}]{ngf06}
{Nakar} E, {Gal-Yam} A, {Fox} DB. 2006.
\newblock \textit{\apj} 650:281--290

\bibitem[{{Nakar} \& {Piran}(2002)}]{np02}
{Nakar} E, {Piran} T. 2002.
\newblock \textit{\mnras} 330:920--926

\bibitem[{{Nakar} \& {Piran}(2011)}]{np11}
{Nakar} E, {Piran} T. 2011.
\newblock \textit{\nat} 478:82--84

\bibitem[{{Narayan}, {Paczynski} \& {Piran}(1992)}]{npp92}
{Narayan} R, {Paczynski} B, {Piran} T. 1992.
\newblock \textit{\apjl} 395:L83--L86

\bibitem[{{Nicuesa Guelbenzu} et~al.(2012{\natexlab{a}}){Nicuesa Guelbenzu},
  {Klose}, {Greiner}, {Kann}, {Kr{\"u}hler} et~al.}]{nkg+12}
{Nicuesa Guelbenzu} A, {Klose} S, {Greiner} J, {Kann} DA, {Kr{\"u}hler} T,
  et~al. 2012{\natexlab{a}}.
\newblock \textit{\aap} 548:A101

\bibitem[{{Nicuesa Guelbenzu} et~al.(2012{\natexlab{b}}){Nicuesa Guelbenzu},
  {Klose}, {Kr{\"u}hler}, {Greiner}, {Rossi} et~al.}]{nkk+12}
{Nicuesa Guelbenzu} A, {Klose} S, {Kr{\"u}hler} T, {Greiner} J, {Rossi} A,
  et~al. 2012{\natexlab{b}}.
\newblock \textit{\aap} 538:L7

\bibitem[{{Nicuesa Guelbenzu} et~al.(2011){Nicuesa Guelbenzu}, {Klose},
  {Rossi}, {Kann}, {Kr{\"u}hler} et~al.}]{nkr+11}
{Nicuesa Guelbenzu} A, {Klose} S, {Rossi} A, {Kann} DA, {Kr{\"u}hler} T, et~al.
  2011.
\newblock \textit{\aap} 531:L6

\bibitem[{{Norris} \& {Bonnell}(2006)}]{nb06}
{Norris} JP, {Bonnell} JT. 2006.
\newblock \textit{\apj} 643:266--275

\bibitem[{{Norris} et~al.(1984){Norris}, {Cline}, {Desai} \&
  {Teegarden}}]{ncd+84}
{Norris} JP, {Cline} TL, {Desai} UD, {Teegarden} BJ. 1984.
\newblock \textit{\nat} 308:434

\bibitem[{{Norris}, {Gehrels} \& {Scargle}(2010)}]{ngs10}
{Norris} JP, {Gehrels} N, {Scargle} JD. 2010.
\newblock \textit{\apj} 717:411--419

\bibitem[{{Norris}, {Gehrels} \& {Scargle}(2011)}]{ngs11}
{Norris} JP, {Gehrels} N, {Scargle} JD. 2011.
\newblock \textit{\apj} 735:23

\bibitem[{{Norris} et~al.(1991){Norris}, {Hertz}, {Wood} \&
  {Kouveliotou}}]{nhw+91}
{Norris} JP, {Hertz} P, {Wood} KS, {Kouveliotou} C. 1991.
\newblock \textit{\apj} 366:240--252

\bibitem[{{Nousek} et~al.(2006){Nousek}, {Kouveliotou}, {Grupe}, {Page},
  {Granot} et~al.}]{nkg+06}
{Nousek} JA, {Kouveliotou} C, {Grupe} D, {Page} KL, {Granot} J, et~al. 2006.
\newblock \textit{\apj} 642:389--400

\bibitem[{{Nuttall} \& {Sutton}(2010)}]{ns10}
{Nuttall} LK, {Sutton} PJ. 2010.
\newblock \textit{\prd} 82:102002

\bibitem[{{Nysewander}, {Fruchter} \& {Pe'er}(2009)}]{nfp09}
{Nysewander} M, {Fruchter} AS, {Pe'er} A. 2009.
\newblock \textit{\apj} 701:824--836

\bibitem[{{Ofek} et~al.(2007){Ofek}, {Cenko}, {Gal-Yam}, {Fox}, {Nakar}
  et~al.}]{ocg+07}
{Ofek} EO, {Cenko} SB, {Gal-Yam} A, {Fox} DB, {Nakar} E, et~al. 2007.
\newblock \textit{\apj} 662:1129--1135

\bibitem[{{Ofek} et~al.(2008){Ofek}, {Muno}, {Quimby}, {Kulkarni}, {Stiele}
  et~al.}]{omq+08}
{Ofek} EO, {Muno} M, {Quimby} R, {Kulkarni} SR, {Stiele} H, et~al. 2008.
\newblock \textit{\apj} 681:1464--1469

\bibitem[{{O'Shaughnessy}, {Belczynski} \& {Kalogera}(2008)}]{obk08}
{O'Shaughnessy} R, {Belczynski} K, {Kalogera} V. 2008.
\newblock \textit{\apj} 675:566--585

\bibitem[{{Paciesas} et~al.(2003){Paciesas}, {Briggs}, {Preece} \&
  {Mallozzi}}]{pbp+03}
{Paciesas} WS, {Briggs} MS, {Preece} RD, {Mallozzi} RS. 2003.
\newblock In \textit{Gamma-Ray Burst and Afterglow Astronomy 2001: A Workshop
  Celebrating the First Year of the HETE Mission}, eds. GR~{Ricker},
  RK~{Vanderspek}, vol. 662 of \textit{American Institute of Physics Conference
  Series}

\bibitem[{{Paczynski}(1986)}]{pac86}
{Paczynski} B. 1986.
\newblock \textit{\apjl} 308:L43--L46

\bibitem[{{Paczynski}(1995)}]{pac95}
{Paczynski} B. 1995.
\newblock \textit{\pasp} 107:1167

\bibitem[{{Paczynski} \& {Rhoads}(1993)}]{pr93}
{Paczynski} B, {Rhoads} JE. 1993.
\newblock \textit{\apjl} 418:L5

\bibitem[{{Palenzuela} et~al.(2013){Palenzuela}, {Lehner}, {Ponce}, {Liebling},
  {Anderson} et~al.}]{plp+13}
{Palenzuela} C, {Lehner} L, {Ponce} M, {Liebling} SL, {Anderson} M, et~al.
  2013.
\newblock \textit{Physical Review Letters} 111:061105

\bibitem[{{Panaitescu} \& {Kumar}(2002)}]{pk02}
{Panaitescu} A, {Kumar} P. 2002.
\newblock \textit{\apj} 571:779--789

\bibitem[{{Panaitescu}, {Kumar} \& {Narayan}(2001)}]{pkn01}
{Panaitescu} A, {Kumar} P, {Narayan} R. 2001.
\newblock \textit{\apjl} 561:L171--L174

\bibitem[{{Pedersen} et~al.(2005){Pedersen}, {El{\'{\i}}asd{\'o}ttir},
  {Hjorth}, {Starling}, {Cer{\'o}n} et~al.}]{peh+05}
{Pedersen} K, {El{\'{\i}}asd{\'o}ttir} {\'A}, {Hjorth} J, {Starling} R,
  {Cer{\'o}n} JMC, et~al. 2005.
\newblock \textit{\apjl} 634:L17--L20

\bibitem[{{Penner} et~al.(2012){Penner}, {Andersson}, {Jones}, {Samuelsson} \&
  {Hawke}}]{paj+12}
{Penner} AJ, {Andersson} N, {Jones} DI, {Samuelsson} L, {Hawke} I. 2012.
\newblock \textit{\apjl} 749:L36

\bibitem[{{Perley} et~al.(2009){Perley}, {Metzger}, {Granot}, {Butler},
  {Sakamoto} et~al.}]{pmg+09}
{Perley} DA, {Metzger} BD, {Granot} J, {Butler} NR, {Sakamoto} T, et~al. 2009.
\newblock \textit{\apj} 696:1871--1885

\bibitem[{{Perley} et~al.(2012){Perley}, {Modjaz}, {Morgan}, {Cenko}, {Bloom}
  et~al.}]{pmm+12}
{Perley} DA, {Modjaz} M, {Morgan} AN, {Cenko} SB, {Bloom} JS, et~al. 2012.
\newblock \textit{\apj} 758:122

\bibitem[{{Perna}, {Armitage} \& {Zhang}(2006)}]{paz06}
{Perna} R, {Armitage} PJ, {Zhang} B. 2006.
\newblock \textit{\apjl} 636:L29--L32

\bibitem[{{Perna} \& {Belczynski}(2002)}]{pb02}
{Perna} R, {Belczynski} K. 2002.
\newblock \textit{\apj} 570:252--263

\bibitem[{{Pian} et~al.(2006){Pian}, {Mazzali}, {Masetti}, {Ferrero}, {Klose}
  et~al.}]{pmm+06}
{Pian} E, {Mazzali} PA, {Masetti} N, {Ferrero} P, {Klose} S, et~al. 2006.
\newblock \textit{\nat} 442:1011--1013

\bibitem[{{Piran}(1992)}]{pir92}
{Piran} T. 1992.
\newblock \textit{\apjl} 389:L45--L48

\bibitem[{{Piran}(2004)}]{pir04}
{Piran} T. 2004.
\newblock \textit{Reviews of Modern Physics} 76:1143--1210

\bibitem[{{Piran}, {Nakar} \& {Rosswog}(2013)}]{pnr13}
{Piran} T, {Nakar} E, {Rosswog} S. 2013.
\newblock \textit{\mnras} 430:2121--2136

\bibitem[{{Piranomonte} et~al.(2008){Piranomonte}, {D'Avanzo}, {Covino},
  {Antonelli}, {Beardmore} et~al.}]{pdc+08}
{Piranomonte} S, {D'Avanzo} P, {Covino} S, {Antonelli} LA, {Beardmore} AP,
  et~al. 2008.
\newblock \textit{\aap} 491:183--188

\bibitem[{{Prieto}, {Stanek} \& {Beacom}(2008)}]{psb08}
{Prieto} JL, {Stanek} KZ, {Beacom} JF. 2008.
\newblock \textit{\apj} 673:999--1008

\bibitem[{{Prochaska} et~al.(2006){Prochaska}, {Bloom}, {Chen}, {Foley},
  {Perley} et~al.}]{pbc+06}
{Prochaska} JX, {Bloom} JS, {Chen} HW, {Foley} RJ, {Perley} DA, et~al. 2006.
\newblock \textit{\apj} 642:989--994

\bibitem[{{Proga} \& {Zhang}(2006)}]{pz06}
{Proga} D, {Zhang} B. 2006.
\newblock \textit{\mnras} 370:L61--L65

\bibitem[{{Qin} et~al.(1998){Qin}, {Wu}, {Chu}, {Fang} \& {Hu}}]{qwc98}
{Qin} B, {Wu} XP, {Chu} MC, {Fang} LZ, {Hu} JY. 1998.
\newblock \textit{\apjl} 494:L57

\bibitem[{{Qin} \& {Chen}(2013)}]{qc13}
{Qin} YP, {Chen} ZF. 2013.
\newblock \textit{\mnras} 430:163--173

\bibitem[{{Rees} \& {Meszaros}(1992)}]{rm92}
{Rees} MJ, {Meszaros} P. 1992.
\newblock \textit{\mnras} 258:41P--43P

\bibitem[{{Rezzolla} et~al.(2011){Rezzolla}, {Giacomazzo}, {Baiotti}, {Granot},
  {Kouveliotou} \& {Aloy}}]{rgb+11}
{Rezzolla} L, {Giacomazzo} B, {Baiotti} L, {Granot} J, {Kouveliotou} C, {Aloy}
  MA. 2011.
\newblock \textit{\apjl} 732:L6

\bibitem[{{Rhoads}(1999)}]{rho99}
{Rhoads} JE. 1999.
\newblock \textit{\apj} 525:737--749

\bibitem[{{Ricker} et~al.(2003){Ricker}, {Atteia}, {Crew}, {Doty}, {Fenimore}
  et~al.}]{rac+03}
{Ricker} GR, {Atteia} JL, {Crew} GB, {Doty} JP, {Fenimore} EE, et~al. 2003.
\newblock In \textit{Gamma-Ray Burst and Afterglow Astronomy 2001: A Workshop
  Celebrating the First Year of the HETE Mission}, eds. GR~{Ricker},
  RK~{Vanderspek}, vol. 662 of \textit{American Institute of Physics Conference
  Series}

\bibitem[{{Roberts} et~al.(2011){Roberts}, {Kasen}, {Lee} \&
  {Ramirez-Ruiz}}]{rkl+11}
{Roberts} LF, {Kasen} D, {Lee} WH, {Ramirez-Ruiz} E. 2011.
\newblock \textit{\apjl} 736:L21

\bibitem[{{Rosswog}(2005)}]{ros05}
{Rosswog} S. 2005.
\newblock \textit{\apj} 634:1202--1213

\bibitem[{{Rosswog}(2007)}]{ros07}
{Rosswog} S. 2007.
\newblock \textit{\mnras} 376:L48--L51

\bibitem[{{Rosswog} et~al.(2000){Rosswog}, {Davies}, {Thielemann} \&
  {Piran}}]{rdt00}
{Rosswog} S, {Davies} MB, {Thielemann} FK, {Piran} T. 2000.
\newblock \textit{\aap} 360:171--184

\bibitem[{{Rosswog} et~al.(1999){Rosswog}, {Liebend{\"o}rfer}, {Thielemann},
  {Davies}, {Benz} \& {Piran}}]{rlt+99}
{Rosswog} S, {Liebend{\"o}rfer} M, {Thielemann} FK, {Davies} MB, {Benz} W,
  {Piran} T. 1999.
\newblock \textit{\aap} 341:499--526

\bibitem[{{Rosswog}, {Piran} \& {Nakar}(2013)}]{rpn13}
{Rosswog} S, {Piran} T, {Nakar} E. 2013.
\newblock \textit{\mnras} 430:2585--2604

\bibitem[{{Rosswog}, {Ramirez-Ruiz} \& {Davies}(2003)}]{rrd03}
{Rosswog} S, {Ramirez-Ruiz} E, {Davies} MB. 2003.
\newblock \textit{\mnras} 345:1077--1090

\bibitem[{{Rosswog} \& {Ramirez-Ruiz}(2002)}]{rr02}
{Rosswog} S, {Ramirez-Ruiz}, E. 2003.
\newblock \textit{\mnras} 336:L7--L11

\bibitem[{{Rowlinson} et~al.(2013){Rowlinson}, {O'Brien}, {Metzger}, {Tanvir}
  \& {Levan}}]{rom+13}
{Rowlinson} A, {O'Brien} PT, {Metzger} BD, {Tanvir} NR, {Levan} AJ. 2013.
\newblock \textit{\mnras} 430:1061--1087

\bibitem[{{Rowlinson} et~al.(2010){Rowlinson}, {Wiersema}, {Levan}, {Tanvir},
  {O'Brien} et~al.}]{rwl+10}
{Rowlinson} A, {Wiersema} K, {Levan} AJ, {Tanvir} NR, {O'Brien} PT, et~al.
  2010.
\newblock \textit{\mnras} 408:383--391

\bibitem[{{Ruderman}(1975)}]{rud75}
{Ruderman} M. 1975.
\newblock In \textit{Seventh Texas Symposium on Relativistic Astrophysics},
  eds. PG~{Bergman}, EJ~{Fenyves}, L~{Motz}, vol. 262 of \textit{Annals of the
  New York Academy of Sciences}

\bibitem[{{Ruffert} \& {Janka}(2001)}]{rj01}
{Ruffert} M, {Janka} HT. 2001.
\newblock \textit{\aap} 380:544--577

\bibitem[{{Salvaterra} et~al.(2010){Salvaterra}, {Devecchi}, {Colpi} \&
  {D'Avanzo}}]{sdc+10}
{Salvaterra} R, {Devecchi} B, {Colpi} M, {D'Avanzo} P. 2010.
\newblock \textit{\mnras} 406:1248--1252

\bibitem[{{Sari}, {Piran} \& {Halpern}(1999)}]{sph99}
{Sari} R, {Piran} T, {Halpern} JP. 1999.
\newblock \textit{\apjl} 519:L17--L20

\bibitem[{{Sari}, {Piran} \& {Narayan}(1998)}]{spn98}
{Sari} R, {Piran} T, {Narayan} R. 1998.
\newblock \textit{\apjl} 497:L17

\bibitem[{{Savaglio}, {Glazebrook} \& {Le Borgne}(2009)}]{sgl09}
{Savaglio} S, {Glazebrook} K, {Le Borgne} D. 2009.
\newblock \textit{\apj} 691:182--211

\bibitem[{{Schmidt}(2001)}]{sch01}
{Schmidt} M. 2001.
\newblock \textit{\apjl} 559:L79--L82

\bibitem[{{Schmidt}, {Higdon} \& {Hueter}(1988)}]{shh88}
{Schmidt} M, {Higdon} JC, {Hueter} G. 1988.
\newblock \textit{\apjl} 329:L85--L87

\bibitem[{{Setiawan}, {Ruffert} \& {Janka}(2004)}]{srj04}
{Setiawan} S, {Ruffert} M, {Janka} HT. 2004.
\newblock \textit{\mnras} 352:753--758

\bibitem[{{Shemi} \& {Piran}(1990)}]{sp90}
{Shemi} A, {Piran} T. 1990.
\newblock \textit{\apjl} 365:L55--L58

\bibitem[{{Shibata} et~al.(2011){Shibata}, {Suwa}, {Kiuchi} \& {Ioka}}]{ssk+11}
{Shibata} M, {Suwa} Y, {Kiuchi} K, {Ioka} K. 2011.
\newblock \textit{\apjl} 734:L36

\bibitem[{{Shin} \& {Berger}(2007)}]{sb07}
{Shin} MS, {Berger} E. 2007.
\newblock \textit{\apj} 660:1146--1150

\bibitem[{{Soderberg} et~al.(2006{\natexlab{a}}){Soderberg}, {Berger},
  {Kasliwal}, {Frail}, {Price} et~al.}]{sbk+06}
{Soderberg} AM, {Berger} E, {Kasliwal} M, {Frail} DA, {Price} PA, et~al.
  2006{\natexlab{a}}.
\newblock \textit{\apj} 650:261--271

\bibitem[{{Soderberg} et~al.(2006{\natexlab{b}}){Soderberg}, {Kulkarni},
  {Nakar}, {Berger}, {Cameron} et~al.}]{skn+06}
{Soderberg} AM, {Kulkarni} SR, {Nakar} E, {Berger} E, {Cameron} PB, et~al.
  2006{\natexlab{b}}.
\newblock \textit{\nat} 442:1014--1017

\bibitem[{{Spergel} et~al.(2013){Spergel}, {Gehrels}, {Breckinridge},
  {Donahue}, {Dressler} et~al.}]{sgb+13}
{Spergel} D, {Gehrels} N, {Breckinridge} J, {Donahue} M, {Dressler} A, et~al.
  2013.
\newblock \textit{ArXiv}:1305.5425

\bibitem[{{Stanek} et~al.(1999){Stanek}, {Garnavich}, {Kaluzny}, {Pych} \&
  {Thompson}}]{sgk+99}
{Stanek} KZ, {Garnavich} PM, {Kaluzny} J, {Pych} W, {Thompson} I. 1999.
\newblock \textit{\apjl} 522:L39--L42

\bibitem[{{Stanek} et~al.(2006){Stanek}, {Gnedin}, {Beacom}, {Gould}, {Johnson}
  et~al.}]{sgb+06}
{Stanek} KZ, {Gnedin} OY, {Beacom} JF, {Gould} AP, {Johnson} JA, et~al. 2006.
\newblock \textit{Acta Astronomica} 56:333--345

\bibitem[{{Stanek} et~al.(2003){Stanek}, {Matheson}, {Garnavich}, {Martini},
  {Berlind} et~al.}]{smg+03}
{Stanek} KZ, {Matheson} T, {Garnavich} PM, {Martini} P, {Berlind} P, et~al.
  2003.
\newblock \textit{\apjl} 591:L17--L20

\bibitem[{{Stone}, {Loeb} \& {Berger}(2013)}]{slb13}
{Stone} N, {Loeb} A, {Berger} E. 2013.
\newblock \textit{\prd} 87:084053

\bibitem[{{Stratta} et~al.(2007){Stratta}, {D'Avanzo}, {Piranomonte}, {Cutini},
  {Preger} et~al.}]{sdp+07}
{Stratta} G, {D'Avanzo} P, {Piranomonte} S, {Cutini} S, {Preger} B, et~al.
  2007.
\newblock \textit{\aap} 474:827--835

\bibitem[{{Sullivan} et~al.(2006){Sullivan}, {Le Borgne}, {Pritchet},
  {Hodsman}, {Neill} et~al.}]{slp+06}
{Sullivan} M, {Le Borgne} D, {Pritchet} CJ, {Hodsman} A, {Neill} JD, et~al.
  2006.
\newblock \textit{\apj} 648:868--883

\bibitem[{{Svensson} et~al.(2010){Svensson}, {Levan}, {Tanvir}, {Fruchter} \&
  {Strolger}}]{slt+10}
{Svensson} KM, {Levan} AJ, {Tanvir} NR, {Fruchter} AS, {Strolger} LG. 2010.
\newblock \textit{\mnras} 405:57--76

\bibitem[{{Tanaka} \& {Hotokezaka}(2013)}]{th13}
{Tanaka} M, {Hotokezaka} K. 2013.
\newblock \textit{ArXiv}:1306.3742

\bibitem[{{Tanvir} et~al.(2013){Tanvir}, {Levan}, {Fruchter}, {Hjorth},
  {Hounsell} et~al.}]{tlf+13}
{Tanvir} NR, {Levan} AJ, {Fruchter} AS, {Hjorth} J, {Hounsell} RA, et~al. 2013.
\newblock \textit{Nature} 500:547--549

\bibitem[{{Taylor} et~al.(2004){Taylor}, {Frail}, {Berger} \&
  {Kulkarni}}]{tfb+04}
{Taylor} GB, {Frail} DA, {Berger} E, {Kulkarni} SR. 2004.
\newblock \textit{\apjl} 609:L1--L4

\bibitem[{{Th{\"o}ne} et~al.(2008){Th{\"o}ne}, {Fynbo}, {{\"O}stlin},
  {Milvang-Jensen}, {Wiersema} et~al.}]{tfo+08}
{Th{\"o}ne} CC, {Fynbo} JPU, {{\"O}stlin} G, {Milvang-Jensen} B, {Wiersema} K,
  et~al. 2008.
\newblock \textit{\apj} 676:1151--1161

\bibitem[{{Tremonti} et~al.(2004){Tremonti}, {Heckman}, {Kauffmann},
  {Brinchmann}, {Charlot} et~al.}]{thk+04}
{Tremonti} CA, {Heckman} TM, {Kauffmann} G, {Brinchmann} J, {Charlot} S, et~al.
  2004.
\newblock \textit{\apj} 613:898--913

\bibitem[{{Troja} et~al.(2008){Troja}, {King}, {O'Brien}, {Lyons} \&
  {Cusumano}}]{tko+08}
{Troja} E, {King} AR, {O'Brien} PT, {Lyons} N, {Cusumano} G. 2008.
\newblock \textit{\mnras} 385:L10--L14

\bibitem[{{Troja}, {Rosswog} \& {Gehrels}(2010)}]{trg10}
{Troja} E, {Rosswog} S, {Gehrels} N. 2010.
\newblock \textit{\apj} 723:1711--1717

\bibitem[{{Tsang} et~al.(2012){Tsang}, {Read}, {Hinderer}, {Piro} \&
  {Bondarescu}}]{trh+12}
{Tsang} D, {Read} JS, {Hinderer} T, {Piro} AL, {Bondarescu} R. 2012.
\newblock \textit{Physical Review Letters} 108:011102

\bibitem[{{Tsutsui} et~al.(2013){Tsutsui}, {Yonetoku}, {Nakamura}, {Takahashi}
  \& {Morihara}}]{tyn+13}
{Tsutsui} R, {Yonetoku} D, {Nakamura} T, {Takahashi} K, {Morihara} Y. 2013.
\newblock \textit{\mnras} 431:1398--1404

\bibitem[{{van den Bergh}(1983)}]{van83}
{van den Bergh} S. 1983.
\newblock \textit{\apss} 97:385--388

\bibitem[{{van den Bergh}(1990)}]{van90}
{van den Bergh} S. 1990.
\newblock \textit{\pasp} 102:1318--1320

\bibitem[{{van den Bergh} \& {Tammann}(1991)}]{vt91}
{van den Bergh} S, {Tammann} GA. 1991.
\newblock \textit{\araa} 29:363--407

\bibitem[{{van Eerten}, {Zhang} \& {MacFadyen}(2010)}]{vzm10}
{van Eerten} H, {Zhang} W, {MacFadyen} A. 2010.
\newblock \textit{\apj} 722:235--247

\bibitem[{{van Paradijs} et~al.(1997){van Paradijs}, {Groot}, {Galama},
  {Kouveliotou}, {Strom} et~al.}]{vgg+97}
{van Paradijs} J, {Groot} PJ, {Galama} T, {Kouveliotou} C, {Strom} RG, et~al.
  1997.
\newblock \textit{\nat} 386:686--689

\bibitem[{{van Paradijs}, {Kouveliotou} \& {Wijers}(2000)}]{vkw00}
{van Paradijs} J, {Kouveliotou} C, {Wijers} RAMJ. 2000.
\newblock \textit{\araa} 38:379--425

\bibitem[{{Villasenor} et~al.(2005){Villasenor}, {Lamb}, {Ricker}, {Atteia},
  {Kawai} et~al.}]{vlr+05}
{Villasenor} JS, {Lamb} DQ, {Ricker} GR, {Atteia} JL, {Kawai} N, et~al. 2005.
\newblock \textit{\nat} 437:855--858

\bibitem[{{Wainwright}, {Berger} \& {Penprase}(2007)}]{wbp07}
{Wainwright} C, {Berger} E, {Penprase} BE. 2007.
\newblock \textit{\apj} 657:367--377

\bibitem[{{Wang}, {Lai} \& {Han}(2006)}]{wlh06}
{Wang} C, {Lai} D, {Han} JL. 2006.
\newblock \textit{\apj} 639:1007--1017

\bibitem[{{Wang} et~al.(2013){Wang}, {Wang}, {Filippenko}, {Zhang} \&
  {Zhao}}]{wwf+13}
{Wang} X, {Wang} L, {Filippenko} AV, {Zhang} T, {Zhao} X. 2013.
\newblock \textit{Science} 340:170--173

\bibitem[{{Watson} et~al.(2006){Watson}, {Hjorth}, {Jakobsson}, {Xu}, {Fynbo}
  et~al.}]{whj+06}
{Watson} D, {Hjorth} J, {Jakobsson} P, {Xu} D, {Fynbo} JPU, et~al. 2006.
\newblock \textit{\aap} 454:L123--L126

\bibitem[{{Waxman}, {Kulkarni} \& {Frail}(1998)}]{wkf98}
{Waxman} E, {Kulkarni} SR, {Frail} DA. 1998.
\newblock \textit{\apj} 497:288

\bibitem[{{Wong}, {Willems} \& {Kalogera}(2010)}]{wwk10}
{Wong} TW, {Willems} B, {Kalogera} V. 2010.
\newblock \textit{\apj} 721:1689--1701

\bibitem[{{Woosley} \& {Bloom}(2006)}]{wb06}
{Woosley} SE, {Bloom} JS. 2006.
\newblock \textit{\araa} 44:507--556

\bibitem[{{Xue} et~al.(2008){Xue}, {Rix}, {Zhao}, {Re Fiorentin}, {Naab}
  et~al.}]{xrz+08}
{Xue} XX, {Rix} HW, {Zhao} G, {Re Fiorentin} P, {Naab} T, et~al. 2008.
\newblock \textit{\apj} 684:1143--1158

\bibitem[{{Yonetoku} et~al.(2004){Yonetoku}, {Murakami}, {Nakamura},
  {Yamazaki}, {Inoue} \& {Ioka}}]{ymn+04}
{Yonetoku} D, {Murakami} T, {Nakamura} T, {Yamazaki} R, {Inoue} AK, {Ioka} K.
  2004.
\newblock \textit{\apj} 609:935--951

\bibitem[{{Yost} et~al.(2003){Yost}, {Harrison}, {Sari} \& {Frail}}]{yhs+03}
{Yost} SA, {Harrison} FA, {Sari} R, {Frail} DA. 2003.
\newblock \textit{\apj} 597:459--473

\bibitem[{{Yu}, {Zhang} \& {Gao}(2013)}]{yzg13}
{Yu} YW, {Zhang} B, {Gao} H. 2013.
\newblock \textit{ArXiv}:1308.0876

\bibitem[{{Zhang}(2013)}]{zha13}
{Zhang} B. 2013.
\newblock \textit{\apjl} 763:L22

\bibitem[{{Zhang} et~al.(2007){Zhang}, {Zhang}, {Liang}, {Gehrels}, {Burrows}
  \& {M{\'e}sz{\'a}ros}}]{zzl+07}
{Zhang} B, {Zhang} BB, {Liang} EW, {Gehrels} N, {Burrows} DN,
  {M{\'e}sz{\'a}ros} P. 2007.
\newblock \textit{\apjl} 655:L25--L28

\bibitem[{{Zhang} et~al.(2009){Zhang}, {Zhang}, {Virgili}, {Liang}, {Kann}
  et~al.}]{zzv+09}
{Zhang} B, {Zhang} BB, {Virgili} FJ, {Liang} EW, {Kann} DA, et~al. 2009.
\newblock \textit{\apj} 703:1696--1724

\bibitem[{{Zheng} \& {Ramirez-Ruiz}(2007)}]{zr07}
{Zheng} Z, {Ramirez-Ruiz} E. 2007.
\newblock \textit{\apj} 665:1220--1226

\bibitem[{{Zrake} \& {MacFadyen}(2013)}]{zm13}
{Zrake} J, {MacFadyen} AI. 2013.
\newblock \textit{\apjl} 769:L29

\end{thebibliography}
\end{document}